\documentclass[final,5p,times,twocolumn]{elsarticle}

\usepackage{mathtools}
\usepackage{amsmath}    
\usepackage{graphicx}   
\usepackage{verbatim}   
\usepackage{color}      
\usepackage{caption}
\usepackage{lipsum}
\usepackage{subfigure}  
\usepackage{hyperref}   
\usepackage[section]{placeins} 
\usepackage{adjustbox}
\usepackage{soul}
\usepackage{enumitem}
\usepackage{flexisym}
\usepackage[compact]{titlesec}

\begin{document}
\let\today\relax
\makeatletter
\def\ps@pprintTitle{%
	\let\@oddhead\@empty
	\let\@evenhead\@empty
	\def\@oddfoot{\footnotesize\itshape
		{} \hfill\today}%
	\let\@evenfoot\@oddfoot
}
\makeatother

\begin{frontmatter}
\title{Anomalous diffusion in single and coupled standard maps with extensive chaotic phase spaces}


\author[1]{Henok Tenaw Moges\fnref{firstfoot}}
\author[2]{Thanos Manos\fnref{secondfoot}}
\author[1]{Charalampos Skokos\fnref{thirdfoot}}

\address[1]{Nonlinear Dynamics and Chaos group, Department of Mathematics and Applied Mathematics, \\ University of Cape Town, Rondebosch, 7701, Cape Town, South Africa}
\address[2]{Laboratoire de Physique Th{\'e}orique et Mod{\'e}lisation, CY Cergy Paris Universit{\'e}, CNRS, UMR 8089, 95302 Cergy-Pontoise cedex, France}
\fntext[firstfoot]{mgshen002@myuct.ac.za}
\fntext[secondfoot]{thanos.manos@cyu.fr}
\fntext[thirdfoot]{haris.skokos@uct.ac.za}

\begin{abstract}
We investigate the long-term diffusion transport and chaos properties of single and coupled standard maps. We consider model parameters that are known to induce anomalous diffusion in the maps' phase spaces, as opposed to normal diffusion which is associated with Gaussian distribution properties of the kinematic variables. This type of transport originates in the presence of the so-called accelerator modes, i.e.~non-chaotic initial conditions which exhibit ballistic transport, which also affect the dynamics in their vicinity. We first systematically study the dynamics of single standard maps, investigating the impact of different ensembles of initial conditions on their behavior and asymptotic diffusion rates, as well as on the respective time-scales needed to acquire these rates. We consider sets of initial conditions in chaotic regions enclosing accelerator modes, which are not bounded by invariant tori. These types of chaotic initial conditions  typically lead to normal diffusion transport. We then setup different arrangements of coupled standard maps and  investigate their global diffusion properties and chaotic dynamics. Although individual maps bear accelerator modes causing anomalous transport, the global diffusion behavior of the coupled system turns out to depend on the specific configuration of the imposed coupling. Estimating the average diffusion properties for ensembles of initial conditions, as well as measuring the strength of chaos through computations of appropriate indicators, we find conditions and systems' arrangements  which systematically favor the suppression of anomalous transport and long-term convergence to normal diffusion rates.
\end{abstract}


\end{frontmatter}


\section{Introduction}\label{sec:intro}

\noindent The investigation of diffusion and transport phenomena in conservative Hamiltonian systems and area-preserving symplectic maps is a topic of intense scientific research for more than 50 years now (see e.g.~\cite{Chirikov1979,RechWhi1980PRL,RechRosWhi1981PRA,CarMeiBha1981PRA,MCGCKA1983PhyD,Kar1983PhyD,MacMeiPer1984aPRL,MacMeiPer1984bPhyD,HorHatIshMor1990PTP,IshHorKobMor1991PTP,OucMorHorMor1991PTP,MorOkaTom1991PTP,Stef_etal1998PRE,Kroetz2016}). In nonequilibrium statistical mechanics macroscopic properties of matter emerge from microscopic chaotic motion, such as the motion of single atoms or molecules (see e.g.~\cite{KlagesBook2007} and references therein). Transport analysis in dynamical systems aims to relate the collective motion of ensembles of trajectories between regions of phase space, with several different physical phenomena, such as, the determination of chemical reaction rates, the mixing rates in a fluid, and the particle confinement times in accelerator or fusion plasma devices (see \cite{Meiss2015} for a recent review).

The classical standard map (SM) \cite{Chirikov1979}, also known as the Chirikov map or the kicked rotor system, is a well-known and studied  model which exhibits a rather rich diffusive dynamics depending on the value of the so-called kick-strength control parameter $K$ (more information is provided in Sect.~\ref{sec:model} below). The system describes the motion of a rotor modeled via the time evolution of two variables, respectively its angle and angular momentum, which typically are bounded in the interval $[0, 2 \pi)$ as both variables are considered to be modulo $2 \pi$. Depending on the value of the kick-strength $K$, the angular momentum variable can become `random' (uncorrelated) performing a Brownian/Gaussian random walk. When chaos dominates the phase space of the system, the motion transport occurs in a fashion termed as `normal diffusion' in the literature (see e.g.~\cite{AltmannKantzRevChap2008}), with the related transport rate referring to the Gaussian distribution of a characteristic quantity of the system, whose variance increases as time evolves.

However, there are intervals of $K$ values where the diffusion deviates from `normal' and the transport of motion evolves by following an `anomalous' rate associated with the presence of the so-called accelerator modes (AMs) \cite{Chirikov1979}. The latter ones are regular regions (islands of stability) surrounding stable periodic orbits of different periods in the model's (compact) phase space. There the dynamics produces leaps (equal to $2\pi$ or integer multiples of $2\pi$) in the angular momentum variable and orbits get transported infinitely in the de-compactified phase space (i.e.~when the modulo requirement is relaxed for the angular momentum). All orbits starting with initial conditions (ICs) inside an AM island are evolving in a ballistic manner, which is linear in time and its diffusion rate is larger than the one observed for normal diffusive transport. Another intermediate in rate (between normal and ballistic) type of diffusion, which is referred to as superdiffusion, can be observed when ICs happen to lie in the vicinity of the AMs' region or get near them and become trapped as time evolves. In such cases orbits experience the so-called effect of stickiness (see e.g.~\cite{Dvorak1998} and references therein) due to surrounding cantori and exhibit anomalous diffusion. On the other hand, orbits whose ICs lie within ordinary islands of stability do not diffuse and exhibit what is termed as subdiffusion (see e.g.~\cite{ZasEde2000Chaos,Veneg2007PRL,Veneg2008PRE,Veneg2008PRL}).

A rather systematic analysis of the general diffusion processes in the SM for a large set of  parameter values generating phase spaces of different diffusion transport processes (normal and anomalous) due to the presence of AMs of different period, was performed in \cite{ManRob2014PRE}. That analysis was accompanied by detailed stability maps (where regular and chaotic regions were identified), a description of the momentum distribution with L\'evy stable distributions, as well as numerical estimations of the diffusion rates. That work was motivated by studies on the impact of the classical SM on the quantized version of the map and its wave function's localization properties \cite{ManRob2013PRE,BatManRob2013}, and was extended in \cite{ManRob2015PRE} by mainly focusing on quantum (short) characteristic time scales, such as the Heisenberg and localization times.

The global/local transport and the related mean diffusion exponents in the case of a single SM in the presence of AMs were investigated in \cite{HC2018}. Furthermore, in that work the respective time scales of orbits to escape the initial ballistic motion and converge to normal diffusion, as a function of their distance from the AM island in the phase space, were also studied. That work was later extended in \cite{HKC2019} where a detailed comparison between the Lyapunov time, the Poincare recurrence time and the escape time was performed. In addition, in \cite{CincottaGiordanoCMDA2018}, the authors investigated the presence of phase correlations using the Shannon entropy to distinguish between strong correlations in the time evolution associated with anomalous diffusion, while in \cite{DiazUnPub2020} mechanisms for controlling the escape of orbits in the SM were studied. A generalized SM, the so-called rational SM, was considered in \cite{CincottaSimoPhysD2020} (see references therein regarding the model's introduction), where an extensive study and comparison between the different dynamical and diffusion trends appearing in the SM were performed.

Although most works focused on the behavior of a single SM, some studies on higher dimensional maps have also been conducted. For example, processes exhibiting strong chaos and anomalous diffusion in coupled SMs have were studied in \cite{AltmannKantzEPL2007}. There the existence of an enhanced trapping regime induced by trajectories performing a random walk inside the area corresponding to regular islands of the uncoupled maps was found. Nevertheless, the main focus of that paper was on the dynamics of regions with relatively large size islands of stability, as well as on the determination of diffusion time scales for orbits trapped by invariant tori. In \cite{ANTONOPOULOS2016} the authors studied the diffusion of weak vs strong chaotic motion in multi-dimensional McMillan coupled maps and demonstrated similarities between such types of symplectic maps and the dynamical properties of disordered Klein–Gordon Hamiltonian systems.

Apart from studies of transport phenomena in systems related to the SM, similar investigations have been performed also for models more directly associated with physical problems. For instance, in \cite{SatoKlagesPRL2019} a study was carried out on the type of dynamics and anomalous diffusion properties in systems where particles are evolved under the influence of a combined chaotic dynamical system generating Brownian motion-like diffusion and a non-chaotic system in which all particles localize. Recently, complex transport properties have also become  relevant in studies of electron motion in low-dimensional  nanosystems describing important physical structures, like for example molecular graphene (see e.g.~\cite{GilGallegosEPJST2018} and references therein). More specifically, the authors of that work  investigated the energy-dependent diffusion in a soft periodic Lorentz gas with Fermi potentials  where macroscopic transport emerges from microscopic chaos and identified different diffusion transport rates at different dynamical regimes.

In this paper, we set out to investigate the long-term (asymptotic) diffusion transport properties of ensembles of ICs in the presence of AMs of different periods in their respective phase space neighborhoods for coupled SMs, extending in some sense the work of \cite{ManRob2014PRE}. To this end, we first begin with a rather systematic and exhaustive investigation of the characteristic diffusion trends and properties in the respective uncoupled system (which practically corresponds to a single SM), which possess AMs. We are mainly interested in studying global averaged diffusion rates of the whole phase space and not just diffusion properties of single ICs. We mostly focus on cases with kick-strength  values for which the phase space of the single SM is almost globally dominated by chaotic orbits with some small stable islands surrounding AMs. This analysis provides us quite detailed information on the independent dynamics of each map of the coupled system, as well as it allows us to better understand the dissimilarities in the diffusion rates and time-scales that different ensembles of ICs require to converge to their respective long-term (asymptotic) diffusion rate value. Such time scales are found to be a function of the relative fraction of chaotic vs stable areas surrounding an AM.

We then study a system of coupled SMs whose individual SM setups are similar to the ones studied in the first part of our work. Initially, all coupled SMs have the same  kick-strength parameter while later on, we also explore SM arrangements of different kick-strengths per map. For these arrangements we investigate (i) the role of the coupling-strength and (ii) the impact of the choice of the ensembles of ICs (different fractions of chaotic regions around the AMs) on each SM, on the global diffusion rates and on the time scales to converge to their long-term values. Finally, we associate these diffusion time scales with the nature of the underlying dynamics, which is quantified through measurements of the extent of the chaotic and regular components. More specifically, we obtain average characterizations of the nature of ICs' ensembles by implementing some well-known and extensively used chaos indicators like the Maximum Lyapunov Exponent (MLE) \cite{benettin:1980a,benettin:1980b,S2010} and the Generalized Alignment Index (GALI) method \cite{SBA2007,SkoBouAntEPJST2008,MSA2012,SM2016}. These chaos detection techniques have already been successfully used in studies of the chaoticity of coupled SMs \cite{MSA2012,BMC2009}.

The paper is organized as follows: in Sect.~\ref{sec:model}, we introduce the SM model and the relevant diffusion rate measurements, while in Sect.~\ref{sec:NumTech} we provide the definitions of the chaos detection techniques used throughout this work. Our results are presented in Sect.~\ref{sec:results} while Sect.~\ref{sec:sum} summarizes our main findings.

\section{Models and diffusion measures}\label{sec:model}

\noindent The two-dimensional (2D) SM is defined by the equations:
\begin{equation}
\begin{aligned}
  x_{n+1} &= x_n + y_{n+1}, \\
  y_{n+1} &= y_n + K\sin(x_n),
\end{aligned}
\label{eq:sm}
\end{equation}
where the variables $x_n$ and $y_n$ are both considered in modulo $2 \pi$ and $n=0,\,1, \,2, \ldots$, counts the number of discrete time steps or in other words the map's iterations. This map represents the evolution of the rotation angle $x_n$ and the angular momentum $y_n$  of a periodically `kicked' pendulum rotating in a free field, where the kick by a nonlinear force takes place at each time unit. The constant $K$ measures the kick's nonlinear intensity. The variables $x_n$, $y_n$ and the parameter $K$ are dimensionless.

Despite the fact that SM is a relatively simple model, analogous to an autonomous Hamiltonian system with 2 degrees of freedom, it exhibits rather rich chaotic and regular dynamics. For $K=0$, the SM is linear and as such only quasiperiodic and periodic motion occurs in the system (depending on the initial value $y_0$ of the angular momentum), filling respectively the 2D phase plane with horizontal straight lines or series of a finite number (equal to the periodicity $p$ of the periodic orbit) of distinct points. For $K > 0$ some of these lines break into pairs of isolated points, corresponding to periodic orbits, half of which are stable and half unstable \cite{Chirikov1979}. The former are surrounded by families of quasiperiodic orbits (closed curves around these stable periodic orbits), while the latter ones are saddle points surrounded by thin chaotic areas. As the $K > 0$ value increases, these chaotic areas grow in size and for relatively large $K$ values tend to occupy the whole phase plane.

The quantification of the SM's diffusion process can be done through the relation (see e.g.~\cite{Chirikov1979,Meiss2015}):
\begin{flalign} \label{eq:yvar}
\langle(\Delta y)^2\rangle = D_{\mu}(K) \, n^{\mu},
\end{flalign}
where $\Delta y = y_{n}-y_0$, with $y_0$ being the IC of the orbit's angular momentum $y_n$. The quantity $\langle(\Delta y)^2\rangle$ represents the average of $(\Delta y)^2$ at each iteration $n$, over an ensemble of ICs. The numerically estimated diffusion exponent $\mu$ is in the interval $[0,2]$ and $D_{\mu}(K)$, for $n\rightarrow \infty$, is the generalized classical diffusion coefficient. When $\mu=1$ we observe \textit{normal diffusion}, and $D_1(K)$ is the so-called normal diffusion coefficient. In cases of anomalous diffusion we encounter two types of behaviors: \textit{subdiffusion} if $0 < \mu < 1$  and \textit{superdiffusion} when $1 <\mu \le 2$, while obviously $\mu=0$ signifies the lack of diffusion. In the most extreme case where $\mu=2$, \textit{ballistic transport} occurs which is strictly associated  with the presence and dominance of AMs in the dynamics of the system.

The theoretically estimated value of $D_1(K)$ for normal diffusion ($\mu=1$) can be found in \cite{I1990} to be:
\begin{equation} \label{eq:Dcl}
 D_{1}(K)=
\begin{cases}
 \frac{K^2}{2}  \left\{ 1- 2J_2(K) \left[ 1-J_2(K) \right] \right\}, & \text{if } \, K \ge 4.5, \\
 0.30(K-K_{cr})^3, & \text{if } \, K_{cr} < K < 4.5,
\end{cases}
\end{equation}
where $K_{cr}\approx 0.9716$ and $J_2(K)$ is a Bessel function. This theoretical result about the dependence of the diffusion coefficient on $K$ fails around the period $p=1$ AM existence intervals:
\begin{equation} \label{eq:acmdint}
2\pi l \le K \le \sqrt{(2\pi l)^2 +16 },
\end{equation}
with $l$ being any positive integer. For example, in the interval obtained from \eqref{eq:acmdint} for $l=1$ there exist two \textit{stable periodic orbits} located at $y=0,\; x= \pi -x^*$ and $y=0,\; x = \pi +x^*$, along with two \textit{unstable periodic orbits} at $y=0,\; x = x^*$ and $y=0,\; x = 2\pi - x^*$, where  $x^* = \arcsin (2\pi/K)$.

In order to characterize the different rate of diffusion, we calculate the \textit{effective} diffusion coefficient \cite{Chirikov1979,Meiss2015,ManRob2014PRE}
\begin{equation}\label{eq:Deff_sm}
D_{\rm eff}=\frac{\langle(\Delta y)^2\rangle}{n},
\end{equation}
numerically for large but finite values of $n$, although the coefficient is theoretically defined for $n \rightarrow \infty$. We note that the quantity $D_{\rm eff}$, in general, is not equal to the coefficient $D_{\mu}$ defined in \eqref{eq:yvar} (see also \cite{ManRob2014PRE} for more details and references therein). For the 2D  SM \eqref{eq:sm} the dependence of the quantity $D_{\rm eff}$  on the value of $K$ has been already presented in many previous papers, like for example in Figs.~1 of \cite{ManRob2014PRE,HC2018}. In those plots we see peaks in the values of $D_{\rm eff}$ at certain intervals of $K$ values, where AMs are predicted to be present. It is worth noting that these values deviate substantially from those predicted in \eqref{eq:Dcl}, as the theoretical prediction is based on the assumption of normal diffusion rate.

In our study we also consider a system of $N$ coupled SMs, leading to a $2N$D map, described by the following equations \cite{KG1988}:
\begin{flalign}\label{eq:csm}
&x^j_{n+1} = x^j_n + y^j_{n+1} \nonumber,\\
&y^j_{n+1} = y^j_n + K_j \sin(x^j_n) - \beta [\sin(x^{j+1}_n - x^j_n) + \sin(x^{j-1}_n - x^j_n)],
\end{flalign}
where $j=1,2,\dots, N$ is the index of each SM, $K_j$ its respective nonlinear kick-strength parameter, and $\beta$ the coupling-strength parameter between neighboring maps. The choice of each $K_j$ value plays a pivotal role since it initializes each  map at the dynamical state of the respective 2D SM \eqref{eq:sm} before the coupling takes place. This in turn is related to the initial amount of regular and chaotic motion, as well as the presence of AMs in its respective phase plane.  In our investigations, we consider periodic boundary conditions, i.e.~$x^0_n=x^{N}_n$ and $x^{N+1}_n=x^1_n$, a choice which preserves the symplectic nature of the coupled system.

In order to investigate the diffusion process for the $N$ coupled SMs, we use a generalized expression similar to Eq.~(\ref{eq:Deff_sm}), defining (as a limit for $n \rightarrow \infty$) the effective diffusion coefficient of the $2N$D map as:
\begin{flalign} \label{Deff_csm}
D_{\rm eff}^N =\frac{1}{n} \sum\limits_{j=1}^{N}\left< (y^j_{n+1}-y^j_0)^2 \right> = \frac{1}{n} \sum\limits_{j=1}^{N}\left< (\Delta y^j)^2 \right>.
\end{flalign}
In this expression the sum is over the total number of maps and each map has the same number of ICs, but the location of each ensemble of ICs may vary  between the 2D SMs.

\section{Chaos measure techniques}\label{sec:NumTech}

\noindent Let us now discuss the main numerical techniques we implement in our study, starting with a brief presentation of the two chaos detection methods we use, namely the MLE and the GALI, emphasizing their application to area-preserving, symplectic maps. In our presentation we refer to the more general case of the $2N$D map \eqref{eq:csm}, which, for simplicity, can be expressed as:
\begin{equation}\label{eq:map_gen}
\mathbf{x}_{n+1}=F(\mathbf{x}_n),
\end{equation}
with $\mathbf{x}_n$ denoting the vector of the map's coordinates at iteration $n$ and $F(\mathbf{x}_n)$ the functional form of the map, i.e.~the expressions at the right hand side of \eqref{eq:csm}.

The regular or chaotic nature of the map's orbits is determined by the evolution of small perturbations to these orbits. Such perturbation defines the so-called deviation vector $\mathbf{w}_n$. The propagation of this vector is governed  by the system's  tangent map:
\begin{equation}\label{eq:w_map}
 \mathbf{w}_{n+1}=
\frac{\partial F}{\partial\mathbf{x}} (\mathbf{x}_n)\cdot\mathbf{w}_n.
\end{equation}

The MLE $\lambda$ is the average expansion (or shrinking) rate of the distance between two neighboring orbits in the phase space of a dynamical system (quantified in a first order approximation by the length of the deviation vector) and it is computed as \cite{benettin:1980a,benettin:1980b,S2010}:
\begin{flalign}\label{eq:MLE}
  \lambda &= \lim\limits_{n\to \infty} \Lambda(n),
\end{flalign}
where $\Lambda(n)$ is the so-called finite-time MLE (ftMLE) given by:
\begin{flalign}\label{eq:ftMLE}
\Lambda(n) &= \frac{1}{n}\ln \frac{\Vert \mathbf{w}_n \Vert}{\Vert \mathbf{w}_0 \Vert},
\end{flalign}
with $\mathbf{w}_0$ and $\mathbf{w}_n$ being respectively deviation vectors from the given orbit at iterations $n=0$ and $n>0$,  and $\| \cdot \|$ denoting any vector norm (in our study we implement the usual Euclidean norm). In the case of chaotic orbits $\lambda$ is positive, while it is zero for regular orbits. Since it is impractical to actually compute $\lambda$ \eqref{eq:MLE}, as it is obtained through a limit when the number of iterations $n$ goes to infinity, we rely on computations of the ftMLE \eqref{eq:ftMLE}. In practice, for chaotic orbits $\Lambda(n)$ tends to a positive value, while it goes to zero as $\Lambda(n) \propto \frac{\ln{n}}{n} \sim n^{-1}$ for regular orbits.

Although the MLE has been extensively used as a chaos indicator one practical problem it often faces is the potential slow convergence of $\Lambda(n)$ to its limiting $\lambda$ value. Over the years several methods which overcome this problem have been developed like the Smaller Alignment Index (SALI) \cite{SkoJPhA2001,SABV2003,SkoAntBouVra2004JPhA} and its generalization the GALI, the Fast Lyapunov Indicator (FLI) \cite{FLG1997,FCL1997,FL2000,LGF2016} and its variants \cite{B2005,B2006,B2016}, the mean exponential growth of nearby orbits (MEGNO) \cite{CS2000,CGS2003,CG2016} and the `0-1' test \cite{GM2004,GM2005,GM2016} to name a few (for a collection of review papers on various modern chaos detection techniques the reader is referred to \cite{SGL2016}). Here we will implement the GALI method, which has been proven to be very efficient and has already been successfully applied in studies of various dynamical systems (see e.g.~\cite{BMC2009,MA2011,MR2011,MBS2013,CPPSM2017}).

For $2N$D maps  the GALI of order $k$ (GALI$_k$), $2 \leq k \leq 2N$, is determined through the evolution of $k$ initially linearly independent deviation vectors $\mathbf{w}_0^i$, $i=1,2,\ldots,k$. To avoid overflow problems, the resulting deviation vectors $\mathbf{w}_n^i$, $i=1,2,\ldots,k$, are continuously normalized, but their directions are kept intact. Then, according to \cite{SBA2007} GALI$_k$ is defined as the volume of the $k$-parallelogram having as edges the $k$ unit deviation vectors $\hat{\mathbf{w}}_n^i=\mathbf{w}_n^i/ \|\mathbf{w}_n^i \|$, $i=1,2,\ldots,k$, determined through the wedge product of these vectors as:
\begin{equation}\label{eq:GALI}
{\rm GALI}_k(n)=\| \hat{\mathbf{w}}_n^1\wedge \hat{\mathbf{w}}_n^2\wedge \cdots \wedge\hat{\mathbf{w}}_n^k \|,
\end{equation}
with $\| \cdot \|$ denoting the usual Euclidean norm. From this definition it is evident that if at least two of the deviation vectors become linearly dependent, the wedge product in \eqref{eq:GALI} becomes zero and GALI$_k$ vanishes. As mentioned earlier, the GALI method is a generalization of the SALI technique, which practically coincides with GALI$_2$ \cite{SBA2007}.

The behavior of the GALI$_k$ for regular and chaotic orbits was theoretically studied in \cite{SBA2007,SkoBouAntEPJST2008}, where it was shown that all GALI$_k(n)$ tend exponentially to zero for chaotic orbits, with exponents which depend on the first $k$ Lyapunov exponents of the orbit. On the other hand, in the case of regular orbits, GALI$_k(n)$ remains practically constant and positive if $k$ is smaller or equal to the dimensionality of the torus on which the motion occurs, otherwise, it decreases to zero following a well-defined power law decay.

\begin{figure*}[ht] \centering
\includegraphics[width=0.33\textwidth,keepaspectratio]{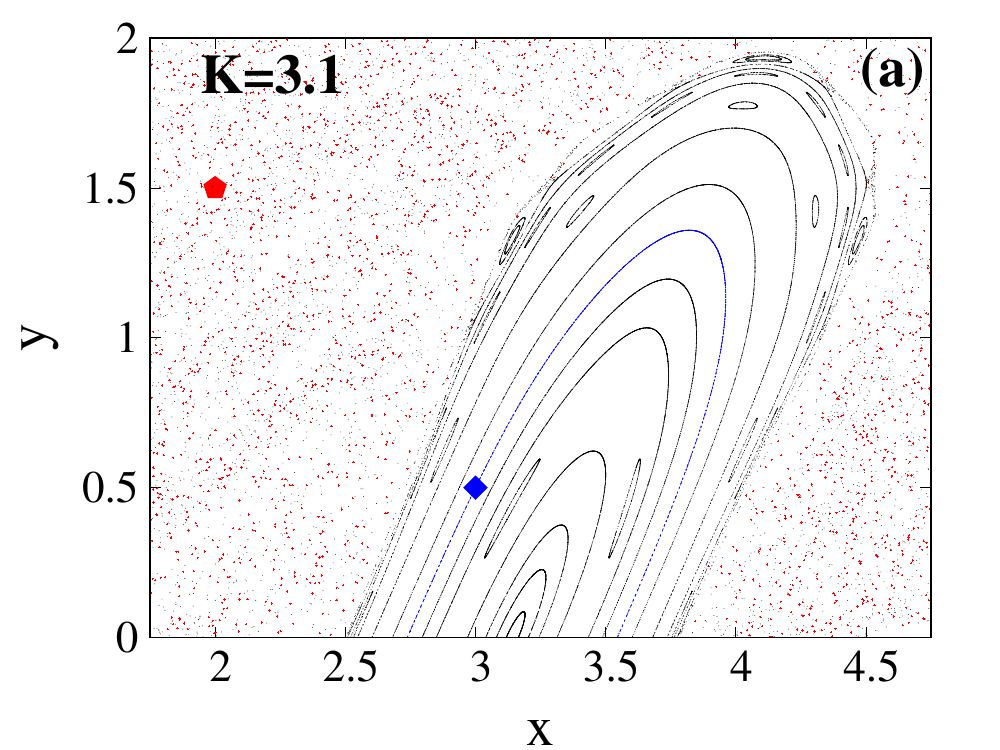}
\includegraphics[width=0.33\textwidth,keepaspectratio]{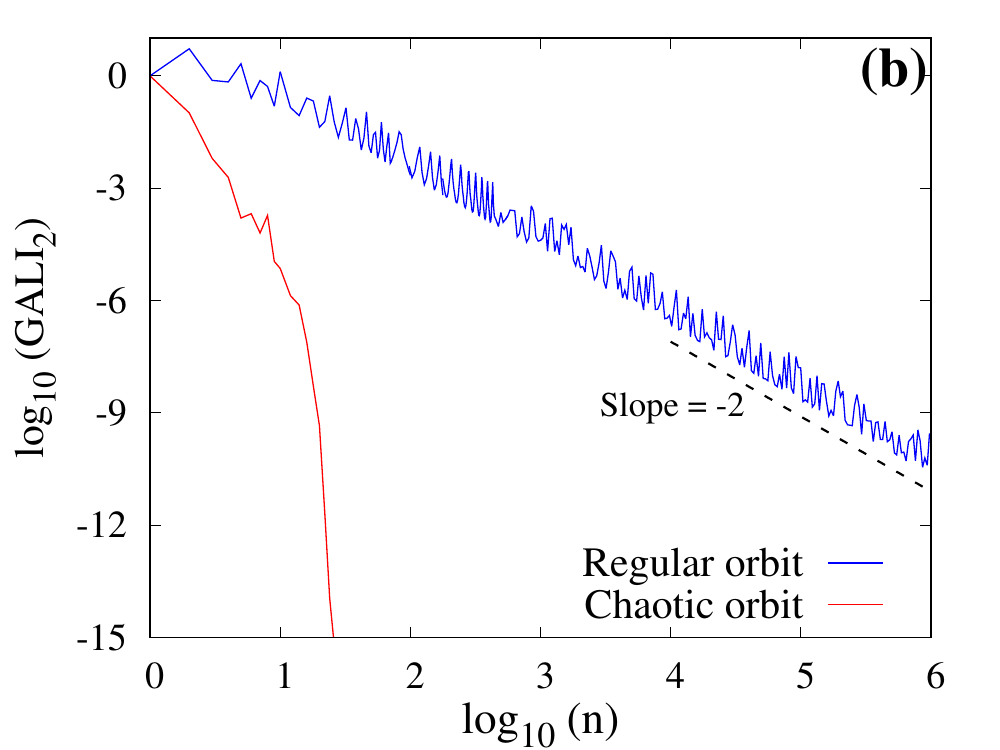}
\includegraphics[width=0.33\textwidth,keepaspectratio]{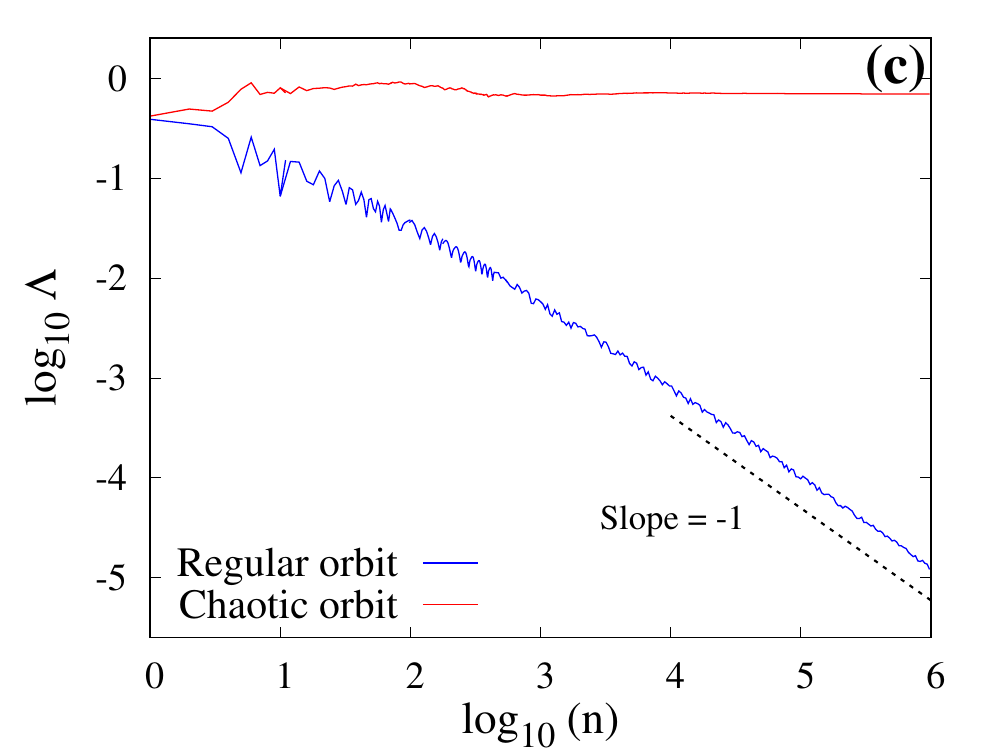} \\
\includegraphics[width=0.33\textwidth,keepaspectratio]{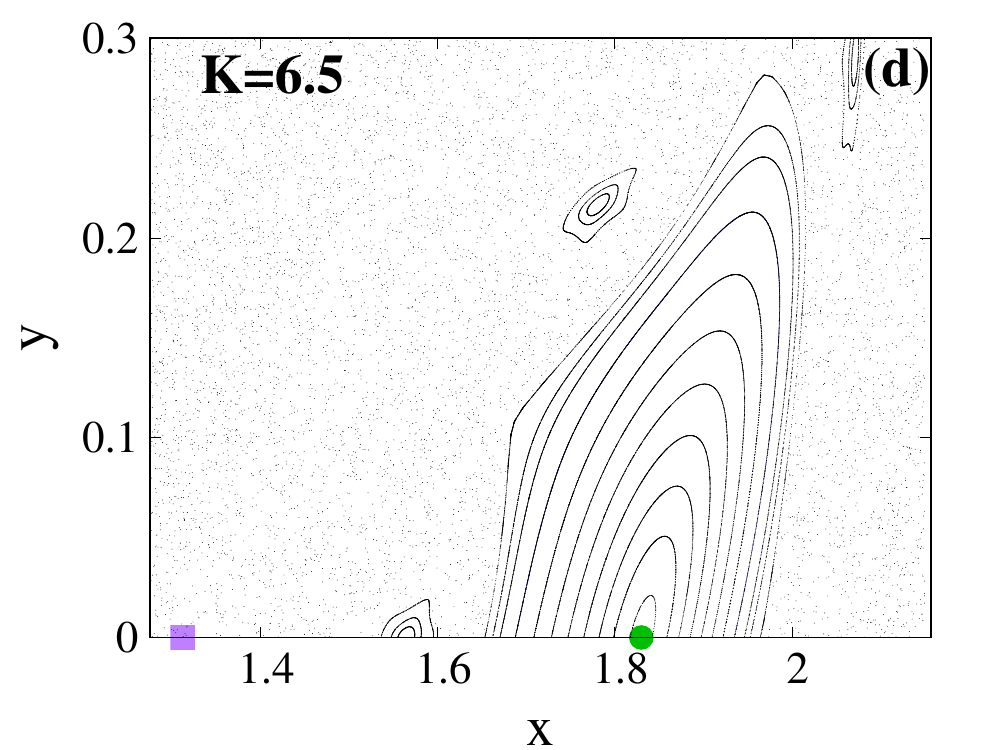}
\includegraphics[width=0.33\textwidth,keepaspectratio]{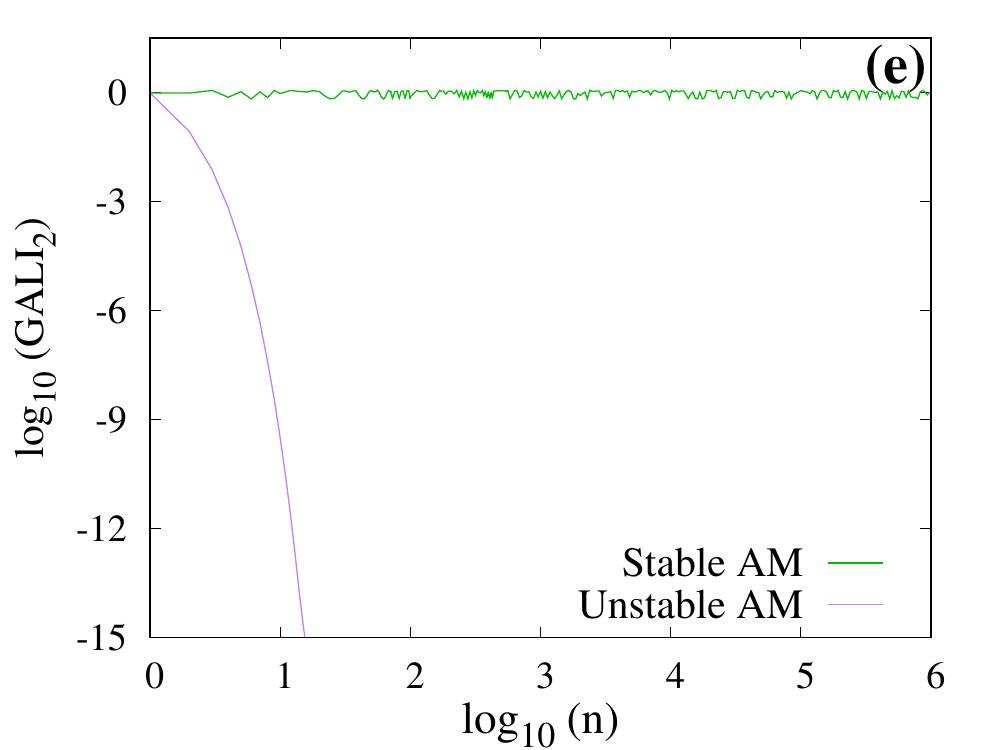}
\includegraphics[width=0.33\textwidth,keepaspectratio]{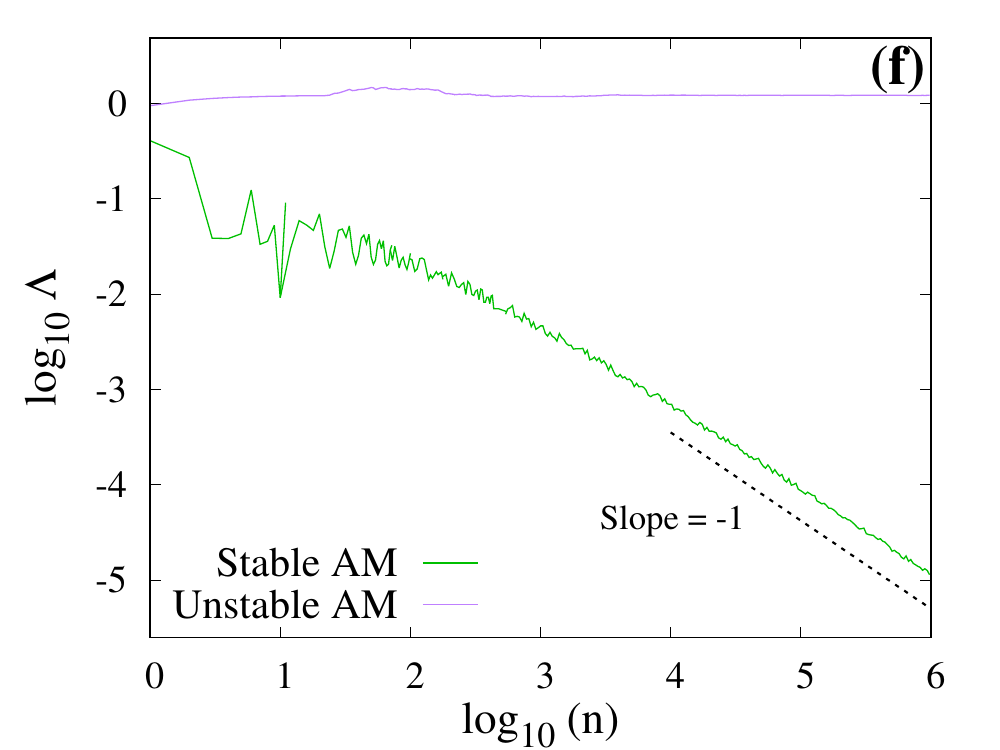}
\caption{Results for the 2D SM \eqref{eq:sm} with [(a), (b) and (c)] $K=3.1$ and [(d), (e) and (f)] $K=6.5$. In (a) and (d) phase space portraits of the SM are shown, which are produced by $10^6$ iterations of several ICs (black points), while colored points are used to identify particular orbits. In (a) the consequents of the chaotic orbit with IC $(x_0,y_0)=(2, 1.5)$ (red pentagon point) are plotted in red, while blue points  correspond to a regular orbit with IC $(x_0,y_0)=(3, 0.5)$ (blue diamond point). In (d) the location of two AMs of period $p=1$ are identified, a stable AM (green circle point) at $(x_0,y_0)=(1.8298, 0)$ and an unstable AM (purple square point) at $(x_0,y_0)=(1.3118, 0)$. The evolution of the GALI$_2$ of all these orbits, with respect to the number of iterations $n$, is illustrated in (b) for the chaotic (red curve) and the regular (blue curve) orbit of (a), and in (e) for the stable (green curve) and the unstable AM (purple curve) of (d). In (b) GALI$_2 (n)$ tends to zero exponentially fast for the chaotic orbit, while it decreases to zero following a power law GALI$_2 \propto n^{-2}$ (indicated by a dashed line) for the regular orbit. In (e) GALI$_2 (n)$ remains practically constant for the stable AM, while it tends to zero exponentially fast for the unstable AM. In (c) and (f) we see the evolution of the ftMLE $\Lambda(n)$ \eqref{eq:ftMLE}, respectively for the chaotic (red curve) and the regular (blue curve) orbit of (a), and for the stable (green curve) and the unstable (purple curve) AM of (d). $\Lambda(n)$ saturates to a positive value in (c) and (f), respectively for the chaotic orbit and the unstable AM, while it tends to zero following a power law $\Lambda(n) \propto n^{-1}$ (indicated by a dotted  line  in both panels) for the regular orbit and the stable AM.}
\label{fig:1}
\end{figure*}

In the particular case of 2D maps, like the SM \eqref{eq:sm}, GALI$_2$ (which is the only GALI that can be defined in this case) tends to zero both for regular and for chaotic orbits, following however completely different time rates, which allow us to distinguish between the two cases \cite{SkoJPhA2001}. More specifically, in the case of a chaotic orbit, any two deviation vectors will be aligned to the direction defined by the MLE $\lambda$, and consequently GALI$_2$ will eventually tend to zero following an exponential decay of the form GALI$_2(n) \propto e^{-2\lambda n}$. In the case of regular orbits, any two deviation vectors tend to fall on the tangent space of the torus on which the motion lies. For a 2D map, this torus is a 1D invariant curve, whose tangent space is also 1D and consequently any two deviation vectors will become aligned. Thus, even in the case of regular orbits in 2D maps  GALI$_2$ tends to zero. This decay follows the power law  GALI$_2(n) \propto 1/n^2$ \cite{SkoJPhA2001,MSA2012}. It is worth noting that, similarly to  chaotic orbits, the GALI$_2(n)$ tends exponentially to zero for unstable periodic orbits, while it remains practically constant, fluctuating around a positive value, for stable periodic orbits \cite{MSA2012}.

The behavior of the GALI$_2$ for all these different types of orbits is seen in Figs.~\ref{fig:1}(b) and (e). More specifically in Fig.~\ref{fig:1}(a) the phase space portrait of SM \eqref{eq:sm} for $K=3.1$ is depicted, where  a chaotic (red points) and a regular (blue points) orbit are also identified. The evolution of the GALI$_2$ for these orbits is shown in Fig.~\ref{fig:1}(b) with the index exhibiting an exponential decay for the chaotic orbit (red curve) and a power law decrease proportional to $n^{-2}$ (indicated by the dashed line) for the regular orbit (blue curve). From the results of Fig.~\ref{fig:1}(c) we see that the ftMLE \eqref{eq:ftMLE} tends to zero for the regular orbit (blue curve), eventually following a $\Lambda(n) \propto n^{-1}$ law (indicated by the dotted line), while it saturates to a positive value for the chaotic orbit (red curve). From the comparison of Figs.~\ref{fig:1}(b) and (c) it is worth noting how much faster the evolution of the GALI$_2$ allows the clear distinction between the regular and chaotic behavior of orbits, with respect to the ftMLE. More specifically, we see that after $n\approx 50$ iterations  the GALI$_2$ of the chaotic orbit becomes practically zero (GALI$_2 \approx 10^{-15}$) attaining values several orders of magnitude smaller than in the case of the regular orbit, while at the same time a clear distinction between the two orbits cannot be securely established from the results of  Fig.~\ref{fig:1}(c), as for example the ftMLE of the regular orbit has not manifested a clear and well-defined tendency to decrease.

In Fig.~\ref{fig:1}(d) the phase space location of a stable (green circle point) and an unstable (purple square point) AM for the 2D SM \eqref{eq:sm} with $K=6.5$ is shown. In accordance to \cite{MSA2012} the GALI$_2$ of the stable AM remains practically constant [green curve in Fig.~\ref{fig:1}(e)], while for the unstable AM [purple curve in Fig.~\ref{fig:1}(e)] it decreases exponentially fast to zero, showing a similar behavior to the one of the chaotic orbit in Fig.~\ref{fig:1}(b), as in the vicinity of the unstable AM chaotic motion occurs. In Fig.~\ref{fig:1}(f) we see that the ftMLE  $\Lambda$ for the stable (green curve) and the unstable (purple curve) AM behaves respectively as in the case of the regular and chaotic orbit of Fig.~\ref{fig:1}(c). We note that for the calculation of the GALI$_2$ and the   ftMLE we impose the modulo $2 \pi$ requirement in order to avoid numerical problems due to the divergence of orbits to infinity.

\section{Results}
\label{sec:results}

\noindent In this study, we set out to ultimately investigate the rate of diffusion and the respective dynamics of ensembles of orbits in coupled SMs. However, and before doing that, we will perform an extensive study of the dynamics of the single SM \eqref{eq:sm} focusing on the effects of AMs (of different periods $p$) in the long term dynamics and transport properties of various sets of ICs.

\subsection{Single standard map} \label{sec:2D_results}

\begin{figure}[h!] \centering
\includegraphics[width=0.85\columnwidth,keepaspectratio]{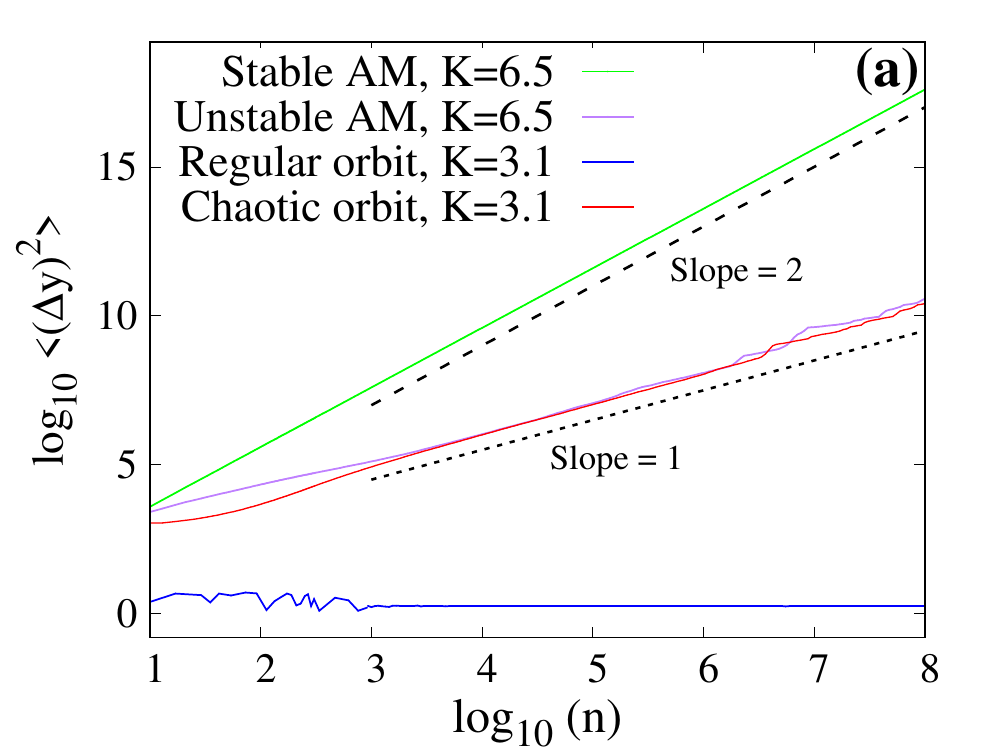}\\
\includegraphics[width=0.85\columnwidth,keepaspectratio]{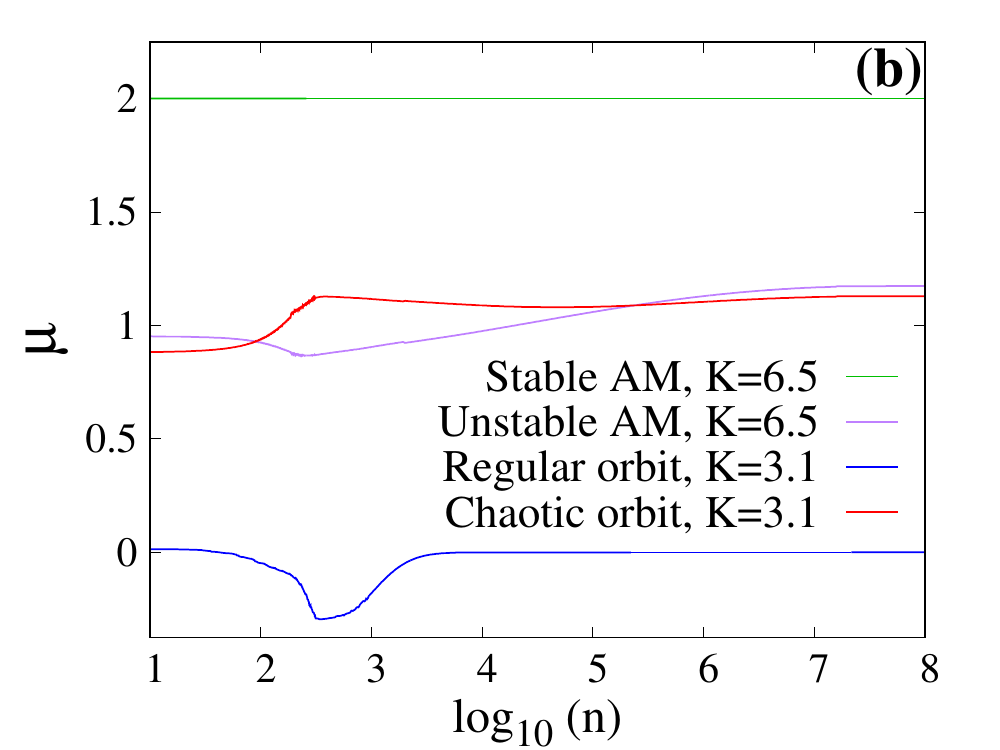}
\caption{(a) The variance $\langle(\Delta y)^2\rangle$ \eqref{eq:yvar} of the angular momentum $y_n$ of the 2D SM \eqref{eq:sm} as a function of the map's iterations $n$, for orbits in four different small regions of the map's phase space. For each region a set of ICs on a $315\times315$ grid of equally spaced  mesh points (i.e.~in total $\approx100,000$ ICs) are taken and the variance is obtained as average over all these orbits. For $K=3.1$ ICs in the region $[2.995,3.005]\times[0.495,0.505]$ (i.e.~$2.995 \leq x_0 \leq 3.005$ and $0.495 \leq y_0 \leq 0.505$) and $[1.995,2.005]\times[1.495,1.505]$ containing respectively the IC of the regular (blue curve) and the chaotic (red curve) orbit of Fig.~\ref{fig:1}(a) are considered, while the green and purple curves respectively correspond to regions $[1.825,1.835]\times[0.0,0.01]$ [containing the stable AM of Fig.~\ref{fig:1}(d)] and $[1.295,1.315] \times [0.0,0.01]$ [surrounding the unstable AM of Fig.~\ref{fig:1}(d)] for the $K=6.5$ SM. The dashed and dotted lines respectively indicate functions proportional to $n^{2}$ and $n^{1}$. Note that the red and purple curves overlap. (b) The numerically computed diffusion exponent $\mu$ of \eqref{eq:yvar} for the curves of (a), where the same color for each case is used.}
\label{fig:2}
\end{figure}

\noindent In order to perform the numerical evaluation of the diffusion exponent $\mu$ of \eqref{eq:yvar} for a specific region of the SM's phase space we consider a grid of $315 \times 315$ equally spaced ICs (i.e.~$\approx100,000$ ICs) and calculate the evolution of the variance $\langle(\Delta y)^2\rangle$ [Eq.~\eqref{eq:yvar}] as a function of the map's iterations $n$. In Fig.~\ref{fig:2}(a) we present the outcome of this approach for sets of orbits with ICs in small regions around the four particular orbits considered in Fig.~\ref{fig:1}. More specifically we perform this computation for a small neighborhood of the regular (blue curve) and the chaotic orbit (red curve) of   Fig.~\ref{fig:1}(a), as well as for regions containing the IC of the unstable (purple curve) and the stable AM (green curve) of Fig.~\ref{fig:1}(d).

From the results of Fig.~\ref{fig:2}(a) we see that ballistic transport appears in the region around the stable AM as the increase of $\langle(\Delta y)^2\rangle$ (green curve) is well described by a function proportional to $n^2$ (dashed line), i.e.~having a diffusion exponent $\mu=2$. On the other hand, the dynamics around the unstable AM (purple curve) and the chaotic orbit (red curve) practically leads to normal diffusion ($\mu=1$) as both curves seem to follow quite well a $\langle(\Delta y)^2\rangle \propto n$ increase represented by a dotted line, while no diffusion is observed in the immediate neighborhood of the considered regular orbit (blue curve). Note that typical regular orbits and stable AMs have similar  properties from a dynamical point of view, see e.g., the evolution of the ftMLEs given by blue and purple curves in Figs.~\ref{fig:1}(c) and (f), however their diffusion rates are different as are respectively presented in Figs.~\ref{fig:2}(a) and (b). The former ones do not practically diffuse ($\mu \approx 0$) while the latter ones diffuse with extreme rate values ($\mu \approx 2$).

The value of $\mu$ can be numerically estimated by performing a linear fitting of the logarithm of $\langle(\Delta y)^2\rangle$  with respect to $\log_{10}n$, but in order to identify potential changes in the dynamical evolution of the ensembles of ICs, which will be reflected in the value of the diffusion exponent, we prefer to  monitor the evolution of the $\mu$ value as $n$ increases by first smoothing the variance of the momentum values through a locally weighted regression smoothing algorithm \cite{CD1988}. Then by implementing the numerical approach followed in \cite{LBKSF2010,BLKSF2011} we compute $\mu$ in a running window in $\log_{10}n$, whose width is much smaller than the total length of the system's evolution. Implementing the latter approach to the results presented in Fig.~\ref{fig:2}(a) we create Fig.~\ref{fig:2}(b), where we clearly observe the presence of ballistic transport ($\mu=2$, green curve) and practically normal diffusion ($\mu \approx 1$, purple and red curves), as well as the absence of it ($\mu=0$, blue curve). From the presented results we see that in all cases no significant changes in the evolution of $\mu$ are observed, because after some transient phase up to $n \approx 10^3$ (apart from the ballistic case where $\mu$ remains constant from the beginning) the values of $\mu$ do not notably vary, as can also be deduced from the results of Fig.~\ref{fig:2}(a). Nevertheless, as we will see below [e.g.~in Fig.~\ref{fig:4}(b) and the last column of Fig.~\ref{fig:7}] plots similar to Fig.~\ref{fig:2}(b) will be extremely helpful in identifying the possible alterations in the diffusion and transport properties of sets of ICs, as well as in determining the number of iterations needed for $\mu$ to approach its asymptotic $\mu=2$ value.

\begin{figure*}[h!] \centering
\includegraphics[width=0.85\columnwidth]{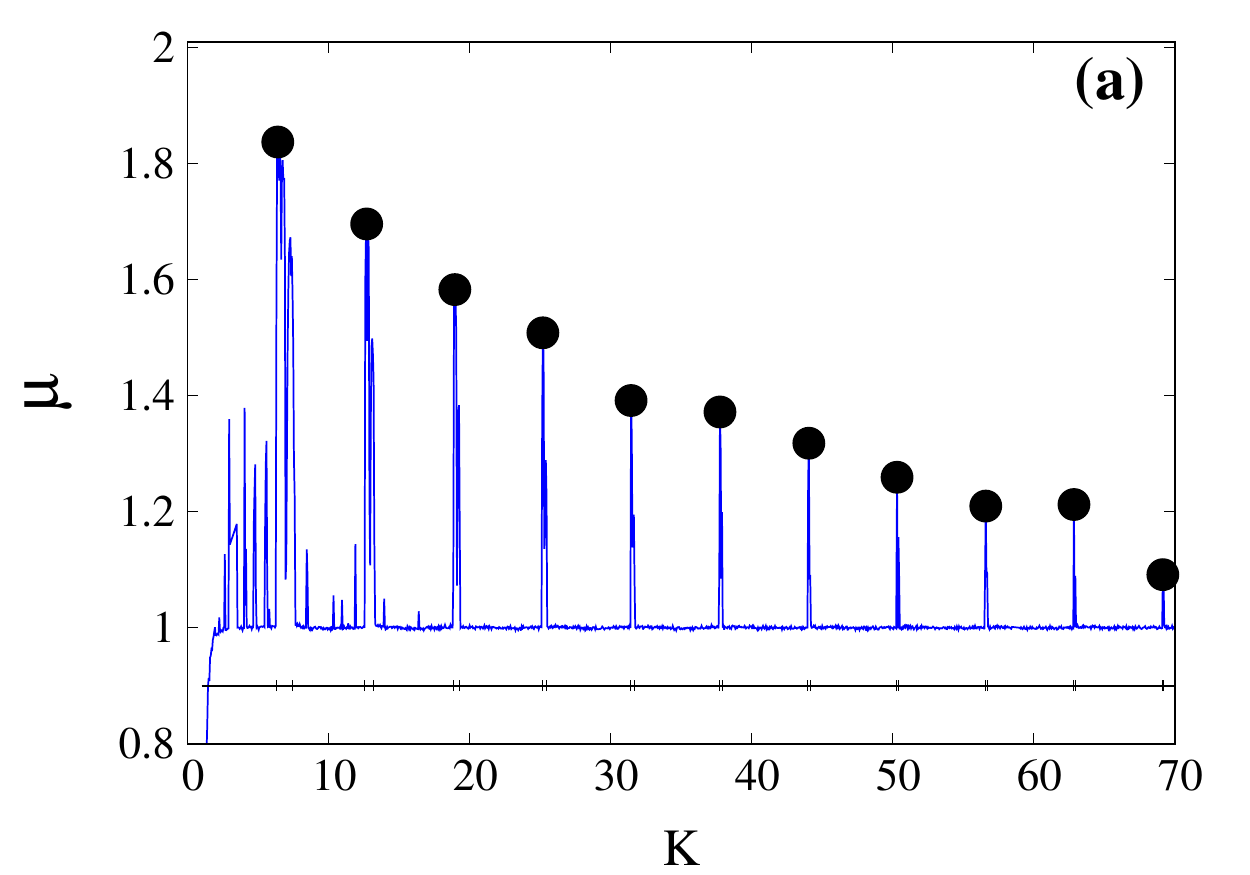}
\includegraphics[width=0.85\columnwidth]{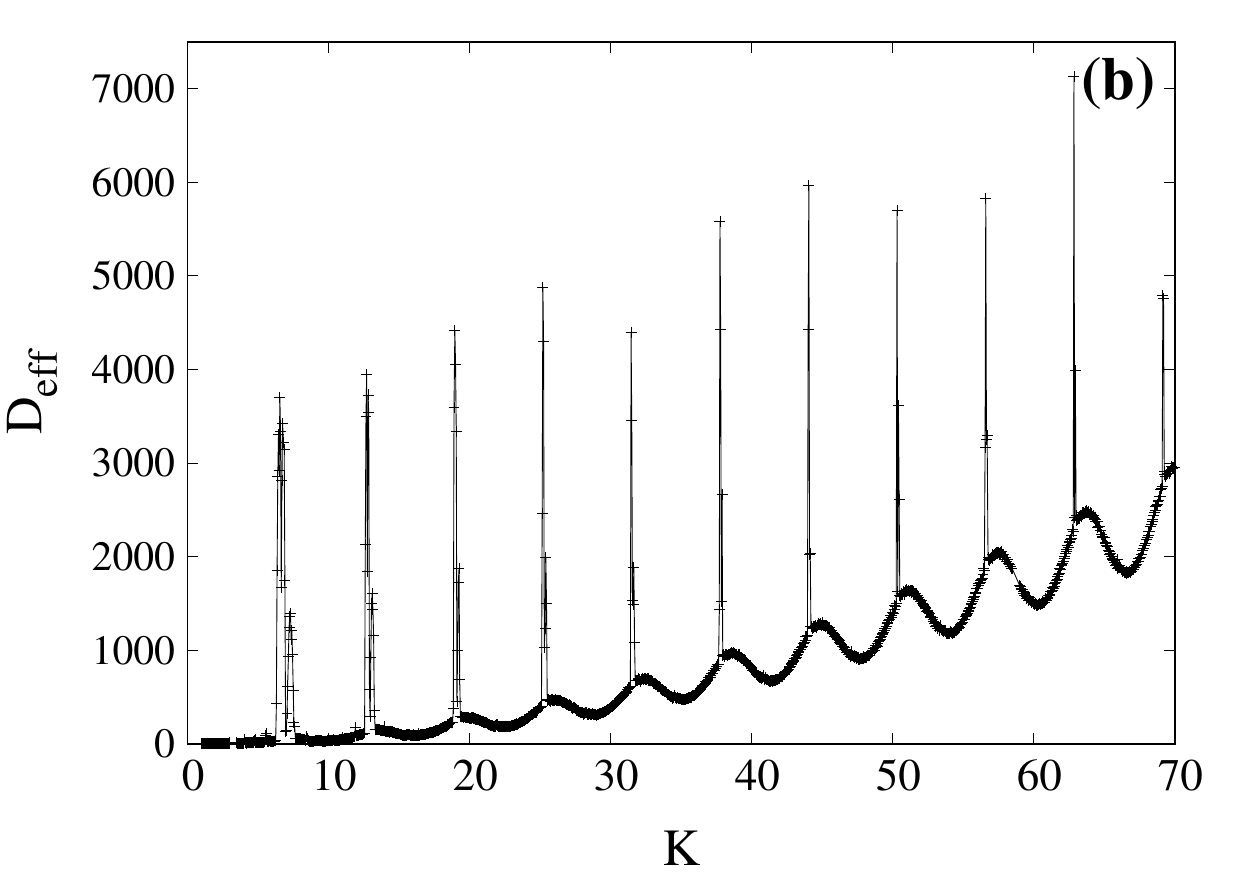}
\includegraphics[width=0.85\columnwidth]{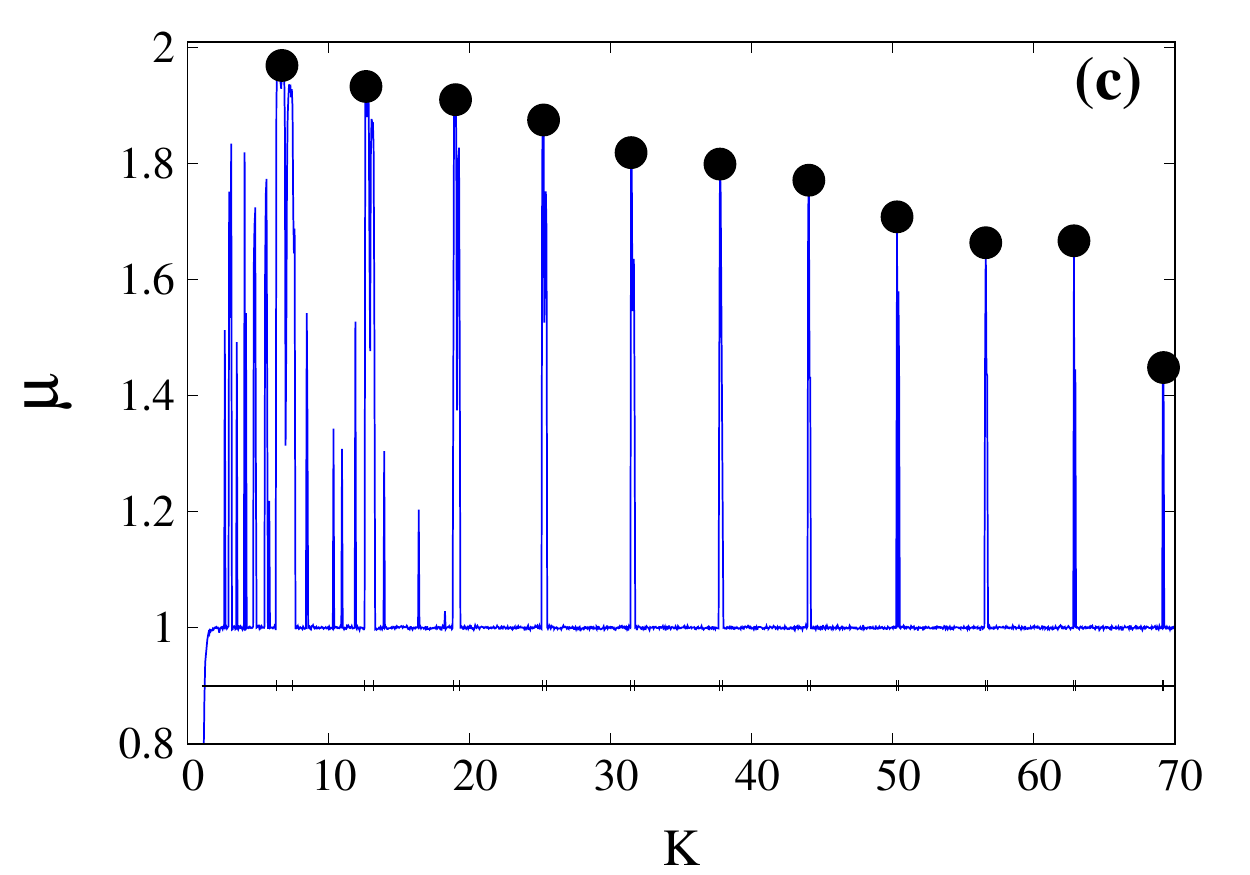}
\includegraphics[width=0.85\columnwidth]{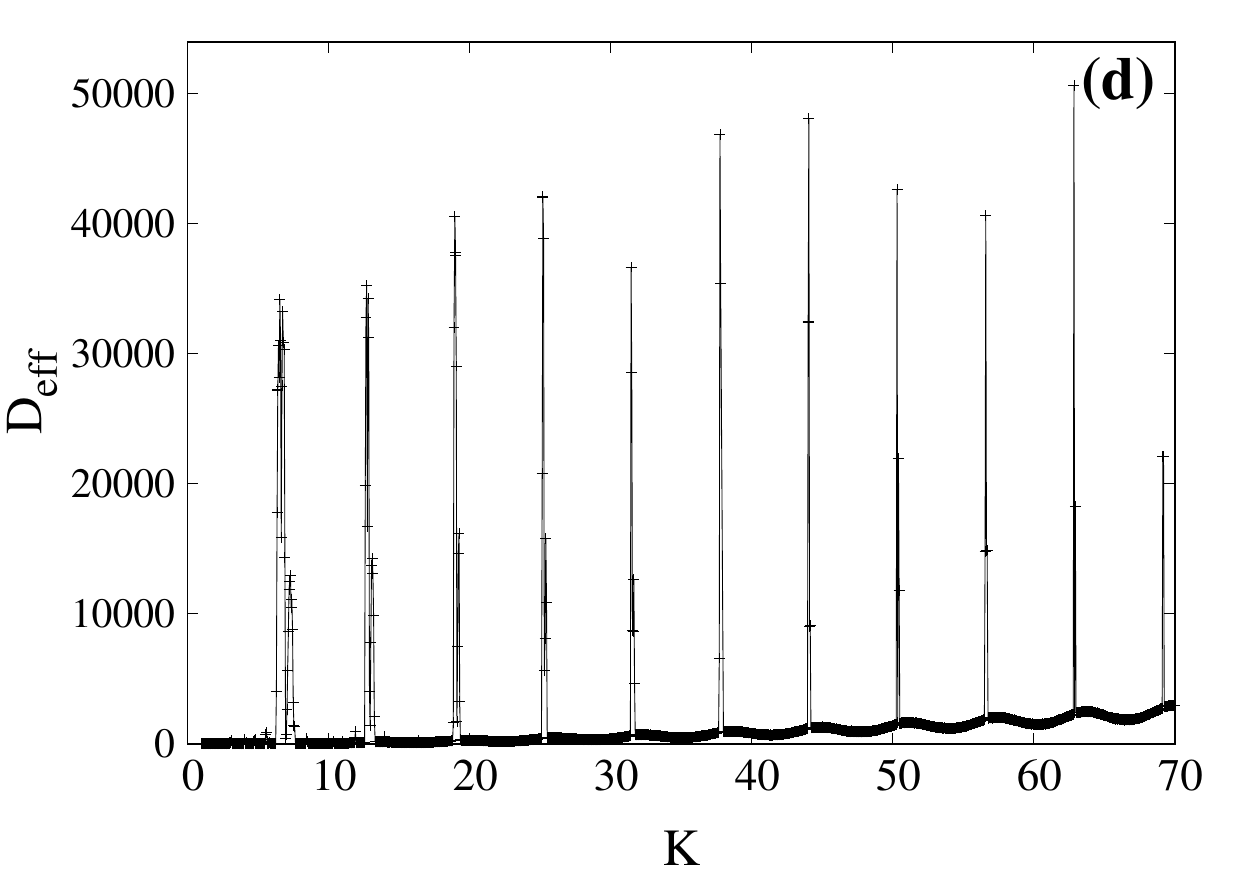}
\caption{Diffusion exponent $\mu$ of \eqref{eq:yvar} [(a) and (c)] and effective diffusion coefficient $D_{\rm eff}$ \eqref{eq:Deff_sm} [(b) and (d)] as a function of $K$ after $n=10^4$ [(a) and (b)] and $n=10^5$ [(c) and (d)] iterations of the 2D SM \eqref{eq:sm} for a set of $315 \times 315$  ICs on a grid of equally spaced mesh points ($\approx100,000$ ICs) on the map's entire phase space $[0, 2 \pi) \times [0, 2 \pi)$. The intervals on the black horizontal line at the bottom of (a) and (c) denote the existence intervals [Eq.~\eqref{eq:acmdint}] of period $p=1$ AMs. The high peaks in (a) and (c) for $K > 2 \pi$ (their highest values $\mu^*$ are marked by black circles) correspond to cases of anomalous diffusion due to the presence of period $p=1$ AMs, while regions with $\mu \approx 1$ tally to normal diffusion processes. Intervals of $K$ values related to anomalous diffusion also generate the peaks in (b) and (d).
}
\label{fig:3}
\end{figure*}

\begin{figure}[h!] \centering
\includegraphics[width=0.85\columnwidth,keepaspectratio]{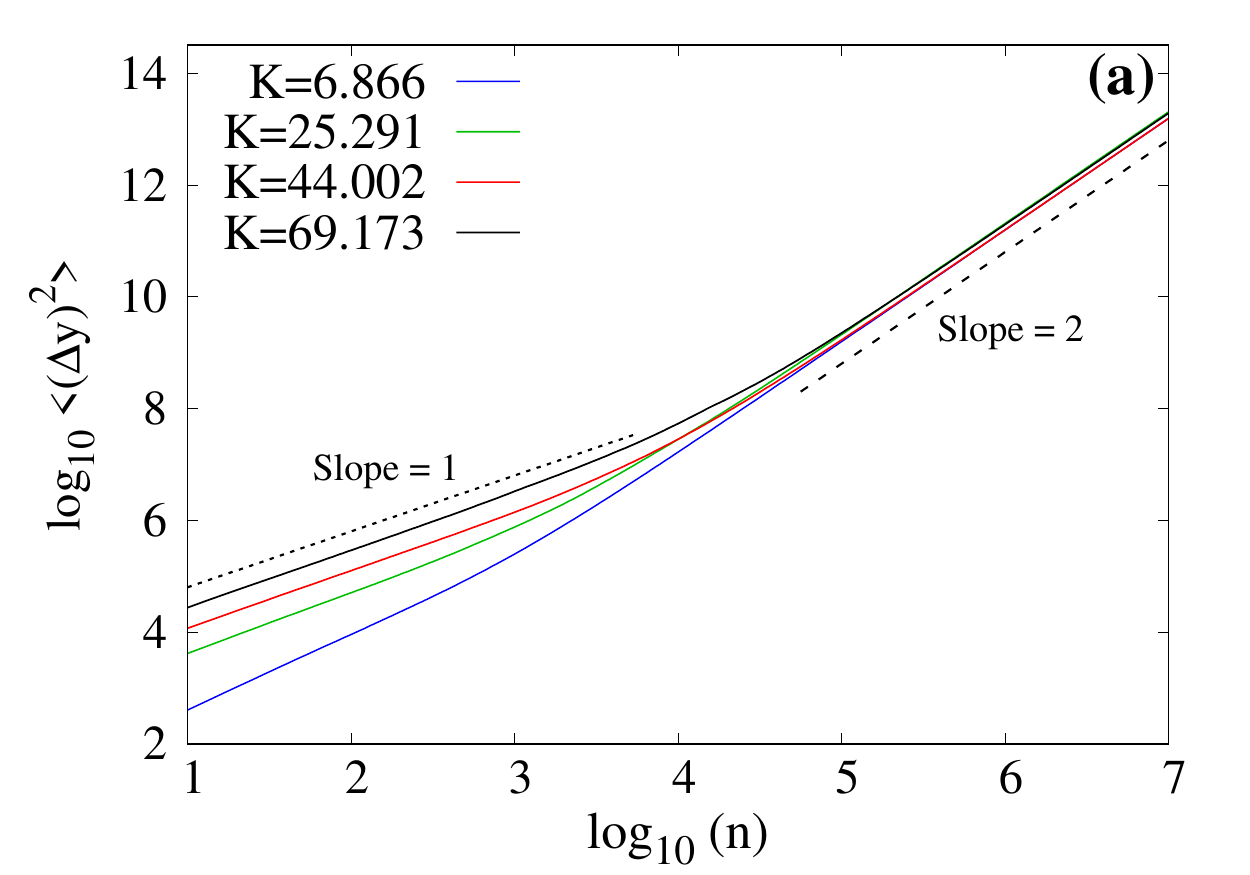}\\
\includegraphics[width=0.85\columnwidth,keepaspectratio]{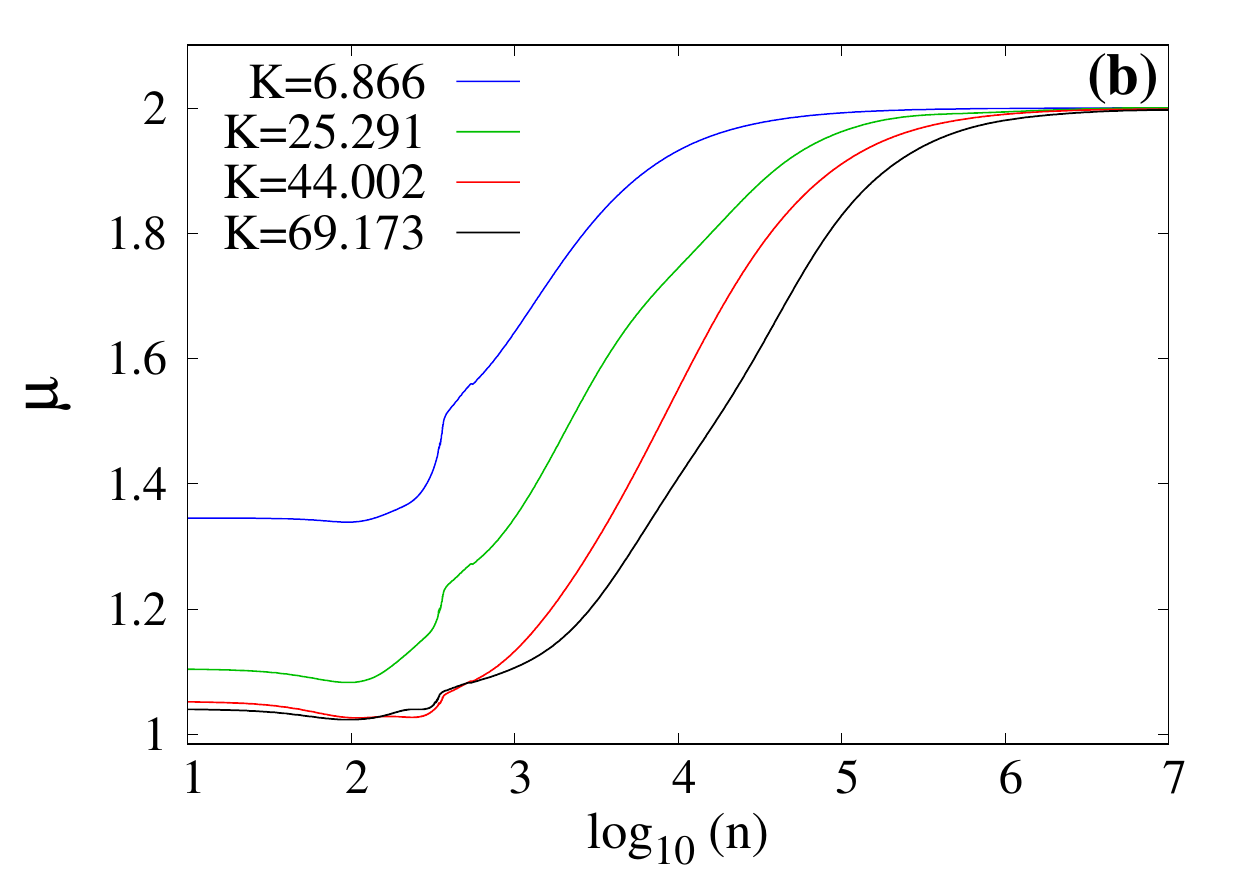}
\caption{The variance $\langle(\Delta y)^2\rangle$ (a) and the corresponding  numerically computed diffusion exponent $\mu$  [Eq.~\eqref{eq:yvar}] (b) of the 2D SM \eqref{eq:sm} as a function of the map's iterations $n$, for an ensemble of orbits with ICs on a $315 \times 315$ grid covering the  map's entire phase space $[0, 2 \pi) \times [0, 2 \pi)$, and four $K$ values, namely $K=6.866$ (blue curves), $K=25.291$ (green curves), $K=44.002$ (red curves) and $K=69.173$ (black curves), which correspond to some of the black circle denoted peaks in Fig.~\ref{fig:3}(a). The black dotted and dashed lines in (a) respectively correspond to $\mu=1$ and $\mu=2$.
}
\label{fig:4}
\end{figure}

Extending the analysis presented in \cite{ManRob2014PRE} we compute the dependence of the diffusion exponent $\mu$  on the value of $K$ for a set of $315 \times 315$  equally spaced ICs on the entire phase space of the SM, i.e.~$x$, $y \in [0,2\pi)$, for many more iterations  than the ones considered in \cite{ManRob2014PRE} (actually up to $n=10^7$), in order to also investigate the dependence of the value of $\mu$ on $n$. In Figs.~\ref{fig:3}(a) and (c) we present $\mu$ as a function of $K$ when, respectively, $n=10^4$ and $n=10^5$ iterations of the considered set of orbits are performed. In particular, $\mu$ is estimated through a linear fitting of $\log_{10} \langle(\Delta y)^2\rangle $ with respect to $\log_{10}n$ in the last two decades of the evolution. This means that, for example in the case of Fig.~\ref{fig:3}(c) the fitting is performed for data with $3 \leq \log_{10}(n) \leq 5$.

At the $K$ intervals [Eq.~\eqref{eq:acmdint}] where AMs of period $p=1$ are expected [also indicated at the horizontal black line at the bottom of Figs.~\ref{fig:3}(a) and (c)] we observe high peaks denoting  anomalous diffusion due to the presence of AMs. We note that for the considered values of $K\leq 70$, 11 such peaks appear. The largest obtained values $\mu^*$ at these peaks are marked by black filled circles. The direct comparison of Figs.~\ref{fig:3}(a) and (c) shows that these peaks, and the related $\mu^*$ values, increase as the number of performed iterations grows, approaching the $\mu=2$ value, which corresponds to ballistic transport. We note that at the same intervals of $K$ values the computed diffusion coefficient $D_{\rm eff}$ \eqref{eq:Deff_sm} also exhibits abrupt spikes [Figs.~\ref{fig:3}(b) and (d)], whose height increases with $n$ as well. The other peaks observed in Figs.~\ref{fig:3}(a) and (c) for $K < 4 \pi$ (whose height also increases for larger $n$) are related to AM of periods $p>1$, while $K$ intervals with $\mu \approx 1$ are associated with normal diffusion processes. We emphasize here that the ballistic motion caused by AM ICs prevails on the system's global diffusion and the effective diffusion coefficient $D_{\rm eff}$ \eqref{eq:Deff_sm} will go to infinity, as was also reported in \cite{HC2018}, while the corresponding, entire phase space  diffusion exponent $\mu$ \eqref{eq:yvar} will eventually converge to the value 2. This is valid even for very large $K$ values, as the AM islands exist even when $K \rightarrow \infty$, having nevertheless smaller and smaller sizes as $K$ grows.

In order to further investigate the dynamical properties of the SM when anomalous diffusion is observed due to the presence of period $p=1$ AMs, we study in more detail the behavior of the map for the 11 $K$ values at which the peaks in  Fig.~\ref{fig:3} appear. In Fig.~\ref{fig:4} we present results for only 4 such cases [namely for $K=6.866$ (blue curves), $K=25.291$ (green curves), $K=44.002$ (red curves) and $K=69.173$ (black curves)] in order to not overload the plots. More specifically, the dependence of $\langle(\Delta y)^2\rangle$ and the related diffusion exponent $\mu$ [Eq.~\eqref{eq:yvar}] on the number $n$ of map's iterations is respectively shown in Figs.~\ref{fig:4}(a) and (b). From these results we see that in all cases $\mu$ eventually becomes $\mu=2$ (indicating asymptotic ballistic transport), but the number of iterations needed to reach this value increases as $K$ becomes larger. This feature is clearly seen in Fig.~\ref{fig:4}(b) as the presented curves saturate to $\mu=2$ at larger $n$ for higher $K$ values.

It is also worth noting that the initial phases of the dynamical evolution are better described as normal diffusion when $K$ has large values (green and in particular red and black curves in both panels of  Fig.~\ref{fig:4}) as for these cases the increase of $\langle(\Delta y)^2\rangle$ is practically proportional to $n$ [dotted line in Fig.~\ref{fig:4}(a)] and the corresponding $\mu$ values are close to $\mu=1$ [Fig.~\ref{fig:4}(b)]. This delayed kick-in of anomalous diffusion for larger $K$ values is also reflected on the lowering of the peak heights in Figs.~\ref{fig:3}(a) and (c) when $K$ increases.

\begin{figure}[h!] \centering
\includegraphics[width=\columnwidth,keepaspectratio]{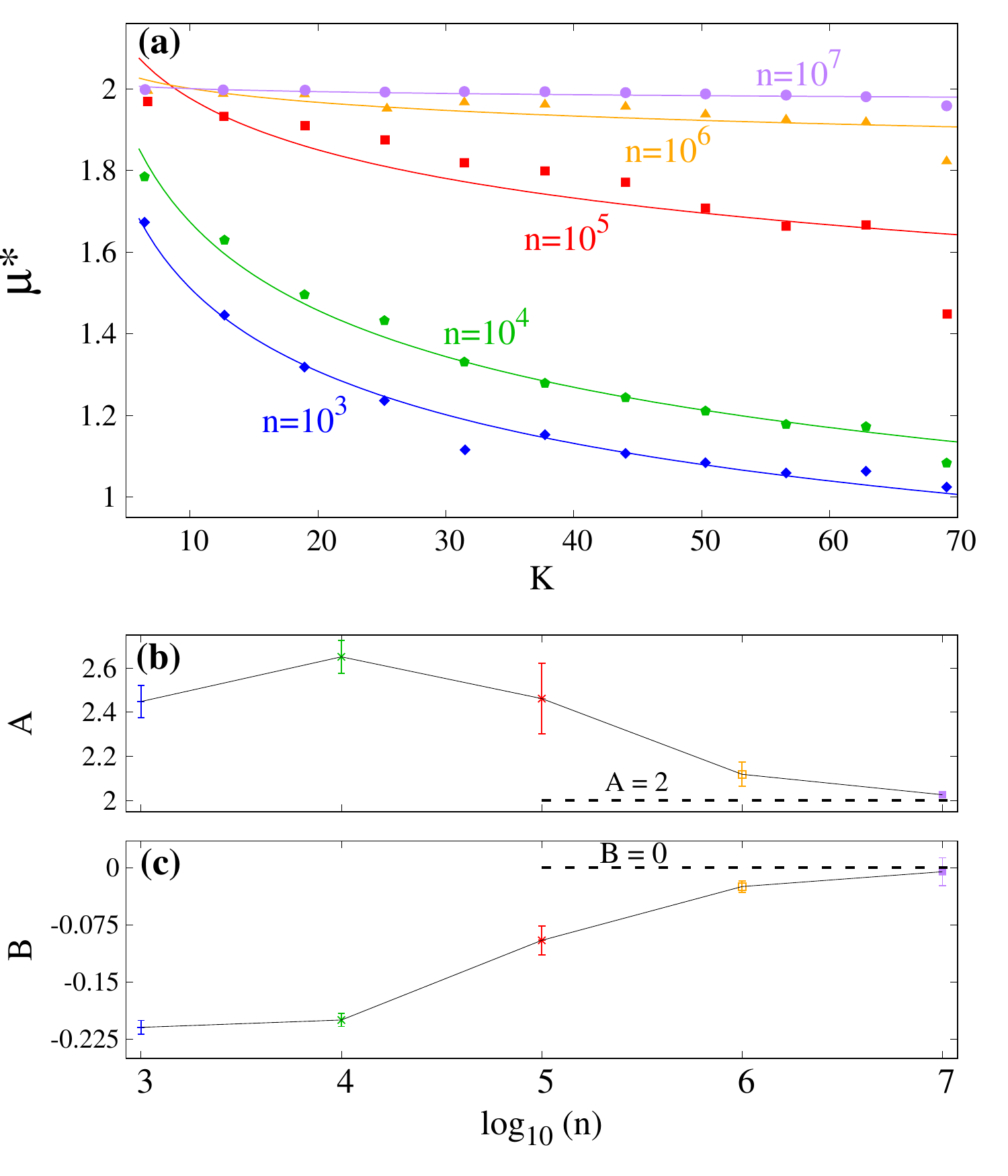}
\caption{(a) The highest values $\mu^*$ of the diffusion exponent $\mu$ [Eq.~\eqref{eq:yvar}] in the first 11 intervals of $K$ values [Eq.~\eqref{eq:acmdint}] for which period $p=1$ AMs exist for the 2D SM \eqref{eq:sm}, as a function of $K$ for $n=10^3$ [blue diamonds], $n=10^4$ [green pentagons; points denoted by black circles in Fig.~\ref{fig:3}(a)], $n=10^5$ [red squares; points denoted by black circles in Fig.~\ref{fig:3}(c)], $n=10^6$ [orange triangles] and $n=10^7$ [purple circles] iterations of the map. In all cases ensembles of orbits with ICs on a $315 \times 315$ grid covering the  map's entire phase space $[0, 2 \pi) \times [0, 2 \pi)$ were considered. The solid curves correspond to fittings of the data with functions $\mu^*=AK^B$ for each presented case. The obtained values of the fitting parameters $A$ and $B$ (along with their uncertainties), as a function of the number of iterations $n$ are respectively given in (b) and (c). The values $A=2$ in (b) and $B=0$ in (c) are indicated by horizontal dashed lines.
}
\label{fig:5}
\end{figure}

The increase of the peak highest values $\mu^*$ of the diffusion exponent in the first 11 intervals [Eq.~\eqref{eq:acmdint}] where $p=1$ AMs exist, and their tendency to approach $\mu=2$ as $n$ grows, is clearly seen in Fig.~\ref{fig:5}(a) where results for $n=10^3$ (blue diamonds), $n=10^4$ (green pentagons), $n=10^5$ (red squares), $n=10^6$ (orange triangles) and $n=10^7$ (purple circles) are plotted. We note that the points for $n=10^4$ and $n=10^5$ are the ones respectively marked by black circles in Figs.~\ref{fig:3}(a) and (c). The solid curves correspond to fittings of each set of data points by a power law of the form:
\begin{equation}\label{eq:fit}
\mu^* = A K^B,
\end{equation}
and the dependence of the fitting parameters $A$ and $B$ on the iterations $n$ of the map is respectively seen in Figs.~\ref{fig:5}(b) and (c).

From the results of Fig.~\ref{fig:5}(a) we vividly see that the highest $\mu$ values in each peak tend asymptotically to $\mu=2$ as $n$ increases. This feature is also reflected in the values of parameters $A$ and $B$, which respectively approach  $A=2$ [dashed line in Fig.~\ref{fig:5}(b)] and $B=0$ [dashed line in Fig.~\ref{fig:5}(c)]. These tendencies denote that for large $n$ the values of $\mu^*$ actually become independent of $K$ and almost equal to $\mu=2$ as \eqref{eq:fit} gives $\mu^*=2$ for $A=2$ and $B=0$.

\begin{figure}[h!] \centering
\includegraphics[width=\columnwidth,keepaspectratio]{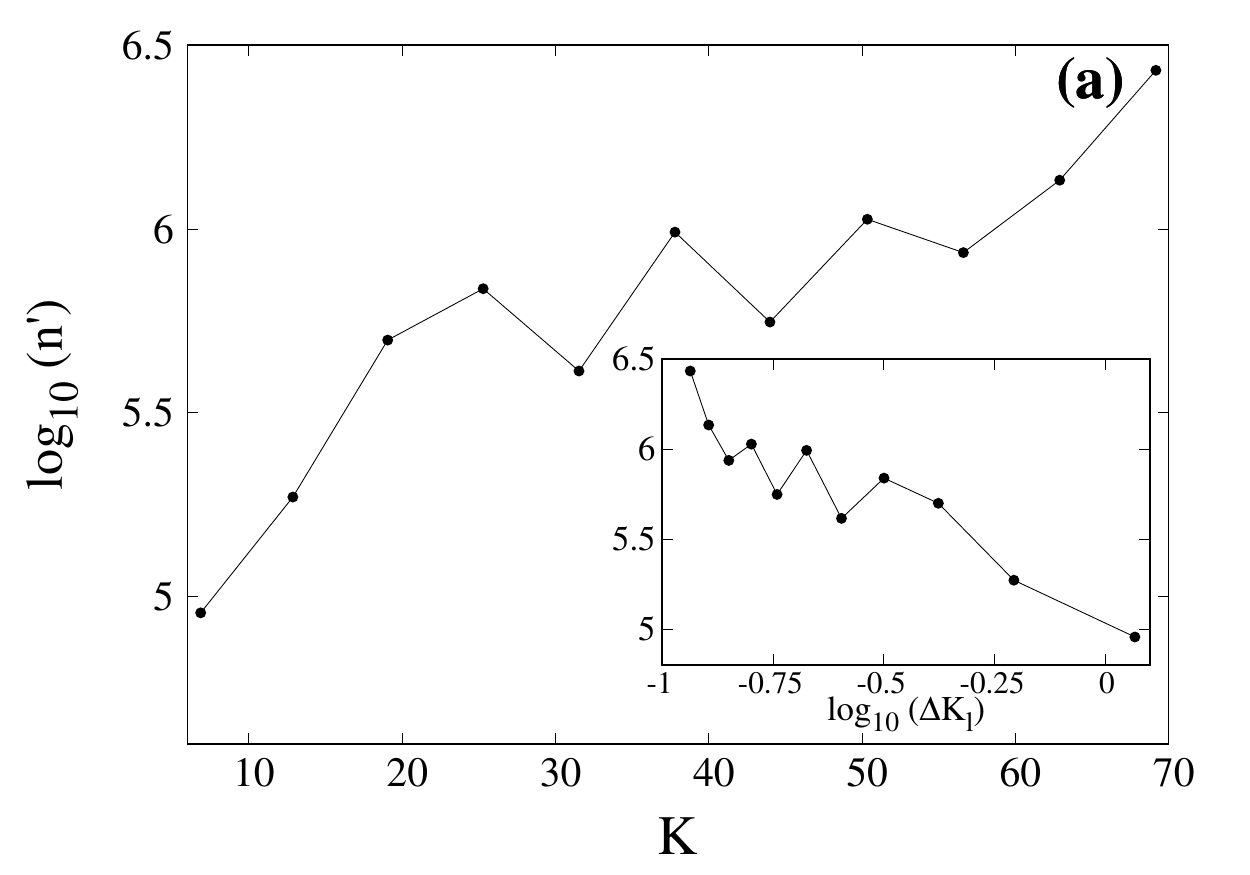}\\
\includegraphics[width=\columnwidth,keepaspectratio]{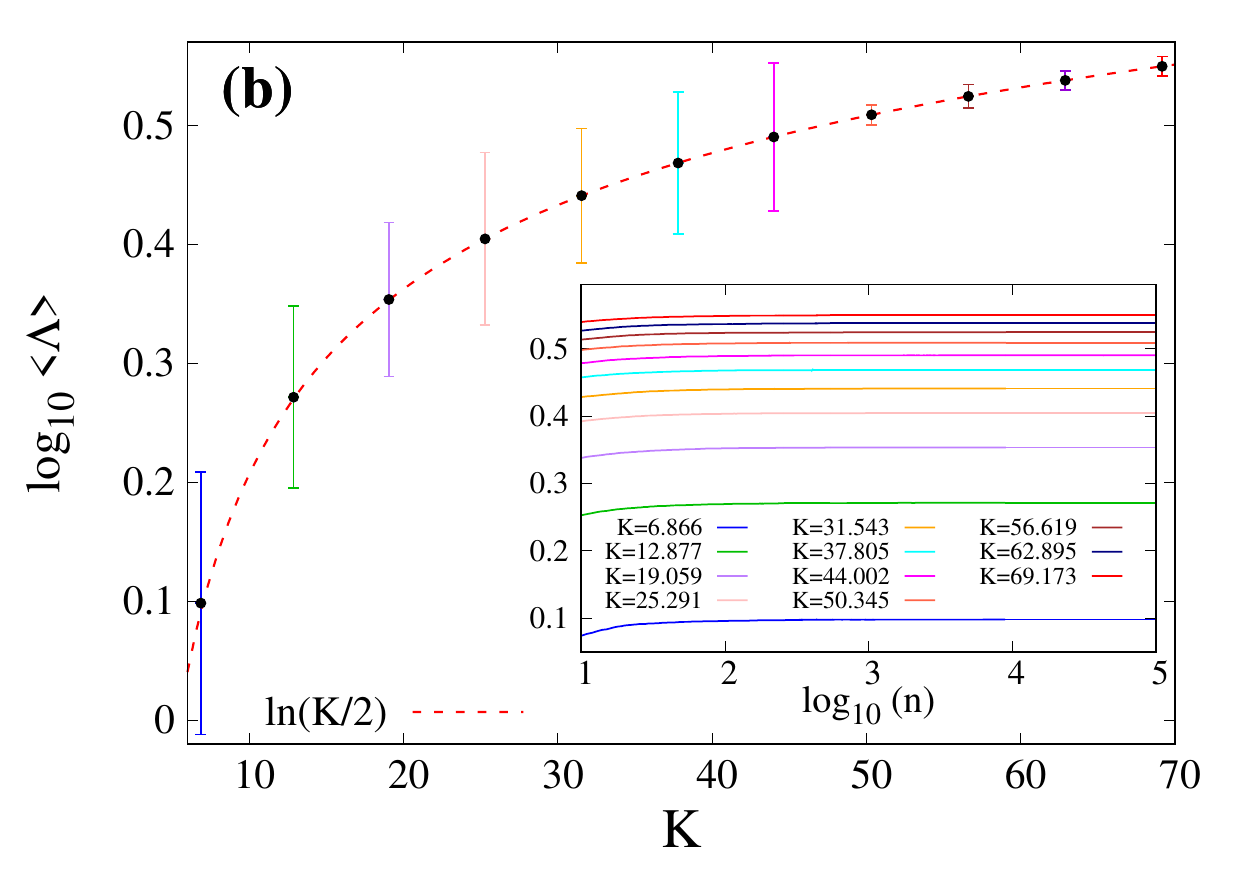}
\caption{(a) The number $n'$ of iterations of the 2D SM \eqref{eq:sm} needed for the diffusion exponent $\mu$ [Eq.~\eqref{eq:yvar}] to reach the $\mu =2$ value (ballistic transport) for $K$ values at which  $\mu$ becomes maximum [see Figs.~\ref{fig:3}(a), (c) and Fig.~\ref{fig:5}(a)] in the  first 11 intervals of existence [Eq.~\eqref{eq:acmdint}] of period $p=1$ AMs, as a function of $K$ (and in the inset as a function of the width, $\Delta K_l=\sqrt{(2\pi l)^2 +16 }-2\pi l$,  of these intervals).
(b) Average value $\langle \Lambda \rangle$ [over 10,000 orbits with uniformly distributed ICs on a $100 \times 100$ grid covering the  map's entire phase space $[0, 2 \pi) \times [0, 2 \pi)$] of the orbits' ftMLE \eqref{eq:ftMLE} after $n=10^5$ iterations, for the same values of $K$ (explicitly given in the inset) as in (a)
The error bars denote one standard deviation in the computation of the average value, while the dashed red curve corresponds to the law $\langle \Lambda \rangle = \ln (K/2)$. Inset: the evolution of $\langle \Lambda \rangle$ with respect to the map's iteration $n$. Higher $\langle \Lambda \rangle$ values correspond to larger $K$ values.
}
\label{fig:6}
\end{figure}

\begin{figure*}[]\centering
\vspace{-0.5cm}
\begin{flushleft} $K=6.5$ (stable AM of period $p=1$) \end{flushleft}
\vspace{-0.45cm}
\includegraphics[width=0.85\textwidth,keepaspectratio]{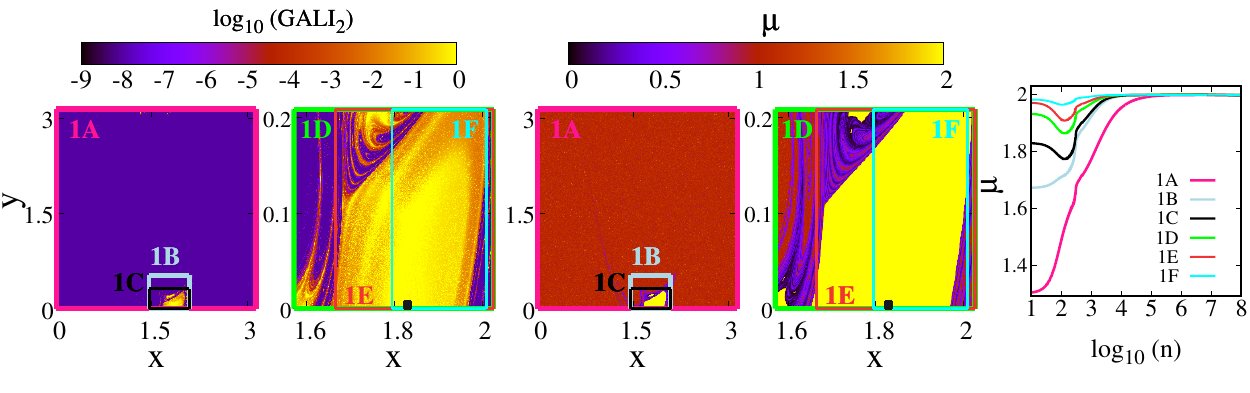}
\vspace{-0.5cm}
\begin{flushleft} $K=4.0844$ (stable AM of period $p=2$) \end{flushleft}
\vspace{-0.45cm}
\includegraphics[width=0.85\textwidth,keepaspectratio]{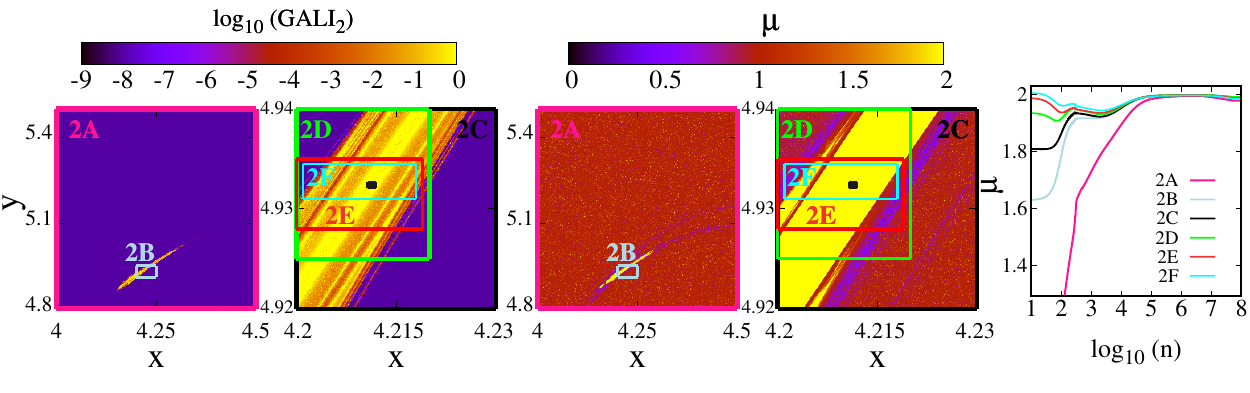}
\vspace{-0.5cm}
\begin{flushleft} $K=6.9115$ (stable AM of period $p=3$) \end{flushleft} \vspace{-0.45cm}
\includegraphics[width=0.85\textwidth,keepaspectratio]{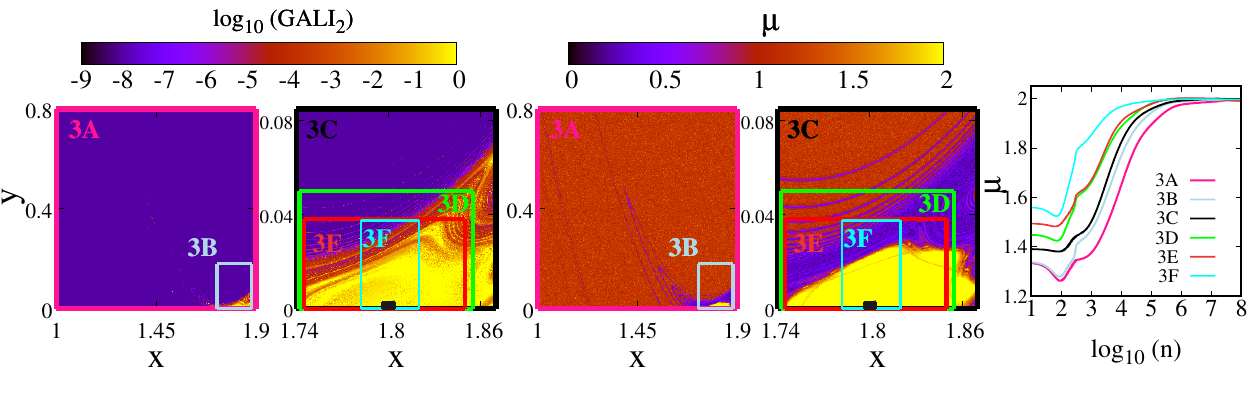}
\vspace{-0.5cm}
\begin{flushleft} $K=3.1$ (stable AM of period $p=4$) \end{flushleft}
\vspace{-0.45cm}
\includegraphics[width=0.85\textwidth,keepaspectratio]{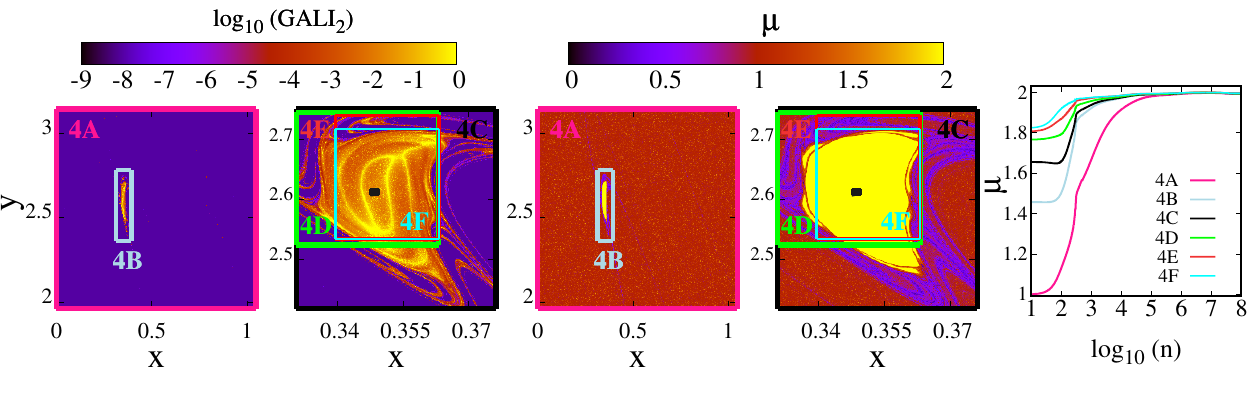}
\vspace{-0.5cm}
\caption{
Phase space portraits of the 2D SM \eqref{eq:sm} colored according to the GALI$_2$ \eqref{eq:GALI} value  (panels in the two left columns) after $n=50$ iterations of orbits with ICs on a dense grid covering the considered phase space region, and the diffusion exponent $\mu$ [Eq.~\eqref{eq:yvar}] (panels in the two middle columns) computed by appropriately fitting the variance $\langle(\Delta y)^2\rangle$ \eqref{eq:yvar} values for  iterations $n$ in the interval $10^3 \leq n \leq 10^4$ (see text for more details). Each row of panels corresponds to a different $K$ value, namely $K=6.5$, $K=4.0844$, $K=6.9115$ and $K=3.1$ (from top to bottom), for which stable AMs of period $p=1,$ 2, 3 and 4 respectively exist. Points are colored according to the color scales on top of the panels. In the GALI$_2$ plots yellow regions refer to regular motion, dark blue and black areas correspond to chaotic motion, while in-between colors to weakly chaotic ICs requiring more iteration to reveal their chaotic nature. In the $\mu$ color plots ICs leading to subdiffusion rates are colored in dark blue ($0 \le \mu < 1$), normal diffusion cases are depicted in dark red ($\mu \approx 1$), superdiffusion  corresponds to lighter shades of red ($1 \le \mu < 2$) and ballistic transport is denoted by yellow ($\mu \approx 2$). The positions of the stable $p=1$ [$(1.8298, 0)$], $p=2$ [$(4.211, 4.9324)$], $p=3$ [$(1.80, 0.0)$] and $p=4$ [$(0.3486, 2.6121)$] AMs are denoted by small black rectangular points in the right GALI$_2$ and $\mu$ panels. In each case 6 different regions of the phase space are considered. The names of these regions, along with their ranges in the $x$ and $y$ coordinates, are given in Table \ref{tb:1}, and their positions are indicated by different rectangles in the color plots accompanied by the names of the regions. These regions are ordered in Table \ref{tb:1} in decreasing content of chaotic orbits (quantified by the percentage $P_C$ of ICs leading to chaotic motion) and they were chosen so that in all cases ensembles with similar names have equivalent $P_C$ values, i.e.~$P_C \approx 99\%$ for `A' regions, $P_C \approx 75\%$ for `B', $P_C \approx 50\%$ for `C', $P_C \approx 25\%$ for `D', $P_C \approx 12.5\%$ for `E' and $P_C \approx 6.5\%$ for `F'.  The panels in the last column depict the diffusion exponent $\mu$ [Eq.~\eqref{eq:yvar}] of the 6 different regions as a function of the map's iterations $n$.
}
\label{fig:7}
\end{figure*}

Nevertheless, as indicated by the results of Figs.~\ref{fig:4}(b) and \ref{fig:5}(a) the diffusion exponents associated with the larger $K$ peaks require more iterations to reach the asymptotic, limiting value $\mu=2$. This behavior is clearly seen in Fig.~\ref{fig:6}(a) where we plot the number $n'$ of iterations needed for $\mu$ to reach $\mu=2$ (actually $n'$ is defined as the number of iterations needed for $\mu$ to take values very close to $\mu=2$ and in particular to become $\mu \geq 1.992$). In the inset of Fig.~\ref{fig:6}(a) we plot $n'$ as a function of the length $\Delta K_l=\sqrt{(2\pi l)^2 +16 }-2\pi l$ of the period $p=1$ AM existence intervals \eqref{eq:acmdint}, and since $\Delta K_l$ decreases for larger $K$  (larger $l$), $n'$ values are higher for smaller $\Delta K_l$.

In order to relate the delay of the diffusion exponent to reach the value $\mu=2$ with the strength of chaos of the SM we compute the average $\langle \Lambda \rangle$ ftMLE \eqref{eq:ftMLE} over a set of orbits covering the entire phase space, as an overall indicator of the system's chaoticity, for the 11 $K$ values  considered in Fig.~\ref{fig:5}, namely $K=6.866$, 12.877,  19.059, 25.291, 31.543, 37.805, 44.002, 50.345, 56.619, 62.895 and 69.173. In particular, we evolve $10,000$ orbits with ICs on a $100 \times 100$ grid on the entire phase space, i.e.~$x$, $y \in [0, 2 \pi)$, for $n=10^5$ iterations, registering the average value $\langle \Lambda \rangle$ of the ftMLE at the end of the simulations, considering also one standard deviation of this process as a good indicator of the error of the computed average. The output of this procedure is seen in Fig.~\ref{fig:6}(b) where a clear increase of $\langle \Lambda \rangle$ as $K$ grows is evident. Actually, the obtained results are in excellent agreement with the relation $\langle \Lambda \rangle = \ln (K/2)$ [dashed red curve in Fig.~\ref{fig:6}(b)] which has been presented in previous works \cite{S04,HKC2019}. Thus, we conclude that in general the existence of period $p=1$ AMs eventually leads to ballistic transport ($\mu=2$) for sets of orbits distributed in the entire phase space of the SM but for stronger chaoticities (larger $K$ values) this behavior is achieved after longer iteration intervals.

We note  that in the computation of the average $\langle \Lambda \rangle$ value we potentially mix regular and chaotic orbits. Nevertheless, although islands of stability always exist in the phase space of the SM, their extent decreases rapidly as $K$ grows and the overall dynamics is practically defined by chaotic motion (see e.g.~\cite{Chirikov1979}). Thus, the possible inclusion of some  regular orbits, whose ftMLE tends to zero, does not practically influence the computation of  $\langle \Lambda \rangle$, as the used ensemble of orbits ($10,000$) is sufficiently large. In addition, the number of iteration used ($n=10^5$) for the evaluation of  $\langle \Lambda \rangle$ is large enough to permit a reliable estimation of a representative value for the system's chaoticity, as $\langle \Lambda \rangle$ has already saturated to its limiting value for all considered cases [see inset of  Fig.~\ref{fig:6}(b)].

Let us now study in more detail the relation between the SM's chaoticity and its diffusion properties in the presence of AMs of various periods. The different behavior of the GALI$_2$ for regular and chaotic orbits, seen in Fig.~\ref{fig:1}(b), allows us to use this index for a fast and clear distinction between regions of chaos and order in the 2D phase space of the SM \eqref{eq:sm}. More specifically, in order to estimate the percentage of chaotic orbits of the SM for a given value of $K$ we implement the subsequent approach: We follow the evolution of orbits whose ICs lie on a 2D grid with equally spaced points on the 2D phase space of the map, and register for each orbit the value of GALI$_2$ after $n=50$ iterations. All orbits having values of GALI$_2$ smaller than $10^{-8}$ at $n=50$ are characterized as chaotic, while all others are considered as non-chaotic (we note that a similar process was also performed in \cite{ManRob2014PRE}). We also point out that for each orbit a different random set of orthonormal initial deviation vectors is used setting GALI$_2(0)=1$. The $10^{-8}$  threshold  value has been broadly and successfully used  in previous works (see for example  \cite{MSAB2008,MSB2008,MSB2009}), and can safely distinguish the two types of orbits as is also evident from the results of Fig.~\ref{fig:1}(b). In addition, we have checked that $n=50$ iterations are sufficient to securely distinguish between the exponential (chaotic motion) and the power law (regular motion) decay of GALI$_2$.

For the single SM \eqref{eq:sm} we start by choosing a $K$ value for which AMs  are present and use the GALI$_2$ method to detect the regular and chaotic regions in the system's phase space, producing color maps which clearly give a global overview of the  dynamics and measuring at the same time the  fraction $P_C$ of regions where chaotic motion occurs. In order to create such detailed color plots we consider ensembles consisting of  $250,000$ ICs on a grid of $500 \times 500$ cells (i.e.~one orbit per cell) covering the entire phase space, i.e.~$(x,y) \in [0,2\pi)\times[0,2\pi)$, or a subspace of it. The evolution of the orbits of this grid is also used to estimate the percentage $P_C$ of chaotic orbits through the computation of their GALIs. The same dense grid is also used to measure the `local' rate of diffusion for each cell and create color plots of the related $\mu$ value [Eq.~\eqref{eq:yvar}], but for such computations  we additionally consider $50 \times 50$ ICs inside each one of the $500 \times 500$ cells (i.e.~2,500 orbits per cell) and numerically calculate  the diffusion exponent $\mu$ for each cell separately by using \eqref{eq:yvar} and performing a linear fit of the variance $\langle(\Delta y)^2\rangle$ as a function of the iterations $n$ in log-log scale for $3 \leq \log_{10} (n) \leq 4$. These processes produce detailed pictures, with fine resolutions, allowing the systematic analysis of both measurements (the GALI index for the system's chaoticity and the diffusion exponent $\mu$ for the diffusion properties of the map) within relatively feasible CPU times and they will be followed whenever we create such color plots in this work. In this way we are able to produce comprehensive plots where regular and chaotic regions are clearly indicated, as well as detect regions characterized by different diffusion rates.

Apart from the global study of  the 1D SM we also focus our attention on subspaces of the entire phase space, and more specifically on regions which contain AMs (where superdiffusion takes place) surrounded by chaotic areas (where normal diffusion occurs). By gradually changing the size of the considered subspace and consequently the percentage $P_C$ of chaos (i.e.~the extent of the chaotic region) around the AMs we investigate the role and impact of this ratio on the diffusion transport properties of the considered subspaces.

The outcomes of these computations for four $K$ values corresponding to the presence of AMs of periods $p=1$, 2, 3 and 4 (respectively for $K=6.5$, $K=4.0844$, $K=6.9115$ and $K=3.1$) are shown in Fig.~\ref{fig:7}. In each case the first two panels correspond to color plots of phase space regions based on the value of GALI$_2$, with the next two panels depicting the same regions colored according to the computed $\mu$ values.  In the GALI$_2$ color plots, regions of regular behavior correspond to large GALI$_2$ values colored in yellow, while dark purple and black areas denote chaotic behavior. In the $\mu$ color plots, regions where ballistic diffusion takes place ($\mu \approx 2$) are colored in yellow, while dark red and purple colored areas respectively correspond to normal diffusion ($\mu \approx 1$) and subdiffusing processes.

For each $K$ case the position of the stable AM is denoted by a small black rectangle point in the right GALI$_2$ and $\mu$ color plots (respectively second and fourth column in Fig.~\ref{fig:7}). Examples of AMs of different periods, as well as the presentation of a method to locate them  can be found in \cite{HC2018}.

For each $K$ case we consider 6 phase space regions whose positions are shown in the GALI$_2$ and $\mu$ color plots of Fig.~\ref{fig:7} by different rectangles. The $x$ and $y$ ranges of these rectangles, along with their names are listed in Table~\ref{tb:1}. The names of these  regions are composed by the value $p$ of the AM's period followed by a letter from `A' to `F'. The various regions were chosen so that the percentage $P_C$ of chaotic orbits they contain decreases as we move from `A' to `F', taking care at the same time that regions with similar names have equivalent $P_C$ values, namely $P_C \approx 99\%$ (A), $75\%$ (B), $50\%$ (C), $ 25\%$ (D), $12.5\%$ (E), $6.5\%$ (F) (the exact $P_C$ values are also reported in Table~\ref{tb:1}). For example, for $K=6.5$ (upper row in Fig.~\ref{fig:7}) regions 1A, 1B and 1C are seen in the left GALI$_2$ and $\mu$ color plots, where they are respectively denoted by dark pink, light blue and black bordered rectangles. Regions 1D, 1E and 1F are respectively depicted in the right GALI$_2$ and $\mu$ panels by green, light red and cyan rectangles.
\begin{table}
\caption{
The number $n'$ of iterations of the 2D SM \eqref{eq:sm} needed for the diffusion exponent $\mu$ [Eq.~\eqref{eq:yvar}] to practically reach its maximum asymptotic $\mu =2$ value (ballistic transport) for several $K$ values and ensembles of ICs in different phase space regions containing various percentages $P_C$ of chaotic orbits. The considered $K$ values are  $K=6.5$, $K=4.0844$, $K=6.9115$, $K=3.1$ and respectively correspond to the appearance of stable AMs of period $p=1,$ 2, 3 and 4. The names of the various regions are composed by the value $p$ of the AM's period followed by a letter from `A' to `F'. The ranges in the $x$ and $y$ coordinates for each region were chosen so that ensembles with the same letter have practically the same $P_C$ value, which decreases as we move from `A' to `F' [$P_C \approx 99\%$ (A), $\approx 75\%$ (B), $\approx 50\%$ (C), $\approx 25\%$ (D), $\approx 12.5\%$ (E), $\approx 6.5\%$ (F)].
}
\label{tb:1}
\begin{adjustbox}{width=\columnwidth,center}
\begin{tabular}{l|l|c|c|c}
\hline $K$ & Ensemble name: $x$ and $y$ ranges & $P_C$ & p & $n'$ \\ \hline
 & 1A: $[0.000,3.142] \times [0.0,3.142]$ & 99.76 &  & 5.45$\times10^{4}$ \\
6.5
 & 1B: $[1.282,2.094] \times [0.0,0.524]$ & 74.98 & 1 & 5.34$\times10^{3}$ \\
 & 1C: $[1.461,2.094] \times [0.0,0.314]$ & 50.03 &  & 5.25$\times10^{3}$ \\
 & 1D: $[1.571,2.027] \times [0.0,0.209]$ & 24.84 &  & 2.61$\times10^{3}$ \\
 & 1E: $[1.665,2.027] \times [0.0,0.209]$ & 12.46 &  & 1.80$\times10^{3}$ \\
 & 1F: $[1.795,2.010] \times [0.0,0.209]$ & 6.53 &  & 4.88$\times10^{2}$ \\
 \hline
 & 2A: $[4.000,4.500] \times [4.800,5.500]$ & 99.23 &  & 1.98$\times10^{5}$ \\
 4.0844
 & 2B: $[4.200,4.250] \times [4.910,4.950]$ & 75.39 & 2 & 1.48$\times10^{5}$ \\
 & 2C: $[4.200,4.230] \times [4.920,4.940]$ & 50.38 &  & 5.25$\times10^{4}$ \\
 & 2D: $[4.200,4.220] \times [4.925, 4.940]$ & 25.38 &  & 4.79$\times10^{4}$ \\
 & 2E: $[4.200,4.210] \times [4.928, 4.935]$ & 12.70 &  & 2.33$\times10^{4}$ \\
 & 2F: $[4.201,4.218] \times [4.931, 4.935]$ & 6.35 &  & 2.70$\times10^{1}$ \\
\hline
& 3A: $[1.0,1.9] \times [0.0,0.80]$ & 99.12 &  & 4.37$\times10^{6}$\\
6.9115
& 3B: $[1.725,1.88] \times [0.0,0.18]$ & 75.31 & 3 & 6.08$\times10^{5}$\\
& 3C: $[1.74,1.87] \times [0.0, 0.085]$ & 50.01 &  & 4.45$\times10^{5}$\\
& 3D: $[1.742, 1.855] \times [0.0, 0.05]$ & 25.38 &  & 1.34$\times10^{5}$\\
& 3E: $[1.745, 1.85] \times [0.0, 0.038]$ & 12.43 &  & 1.27$\times10^{5}$ \\
& 3F: $[1.782, 1.82] \times [0.0,0.0375]$ & 6.28 &  & 9.12$\times10^{4}$ \\
\hline
& 4A: $[0.000,1.047] \times [1.964,3.142]$ & 99.28 &  & 6.03$\times10^{5}$ \\
3.10
& 4B: $[0.314,0.393] \times [2.362,2.780]$ & 75.20 & 4 & 1.99$\times10^{5}$ \\
& 4C: $[0.331,0.376] \times [2.417,2.732]$ & 50.43 &  & 1.22$\times10^{5}$ \\
& 4D: $[0.330,0.363] \times [2.523,2.746]$ & 24.93 &  & 4.21$\times10^{4}$ \\
& 4E: $[0.340,0.363] \times [2.534,2.741]$ & 12.31 &  & 2.33$\times10^{4}$ \\
& 4F: $[0.310,0.363] \times [2.534,2.718]$ & 6.46 &  & 1.30$\times10^{4}$ \\
\hline
\end{tabular}
\end{adjustbox}
\end{table}

The panels in the rightmost column of Fig.~\ref{fig:7} show the diffusion exponent $\mu$ [Eq.~\eqref{eq:yvar}] as a function of the number of iterations $n$ for every considered phase space region of each $K$ case. In all these panels results for similar $P_C$ values are presented by curves with the same color. More specifically, dark pink, light blue, black, green, light red and cyan curves respectively correspond to `A', `B', `C', `D', `E' and `F' named regions. From these results it becomes again clear that the diffusion exponent $\mu$ of ensembles containing more chaotic orbits (larger $P_C$ values) requires more iterations $n'$  to approach its limiting value $\mu=2$. The corresponding $n'$ values are reported in the last column of Table~\ref{tb:1}.

We note here that the GALI$_2$ color plots in Fig.~\ref{fig:7} were created for $n=50$ iterations, which, as we have already argued, are sufficient to distinguish between regular and chaotic orbits, while the $\mu$ color plots are obtained by the analysis of data for $10^3 \leq n \leq 10^4$. Thus, both types of plots capture the system's dynamical features for these specific number of iterations. As $n$ increases the dynamics evolves and some structures in these color plots will alter. For example, the red colored points inside the purple colored (chaotic) regions of the GALI$_2$ plots, as well as the purple points inside the red colored areas of the $\mu$ plots correspond to stickiness effects along unstable asymptotic curves emanating from unstable periodic orbits around the island of stability (the so-called \textit{stickiness in chaos} phenomenon \cite{CH08,CH10}), but for a very large number of iterations they will clearly reveal their chaotic nature obtaining very small GALI$_2$ values, also resulting to $\mu \approx 1$. More generally, for large $n$ GALI$_2$ plots will practically be covered by two colors: yellow (regular motion) and black (chaotic motion), while in the $\mu$ plots we will actually be able to see only regions leading to ballistic (yellow points, $\mu \approx 2$) or normal (red points, $\mu \approx 1$) diffusion.

\subsection{Coupled standard maps} \label{sec:C_SMs}

\noindent So far, we have performed a detailed and systematic  investigation of diffusion trends, characteristic time scales and properties of single 2D SMs, which possess AMs of different periods, using ensembles of ICs which are dominated by the presence of chaotic orbits (of various fractions) around these AMs. Now we set out to investigate the way that the average diffusion rate of sets of ICs and the average chaoticity of a system of coupled SMs \eqref{eq:csm} is affected by the presence of AMs. In more detail, we aim to better understand how the diffusion and chaoticity are influenced  by the individual dynamics of each coupled map (defined by the values of the respective kick-strength parameters, $K_j$, $j=1,2,\ldots, N$), coupling-strength (quantified by the value of the $\beta$ parameter), as well as different maps' arrangements. To this end, we consider a relatively small in size ($N=5$) system of coupled 2D SMs with different setups, i.e.~uniformly equal kick-strengths for all maps, or  maps with different kick-strengths. This choice of the system's size allows us to study sufficiently large sets of ICs, which are needed for the reliable diffusion rate estimation, within feasible CPU times.

For the system of $N=5$ coupled SMs \eqref{eq:csm} we, in general, consider ensembles of ICs whose projections in each one of the $N$ 2D SMs are on a $315 \times 315$ equally spaced grid (i.e.~$\approx$100,000 for each one of the coupled 2D SMs). Due to the type of the interaction between neighboring maps [see Eq.~\eqref{eq:csm}] we pay special attention in avoiding the zeroing of the coupling interaction, which appears by default whenever the angle coordinates $x^j_0$, $j=1,2,\ldots, N$, of the IC are set to be equal in neighboring 2D SMs, i.e.~whenever $x^j_0=x^{j+1}_0$. Several strategies can be implemented to avoid this situation whenever the same range $[x_{\rm min}, \,x_{\rm max}]$ of $x^j_0$ values is considered in adjacent maps. In our study, we impose a shift in the angular position variable $x$ of the form $x^{j+1}_0 = x^j_0 + d$, where $d=(x_{\rm max}-x_{\rm min})/N$ is a displacement defined by dividing the angular value range of the chosen ensemble of ICs by the total number $N$ of maps in the system. If the computed $x^{j+1}_0$ value falls outside the interval $[x_{\rm min}, \,x_{\rm max}]$, i.e.~$x^{j+1}_0 > x_{\rm max}$, then this value is reduced by $(x_{\rm max}-x_{\rm min})$ so that it is moved inside the appropriate value range.

For the numerical estimation of the generalized diffusion exponent $\mu$ we use a similar expression to \eqref{eq:yvar} (which is valid for a single 2D SM), where the left-hand side of the equation is replaced by the quantity $\sum_{j=1}^{N}\left< (\Delta y^j)^2 \right>$ appearing in \eqref{Deff_csm}, i.e.
\begin{equation}\label{eq:yvarN}
\sum_{j=1}^{N}\left< (\Delta y^j)^2 \right>= D_{\mu}^N n^{\mu},
\end{equation}
with $D_{\mu}^N$ being (for $n \rightarrow \infty)$ a diffusion coefficient analogous to $D_{\mu}$ in \eqref{eq:yvar}. We also remark that we have checked that the considered sizes of ensembles of ICs used in our investigations manage to correctly capture the systems' dynamics in feasible CPU times.

\begin{figure}[h!]\centering
\includegraphics[width=\columnwidth]{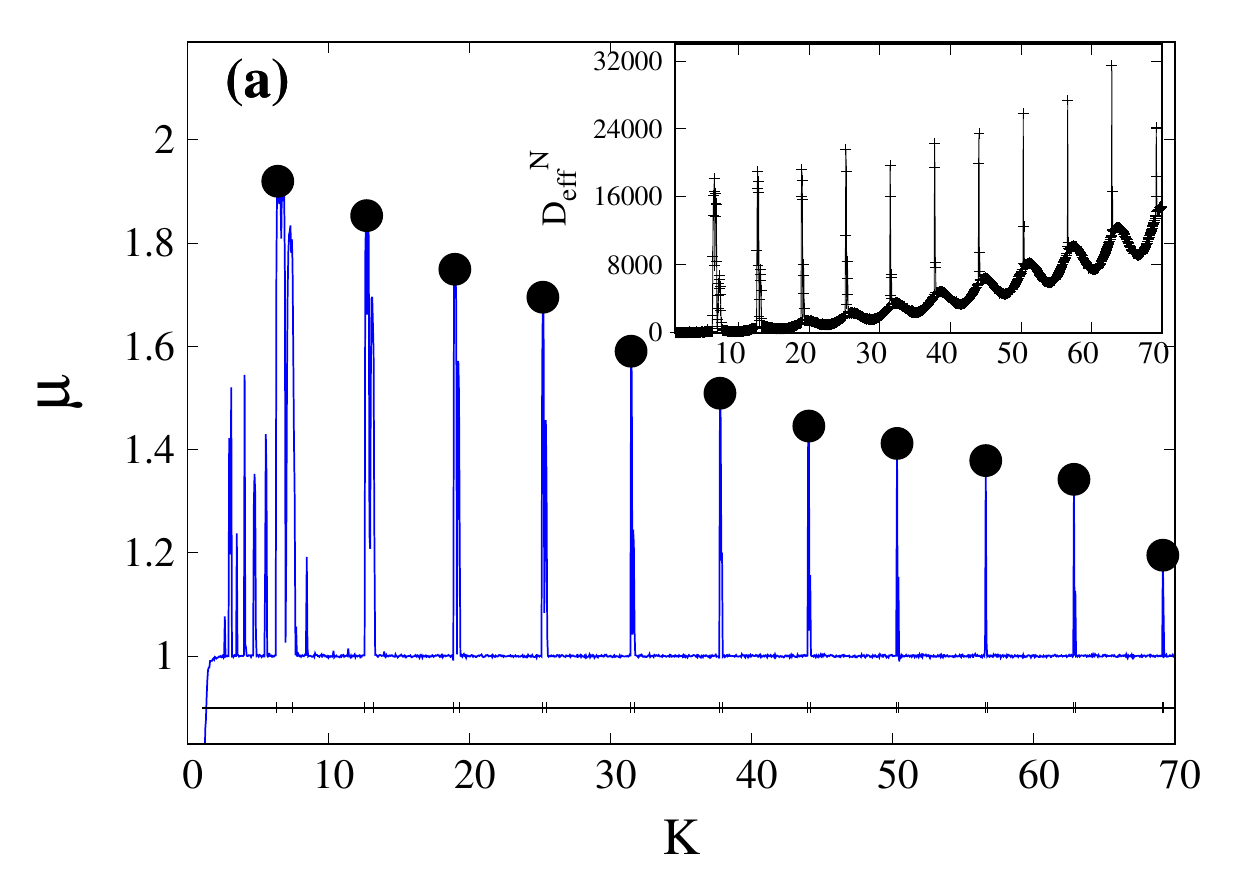}
\includegraphics[width=\columnwidth]{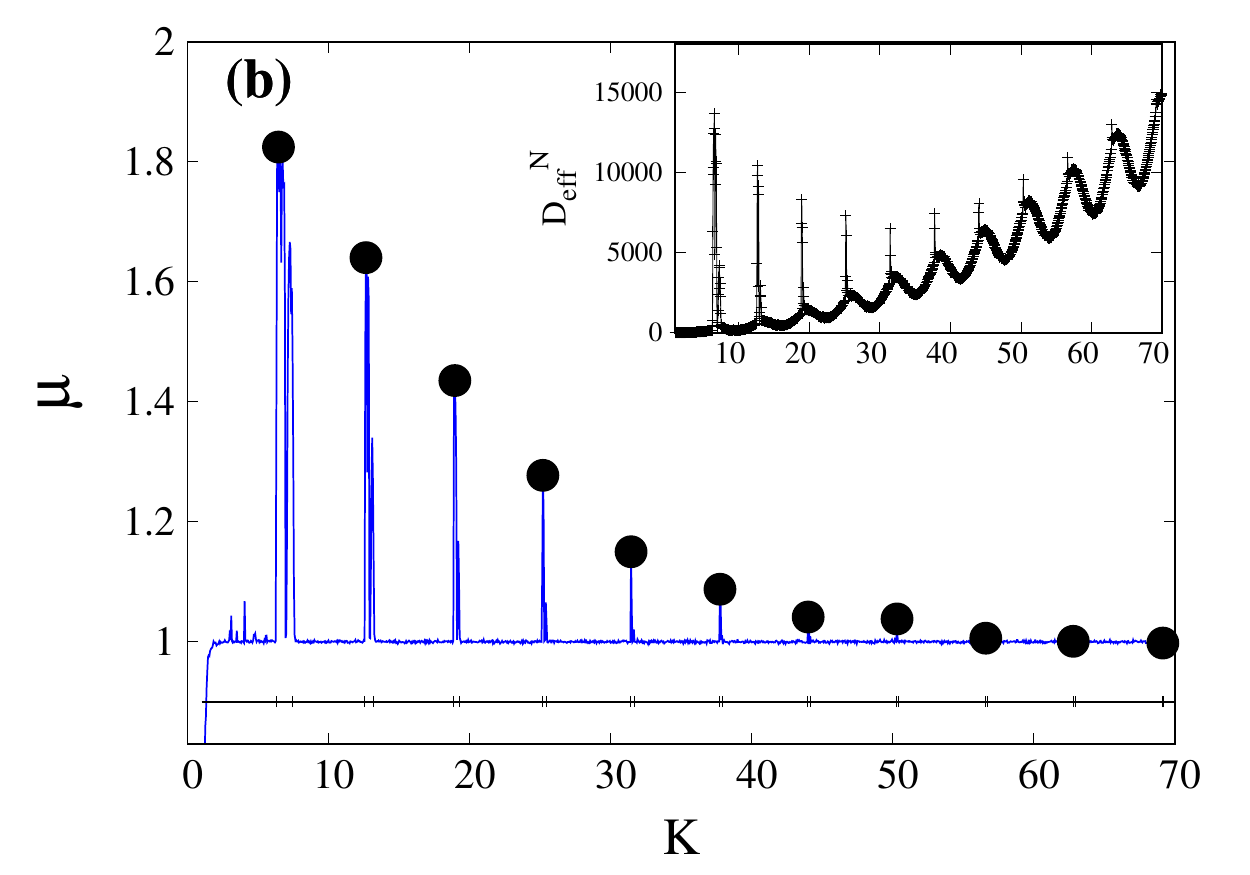}
\caption{Similar to Fig.~\ref{fig:3} but for the $N=5$ coupled system \eqref{eq:csm} with identical kick-strength values $K_j=K$, $j=1,2,\ldots, 5$. (a) Diffusion exponent $\mu$ [Eq.~\eqref{eq:yvarN}] as a function of $K$ for fixed coupling-strength $\beta=10^{-4}$ after $n=10^4$ iterations. The considered ensemble of orbits consists of ICs whose projections in each one of the $5$ 2D SMs are on a $315 \times 315$ equally spaced grid covering the whole map (see text for more details). In the inset, we plot the effective diffusion coefficient $D_{\rm eff}^{N}$ \eqref{Deff_csm} as a function of $K$. (b) Same as panel (a) but for a stronger coupling, $\beta = 10^{-3}$.
}
\label{fig:8}
\end{figure}

\subsubsection{Coupling identical 2D standard maps} \label{sec:id_C_SMs}

\begin{figure}[h!]\centering
\includegraphics[width=\columnwidth]{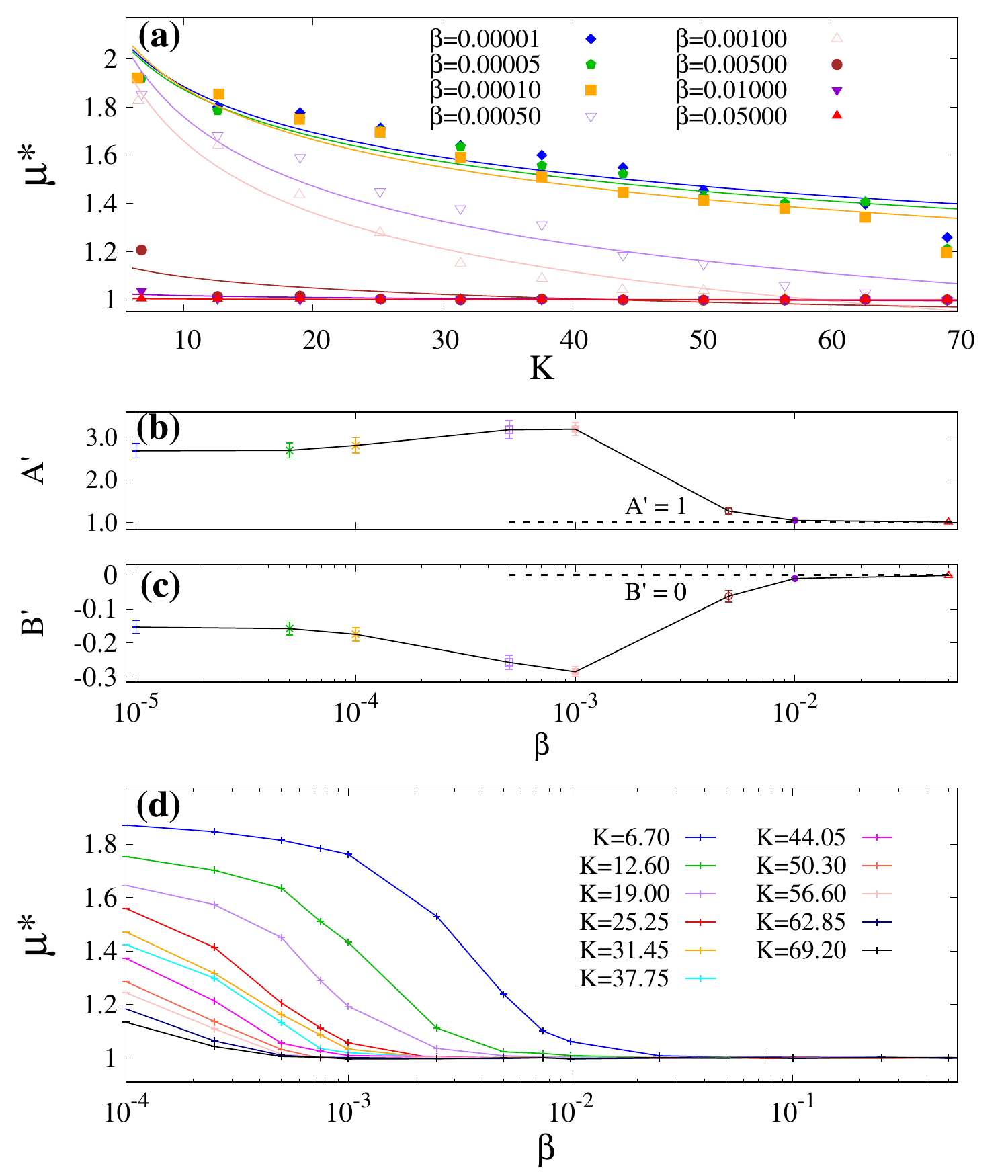}
\caption{
(a) The highest values $\mu^*$ of the diffusion exponent $\mu$ [Eq.~\eqref{eq:yvarN}] in the first 11 intervals (see also Fig.~\ref{fig:8}) of $K$ values  [Eq.~\eqref{eq:acmdint}] for which period $p=1$ AMs exist for the 2D SM \eqref{eq:sm}, as a function of $K$ for various coupling-strength $\beta$ values (see legend), computed for $n=10^4$ iterations of the multidimensional map \eqref{eq:csm} with $N=5$ (see text for more details on the used ensemble of ICs). The solid curves correspond to fittings of the data with functions $\mu^*=A\textprime K^{B\textprime}$ for each presented case. The obtained values of the fitting parameters $A\textprime$ and $B\textprime$ (along with their uncertainties), as a function of $\beta$ are respectively given in (b) and (c). The values $A\textprime=1$ in (b) and $B\textprime=0$ in (c) are indicated by horizontal dashed lines. (d) The highest values $\mu^*$ of the diffusion exponent $\mu$ as a function of the coupling-strength parameter $\beta$ for different (but fixed in all maps) values of the kick-strength parameter $K_j=K$, $j=1,2,\ldots, 5$, for which pronounced superdiffusion transport takes place (corresponding to the $K$ values indicated by black filled circles in Fig.~\ref{fig:8}).
}
\label{fig:9}
\end{figure}

\noindent Following a similar approach as for single SMs (see Fig.~\ref{fig:3}), we start by calculating the diffusion exponent $\mu$ [Eq.~\eqref{eq:yvarN}]  and the effective diffusion coefficient $D_{\rm eff}^N$ \eqref{Deff_csm} for the coupled system of SMs. Our aim is to detect conditions and/or parameter values under which one may expect to observe global diffusion properties similar to those exhibited for single SMs. In other words, we want to investigate the impact of the coupling between SMs in the long-term diffusion  properties of the system's phase space, as well as the  time (iteration) intervals for which the coupled system may still be influenced by the presence of AMs in one or more of the single 2D SMs.
\begin{figure*}[h!]\centering
\includegraphics[width=\columnwidth]{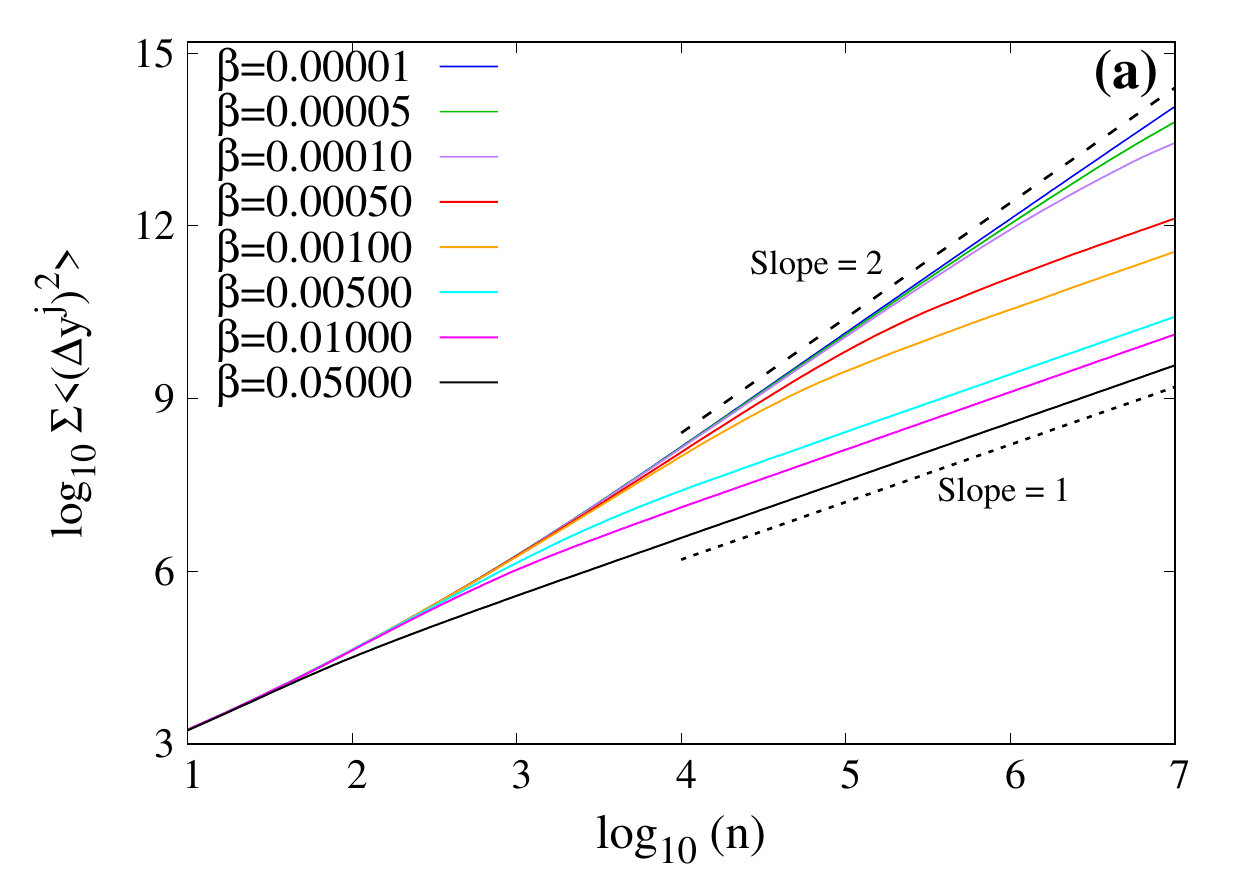}
\includegraphics[width=\columnwidth]{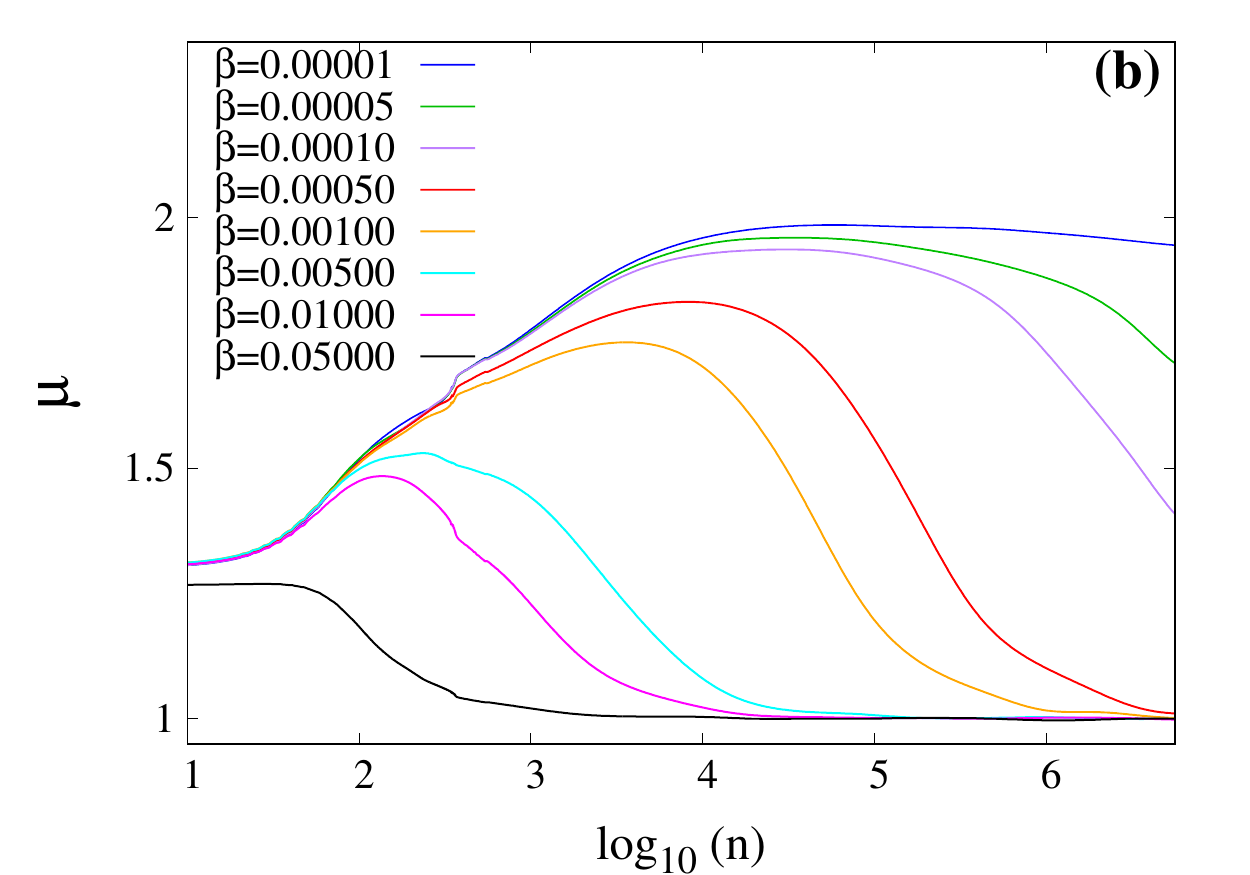}
\includegraphics[width=\columnwidth]{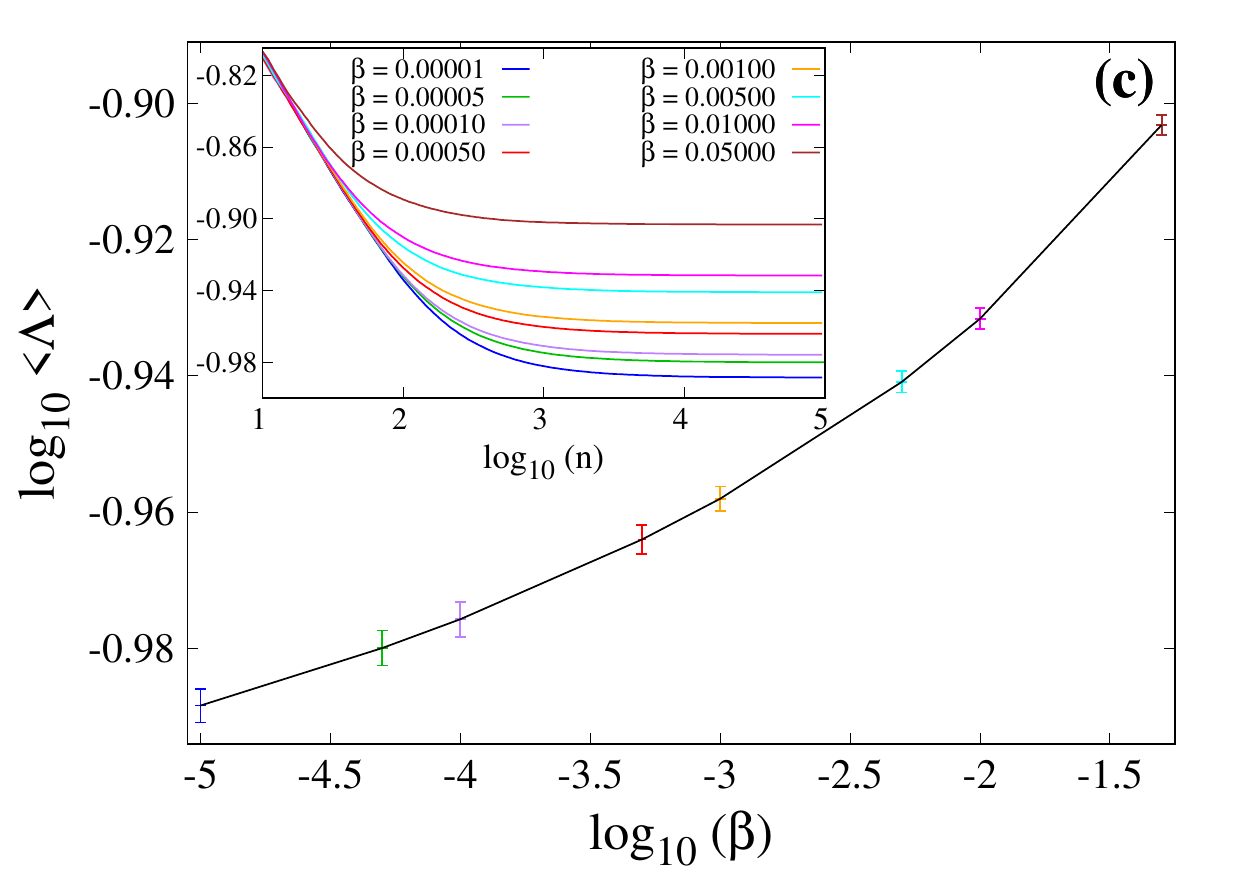}
\includegraphics[width=\columnwidth]{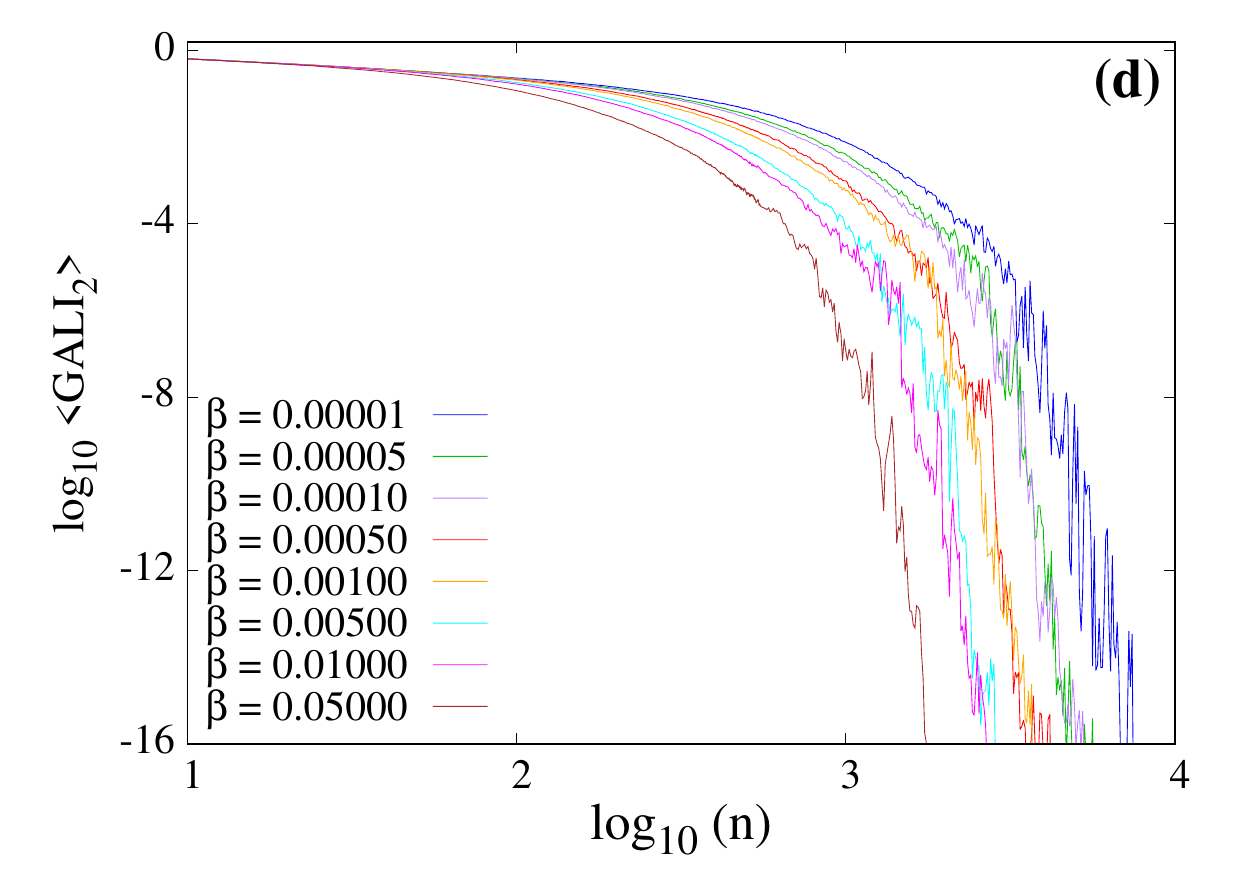}
\caption{(a) The variance $\sum_{j=1}^{N} \left< (\Delta y^j)^2 \right> $ [Eq.~\eqref{Deff_csm}] of the coupled SMs system \eqref{eq:csm} with $N=5$ and $K_j=K=6.5$, $j=1,2,\ldots, 5$, for various values of the coupling-strength parameter $\beta$ (given in the legend) for the same setup and ensemble of ICs described in Figs.~\ref{fig:8} and~\ref{fig:9}, as a function of the system's iterations $n$. The black dotted and dashed lines  respectively correspond to $\mu=1$ and $\mu=2$. (b) The corresponding numerically computed diffusion exponent $\mu$ [Eq.~\eqref{eq:yvarN}] as a function of the map's iterations $n$. For each $\beta$ value  the coupled system gradually shifts from superdiffusion ($\mu > 1$) to normal diffusion ($\mu = 1$). (c) Average (over $\approx100,000$ ICs) value $\left< \Lambda \right>$ of the ftMLE \eqref{eq:ftMLE} as a function of $\beta$  after $n=10^5$ iterations. The error bars denote one standard deviation in the computation of the average value. Inset: the evolution of $\left< \Lambda \right>$ with respect to the number of iterations $n$. Higher $\langle \Lambda \rangle$ values correspond to larger $\beta$, for which  normal diffusion rates are also observed. (d) Average $\left< \rm{GALI}_{2} \right>$  values \eqref{eq:GALI}  for the same set of ICs  as in (c). All $\left< \rm{GALI}_{2} \right>$ decay exponentially fast to zero as the collective dynamics of the particular ensembles of ICs is chaotic. The respective standard deviations (not plotted here) are very small and hardly visible.
}
\label{fig:10}
\end{figure*}

We first consider a system of coupled SMs with similar individual setups for each one of its 2D maps, i.e.~keeping  equal kick-strength parameters  for all maps to $K_j=K$, $j=1,2,\dots,5$. In Fig.~\ref{fig:8}(a) and in its inset, we respectively plot the diffusion exponent $\mu$ [Eq.~\eqref{eq:yvarN}] and the effective diffusion coefficient $D_{\rm eff}^{N}$ \eqref{Deff_csm}  as functions of $K$, for  map \eqref{eq:csm} with $N=5$ and  $\beta=10^{-4}$ (a moderate coupling value allowing dynamical effects to take place within feasibly CPU times). These results are obtained for a set of ICs on a $315 \times 315$ grid covering the whole phase space of each one of the coupled 2D SMs. We note that, after experimenting with different ensemble sizes (total number of ICs), we found that the considered number of ICs is sufficiently large to correctly capture the basic dynamical features of the system, and at the same time small enough to allow the performance of extensive numerical computations in feasible computational times. Both the diffusion exponent $\mu$ and the effective diffusion coefficient $D_{\rm eff}^{N}$ are computed after $n=10^4$ iterations. Fig.~\ref{fig:8}(b) depicts a similar analysis to the one of Fig.~\ref{fig:8}(a) but for a larger coupling-strength value, namely $\beta = 10^{-3}$. It becomes evident that this increase  affects the diffusion exponent $\mu$, especially in intervals of $K$ values where AMs are present. In particular, $\mu$ gradually decreases to values closer to $\mu=1$ (normal diffusion) and this decline is more pronounced at larger $K$ values.

Keeping intact the general system's setup (i.e.~the kick-strength parameters being equal in all coupled SMs) and using the same ensemble of ICs, we probed the decaying scaling laws of the highest ($\mu^*$) values of the diffusion exponent $\mu$ [Eq.~\eqref{eq:yvarN}] (the pronounced peak locations denoted by black filled circles in Fig.~\ref{fig:8}), as a function of $K$ for increasing $\beta$ values. These $\mu^*$ values lie in the first 11 intervals of the kick-strength parameter values $K$  [Eq.~\eqref{eq:acmdint}] for which period $p=1$ AMs exist for the 2D SM \eqref{eq:sm}. The obtained results are shown in Fig.~\ref{fig:9}. The solid curves in Fig.~\ref{fig:9}(a) correspond to fittings of the data, obtained after $n=10^4$ iterations, with functions $\mu^*=A\textprime K^{B\textprime}$, for each presented case  [the considered $K$ values are given in the legend of Fig.~\ref{fig:9}(d)]. From these curves it becomes evident that strong coupling between neighboring 2D maps tends to evoke a global normal diffusive transport, even when initially the 2D maps may include ensembles of ICs leading to superdiffusive rates due to the presence of AMs. We can also observe that for coupling-strength values $\beta \gtrsim 0.005$ the system acquires rather rapidly global normal diffusion rates even for small $K$ values [Fig.~\ref{fig:9}(a)]. The obtained values of the fitting parameters $A\textprime$ and $B\textprime$ (and their standard deviations), as a function of the coupling-strength parameter $\beta$, are respectively given in Figs.~\ref{fig:9}(b) and (c). A clear tendency of $\mu^*$ to become $\mu^*=1$ and independent of $K$ is seen, as $A\textprime$ and $B\textprime$ respectively become $A\textprime=1$ and $B\textprime=0$ [these values are denoted by dashed lines in Figs.~\ref{fig:9}(b) and (c)] for large enough $\beta$ values.

\begin{figure}[h!] \centering
\includegraphics[width=\columnwidth,keepaspectratio]{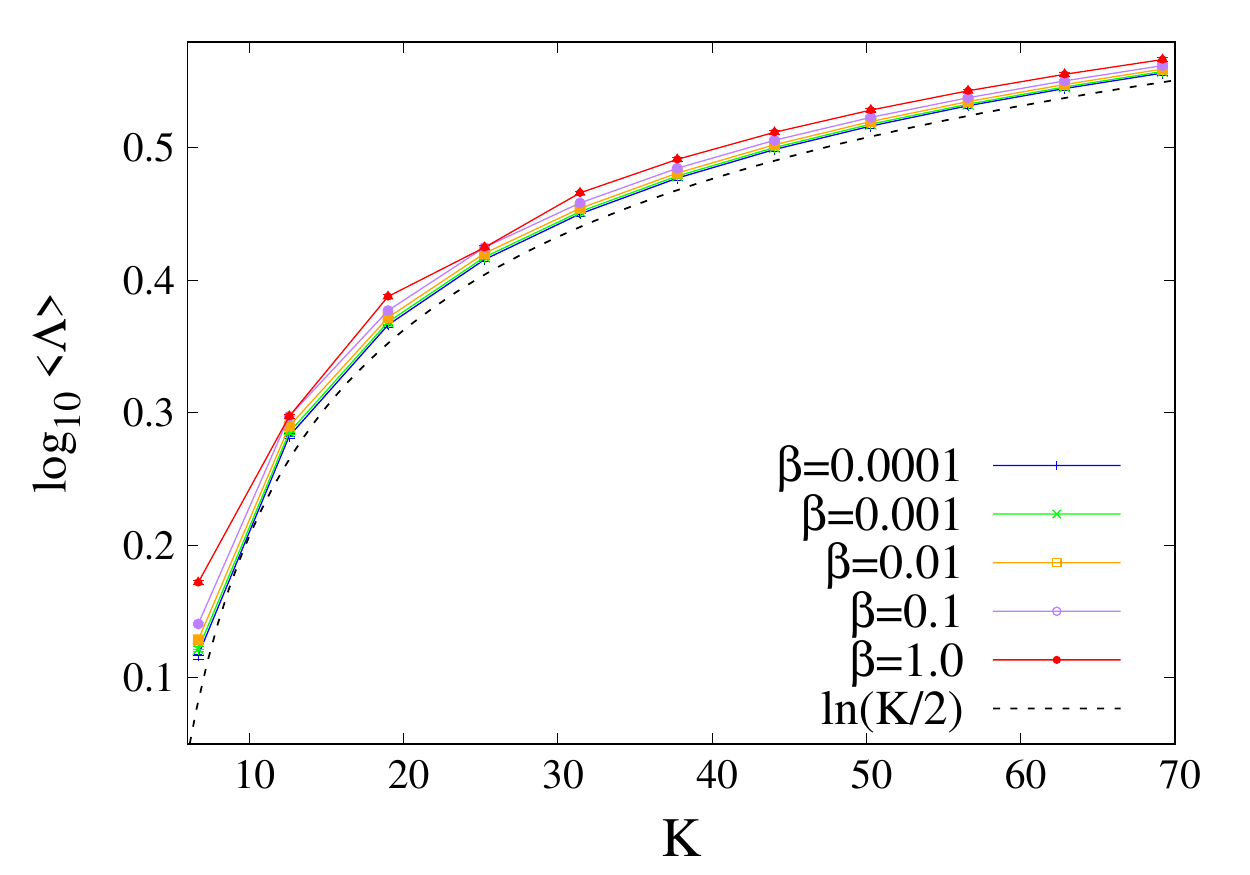}
\caption{
Average [over 10,000 ICs on a $100 \times 100$ grid covering the system's entire phase space (see text for more details)] value $\langle \Lambda \rangle$  of the ftMLE \eqref{eq:ftMLE} of the coupled SMs system \eqref{eq:csm} with $N=5$ and $K_j=K$, $j=1,2,\ldots, 5$, after $n=10^5$ iterations, for the $K$ values used in Fig.~\ref{fig:9} and for different coupling-strengths $\beta=10^{-4}$, $10^{-3}$, $10^{-2}$, $10^{-1}$, $1$ (respectively blue, green, orange, purple and red points/curves). The dashed curve corresponds to the law $\langle \Lambda \rangle = \ln (K/2)$ seen in the case of a single 2D SM (see Fig.~\ref{fig:6}(b)].}
\label{fig:5SM_Lyap}
\end{figure}

Furthermore, and always for the same system setup, we explored the effect of the coupling-strength $\beta$, on the highest values $\mu^*$ of the diffusion exponent $\mu$ for  these 11  particular $K$ values associated with strong superdiffusion. These results are presented in Fig.~\ref{fig:9}(d), where it becomes clear that as the coupling-strength $\beta$ gets stronger the system tends to suppress the superdiffusion and gradually acquires a transport rate $\mu^{*} \approx 1$. This behavior is typically observed in purely chaotic regions in the case of a single SM and for ICs on grids not containing AMs, for which the diffusion is normal and is characterized by $\mu \approx 1$

\begin{figure*}[h!]\centering
\includegraphics[width=\columnwidth]{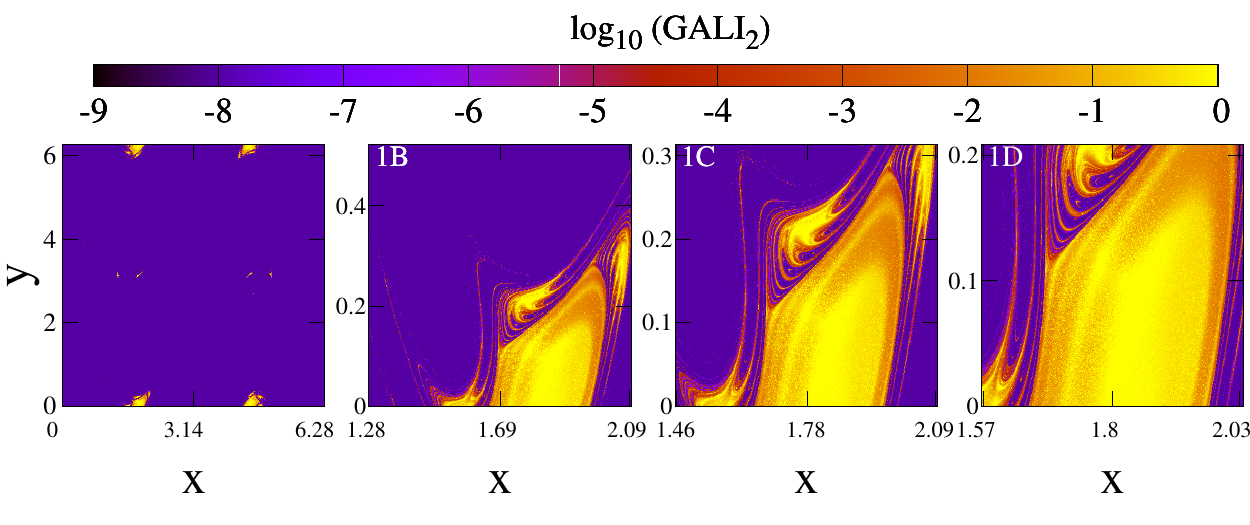}
\vspace{-0.5cm}
\includegraphics[width=\columnwidth]{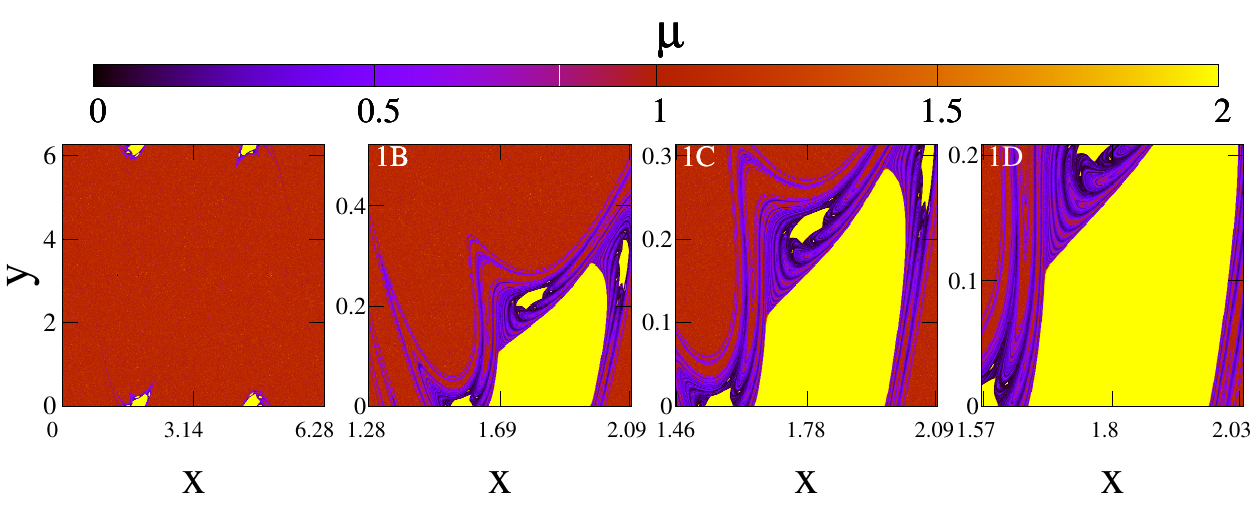}
\caption{Phase space portraits of the 2D SM \eqref{eq:sm} with $K=6.5$ (for which a $p=1$ AM exists; see Table \ref{tb:1}) colored according to the GALI$_2$ \eqref{eq:GALI} values (left four panels), and the diffusion exponent $\mu$ [Eq.~\eqref{eq:yvar}] (right four panels). The left, unlabeled panel in each quartet of figures  depicts an ensemble of ICs on the entire phase space, $[0, 2\pi) \times [0, 2\pi)$, exhibiting extensive chaos ($P_C \approx 99\%$). Panels labelled as 1B, 1C and 1D (as in Table \ref{tb:1}) refer to ensembles of ICs around the AM with respectively $P_C\approx75\%$, $P_C\approx50\%$ and $P_C\approx25\%$.
}
\label{fig:11}
\end{figure*}

\begin{table*} \caption{Nomenclature of different coupled SMs arrangements. In Fig.~\ref{fig:12} we present  results for the coupled system \eqref{eq:csm} with $N=5$ for which each individual 2D map has a \textit{single} (S)  kick-strength value $K_j=K=6.5$, $j=1,2,\ldots, 5$. Note that $K=6.5$ corresponds to the appearance of a $p=1$ AM. In all arrangements (named \textit{Arrangements A}) the considered ICs on the $j=3$, 2D map cover the whole phase space (left panels in the two quartets of panels in Fig.~\ref{fig:11}), while in the other four maps ($j=1,2,4,5$) ICs lie on the phase space regions named 1B, 1C, 1D (see Fig.~\ref{fig:11}, upper row of Fig.~\ref{fig:7} and Table \ref{tb:1}) exhibiting different fractions of chaotic motion: respectively $P_C\approx75\%$, $P_C\approx50\%$ and $P_C\approx25\%$. In Fig.~\ref{fig:13} we show results for a different set of arrangements (named \textit{Arrangements B}) for which the central $j=3$ map corresponds to cases 1B, 1C or 1D and the remaining $j=1,2,4,5$ maps contain ICs on the whole phase space. For all the 6 presented cases the coupling-strength parameter is set to $\beta=0.001$.
}
\label{tb:2}
\centering
\begin{tabular}{l|c|c|c|c|c|c}
\hline	& Case name & Map 1 &  Map 2 & Map 3 & Map 4 & Map 5 \\
\hline  & S75A & 1B & 1B & & 1B & 1B \\
Fig.~\ref{fig:12} & S50A & 1C & 1C & $[0, 2\pi) \times [0, 2\pi)$ & 1C & 1C \\
& S25A & 1D & 1D & & 1D & 1D \\
\hline  & S75B & & & 1B & &  \\
Fig.~\ref{fig:13} & S50B & $[0, 2\pi) \times [0, 2\pi)$ & $[0, 2\pi) \times [0, 2\pi)$ & 1C & $[0, 2\pi) \times [0, 2\pi)$ & $[0, 2\pi) \times [0, 2\pi)$ \\
& S25B & & & 1D & & \\
\hline
\end{tabular}
\end{table*}

\begin{figure*}[]\centering
\includegraphics[width=\columnwidth]{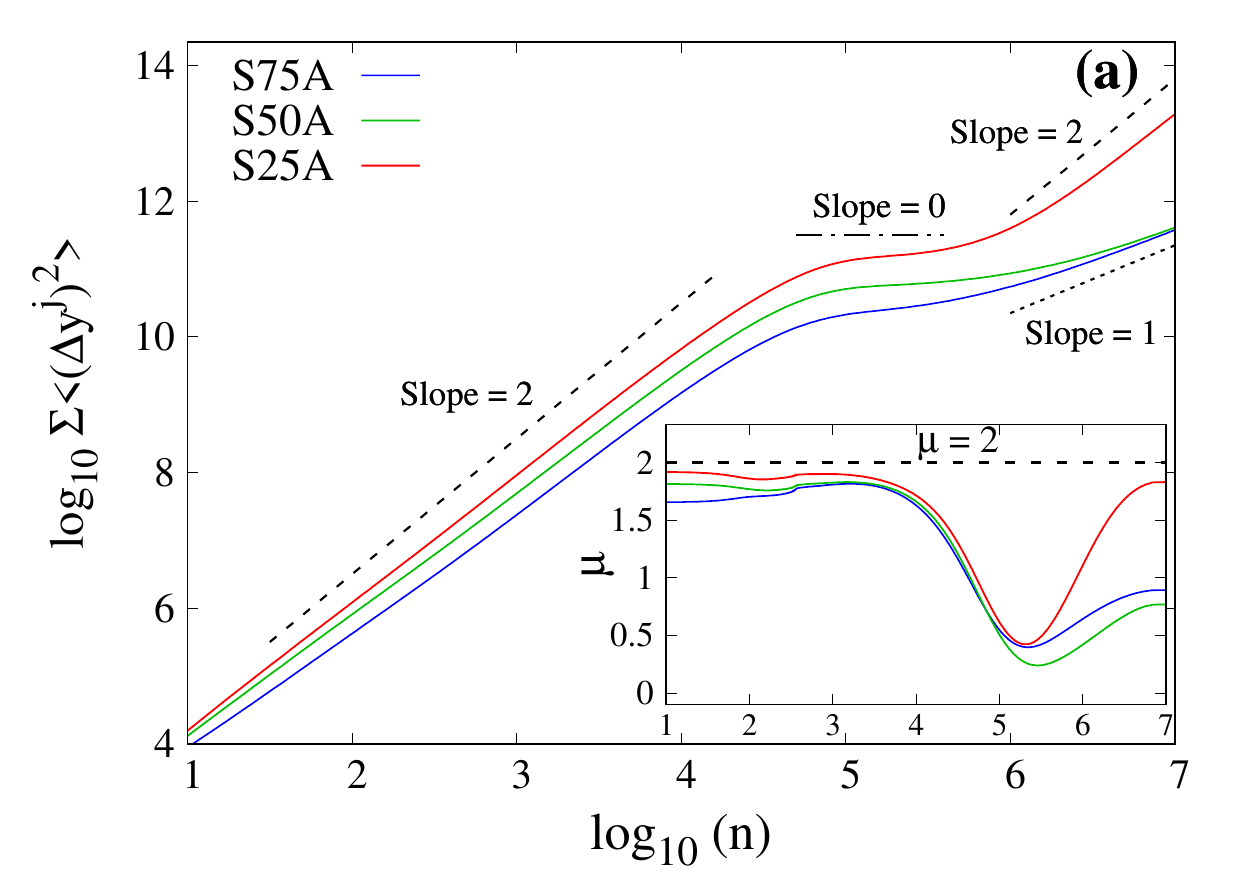}
\includegraphics[width=\columnwidth]{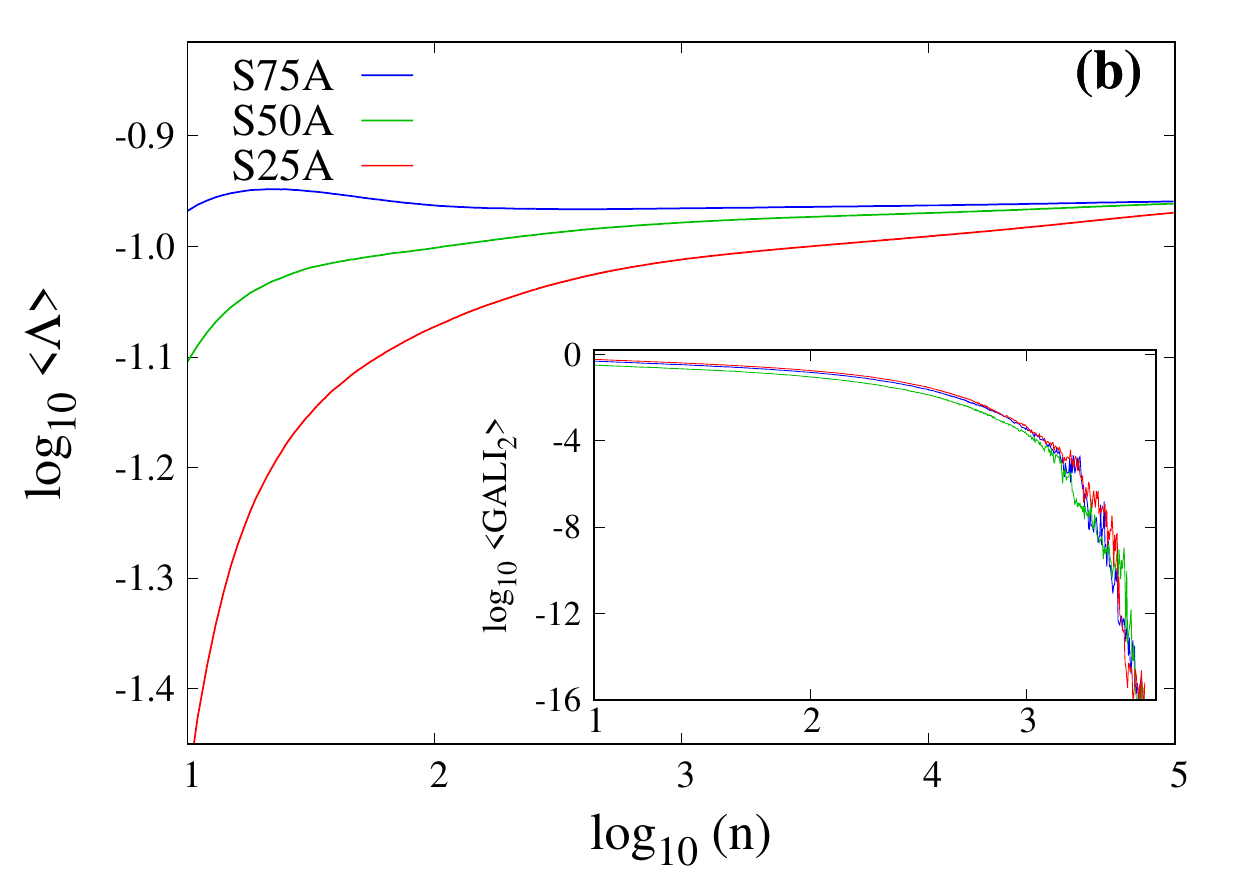}
\caption{Cases S75A (blue curves), S50A (green curves) and S25A (red curves) of the coupled SMs system (see Table \ref{tb:2}), (a) The variance  $\sum_{j=1}^{5} \left< (\Delta y^j)^2 \right>$ of an ensemble  of $\approx 100,000$ ICs, as a function of the system's iterations $n$. Only case S25A shows a tendency to recover ballistic diffusion rate (as a slope $\approx 2$ at the later stages of the evolution indicates), while cases S75A and S50A seem to converge to a normal diffusion rate (slope $\approx 1$). Inset:  the evolution of the diffusion exponent $\mu$ [Eq.~\eqref{eq:yvarN}] as a function of the iterations $n$ for the same cases.
(b) The evolution of the average ftMLE $\left< \Lambda \right>$ as a function of the iterations $n$ for the cases of (a). In all cases  $\left< \Lambda \right>$ converges to approximately the same non-zero value. Inset: the time evolution of the corresponding $\left< \rm{GALI}_{2} \right>$ which decays exponentially to zero as the collective dynamics of these ensembles is chaotic.
}
\label{fig:12}
\end{figure*}

From the analysis of Figs.~\ref{fig:8} and \ref{fig:9} we observe that the coupling-strength $\beta$ operates as a control mechanism, being able to suppress superdiffusion rate ($\mu >1$) into normal diffusion ($\mu \approx 1$). In Sect.~\ref{sec:2D_results}, we performed an investigation of the relation between the delay to reach ballistic transport ($\mu =2$) and the kick-strength parameter $K$ (which affects the fraction of chaotic ICs) in single SMs [see Fig.~\ref{fig:6}(a)]. Here we attempt a similar investigation for the coupled 2$N$D system \eqref{eq:csm} trying to associate the number of iterations required for the diffusion exponent $\mu$ [Eq.~\eqref{eq:yvarN}] to reach the value $\mu =2$ with the  value of  $\beta$.

We first plot in Fig.~\ref{fig:10}(a) the variance  $\sum_{j=1}^{N} \left< (\Delta y^j)^2 \right>$ [Eq.~\eqref{eq:yvarN}] for the same case of coupled SMs we considered so far in Figs.~\ref{fig:8} and \ref{fig:9}  (i.e.~$N=5$, fixed $K_j=K=6.5$) for increasing values of the coupling-strength parameter $\beta$. Fig.~\ref{fig:10}(b) shows the corresponding numerically computed diffusion exponent $\mu$ [Eq.~\eqref{eq:yvarN}] as a function of the map's iterations $n$. One can notice that for all $\beta$ values  the coupled system gradually shifts from superdiffusion ($\mu > 1$) to normal diffusion ($\mu =1$). This transition happens faster as $\beta$ grows.

\begin{figure*}[h!]\centering
\includegraphics[width=\columnwidth]{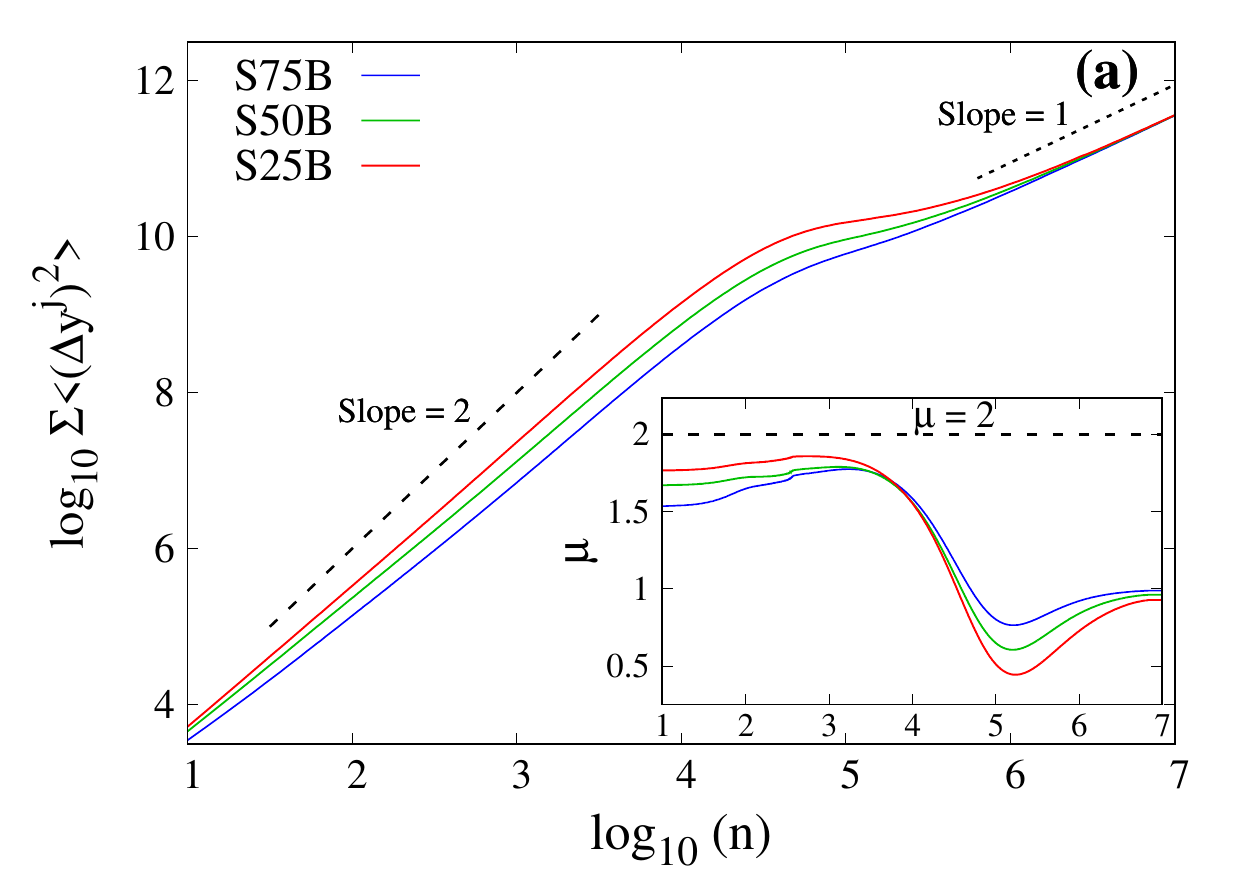}
\includegraphics[width=\columnwidth]{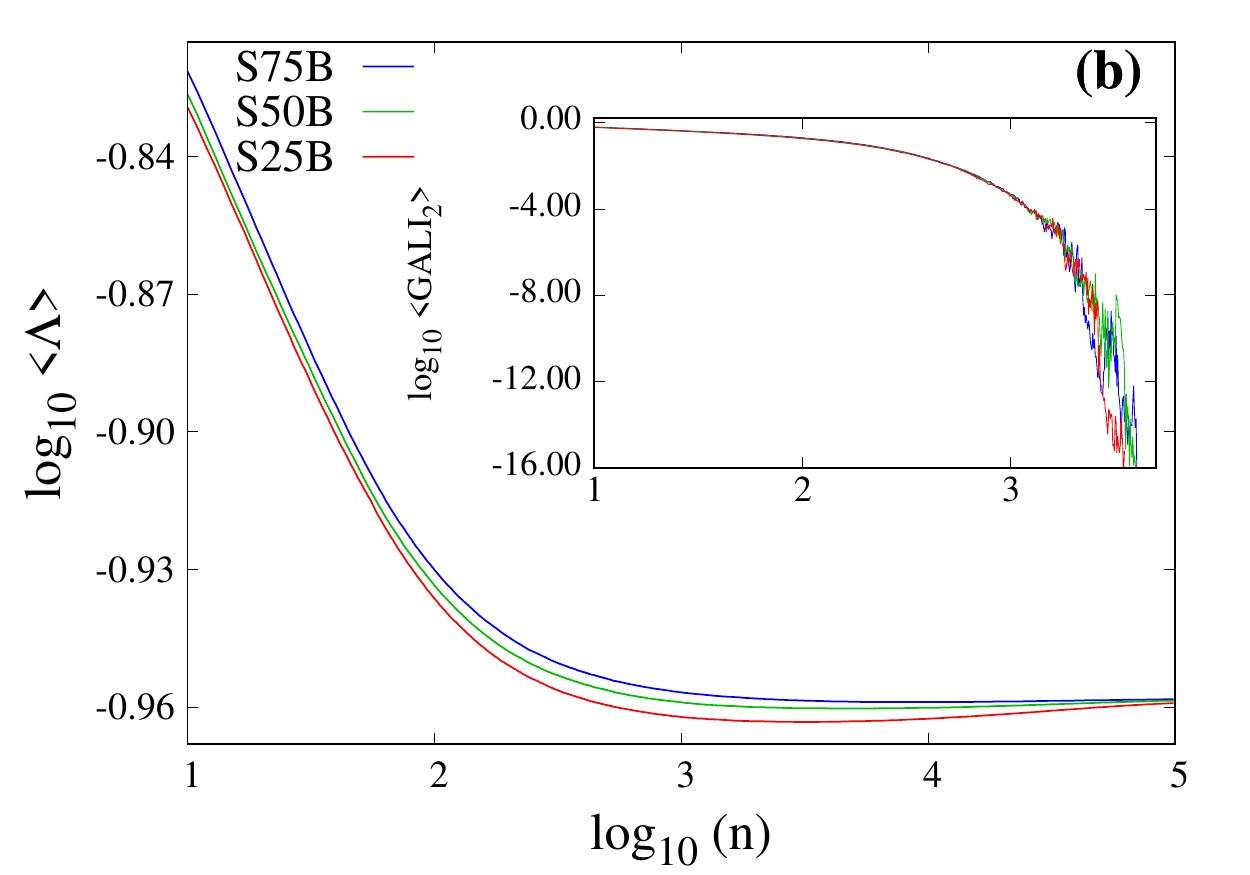}
\caption{Similar to Fig.~\ref{fig:12} but for cases S75B (blue curves), S50B (green curves) and S25B (red curves) of Table \ref{tb:2}.
}
\label{fig:13}
\end{figure*}

In order to investigate the system's chaoticity we present in Fig.~\ref{fig:10}(c)  the average (over all considered ICs) value $\left< \Lambda \right>$ of the ftMLE \eqref{eq:ftMLE} for increasing values of $\beta$, after $n=10^5$ iterations, while the index's evolution is depicted in the figure's inset. Note that larger $\langle \Lambda \rangle$ values correspond to stronger coupling-strength cases, which exhibit normal diffusion rates. In other words, for this particular setup of the coupled SMs and ensembles of ICs, stronger coupling leads to stronger chaos as measured by the average ftMLE. This is also evident by calculating the average $\left< \rm{GALI}_{2} \right>$ values [Fig.~\ref{fig:10}(d)], as all $\left< \rm{GALI}_{2} \right>$  exponentially decay  to zero, indicating collective chaotic behavior, which is also associated with the positive average ftMLE. Moreover, we observe that this $\left< \rm{GALI}_{2} \right>$ rate of decay is related to the $\left< \Lambda \right>$ values, i.e.~the smaller  $\left< \Lambda \right>$ is the slower  $\left< \rm{GALI}_{2} \right>$ tends to zero [compare for example the blue ($\beta=0.00001$) and the  brown ($\beta=0.05$) curves in Figs.~\ref{fig:10}(c) and \ref{fig:10}(d)]. We note that the relation between the rate of exponential decay of the GALI$_2$ and the ftMLE was also mentioned in Sect.~\ref{sec:NumTech}  (see also \cite{SBA2007} for more details).

Trying to further investigate the effect of the coupling-strength parameter $\beta$ on the chaotic behavior of the coupled SMs we present in Fig.~\ref{fig:5SM_Lyap} the system's $\left< \Lambda \right>$ for the $K$ values considered in Fig.~\ref{fig:9} and for different $\beta$ values covering a spectrum of 4 orders of magnitude (from $\beta=10^{-4}$ to $\beta=1$). We note that $\left< \Lambda \right>$ is computed as the average ftMLE over 10,000 ICs (i.e.~points on a  $100 \times 100$ grid covering the system's entire phase space) after $n=10^5$ iterations. From the results of Fig.~\ref{fig:5SM_Lyap} we see that $\left< \Lambda \right>$ increases as $K$ grows for fixed $\beta$ values, following a trend similar to the one seen in Fig.~\ref{fig:6}(b) for the single 2D SM (denoted by the dashed curve  in Fig.~\ref{fig:5SM_Lyap}), having values (in $\log_{10}$ scale) slightly larger than the ones obtained in that case. Furthermore, a steady increase of $\log_{10} \left< \Lambda \right>$ for increasing $\beta$ values is observed for each considered $K$ value. Thus, for relative large $K$ values in the range $6.5 \lesssim K \lesssim 70$, increasing the coupling between identical 2D SMs results to chaotic motion described by slightly growing ftMLEs, with respect to the ones observed for the single 2D SM.

\subsubsection{Coupled 2D SMs with equal kick-strengths  and different fractions of chaos around their AMs} \label{sec:C_SMs_eq_K}

\noindent In what follows, we examine the role of the total amount (percentage) of chaotic ICs in the evolved ensembles on the long-term diffusion rate properties for systems of coupled SMs with homogeneous kick-strengths $K_j=K$, $j=1,2, \dots, 5$. The chaoticity  and diffusion rate of these particular ensembles of ICs are shown in Fig.~\ref{fig:11} for the case of $K=6.5$, for which an AM of period $p=1$ exists (see Table~\ref{tb:1}). The first four left panels show phase space color plots based on GALI$_2$ calculations, while the remaining four panels are color plots based on the value of the diffusion exponent $\mu$ [Eq.~\eqref{eq:yvar}], in a similar fashion to  Fig.~\ref{fig:7}. The first, left, unlabeled panel in each group depicts an ensemble of ICs covering the entire phase space $[0, 2\pi) \times [0, 2\pi)$, for which extensive chaos is present ($P_C\approx 99\%$). Panels 1B, 1C and 1D refer to ensembles of ICs (the explicit ranges of these regions are given in Table~\ref{tb:1}) around an AM of period $p=1$, containing respectively  $P_C\approx 75\%$, $P_C\approx 50\%$ and $P_C \approx 25\%$  chaotic orbits.

We use these ensembles of ICs to form different coupled SMs \eqref{eq:csm}  \textit{Arrangements} and we measure again the resulting diffusion exponent $\mu$ [Eq.~\eqref{eq:yvarN}] and average chaos indicators $\left< \Lambda \right>$ and $\left< \rm{GALI}_{2} \right>$, by setting the coupling-strength parameter at a moderate fixed value, $\beta=0.001$. These choices result to  coupled SMs setups which allow us to investigate, in feasible CPU times, the effect of ICs with different chaos percentages on the diffusion properties in the presence of $p=1$ AMs in the 2D map components of the coupled systems.

The first setup we considered (named \textit{Arrangement A}) comprises of coupled 2D SMs with \textit{single} (S) (i.e.~the same) $K$ values, $K_j=K=6.5$, $j=1,2,\ldots,5$, for which an AM of period $p=1$ exists, and the same  central ($j=3$) map  arrangement for all cases, having ICs covering the entire phase space (see left panels in each quartet of figures in Fig.~\ref{fig:11}). In the other four maps the ICs lie on regions containing  different fractions of chaotic motion, namely $P_C \approx75 \%$, $P_C \approx 50 \%$ and $P_C \approx 25\%$ (these numerical values are used in defining the name of each set up in Table \ref{tb:2}). The second setup type (\textit{Arrangement B}) consists of coupled 2D SMs with again \textit{single} (S)  $K$ values ($K_j=K=6.5$) for which the $j=3$ map corresponds to regions  with various fractions of chaotic motion ($P_C \approx75 \%$, $P_C \approx 50 \%$ and $P_C \approx 25\%$), while the remaining  maps are identical having ICs chosen on a grid covering the entire phase space of the 2D map.

\begin{figure*}[h!]\centering
\includegraphics[width=\columnwidth]{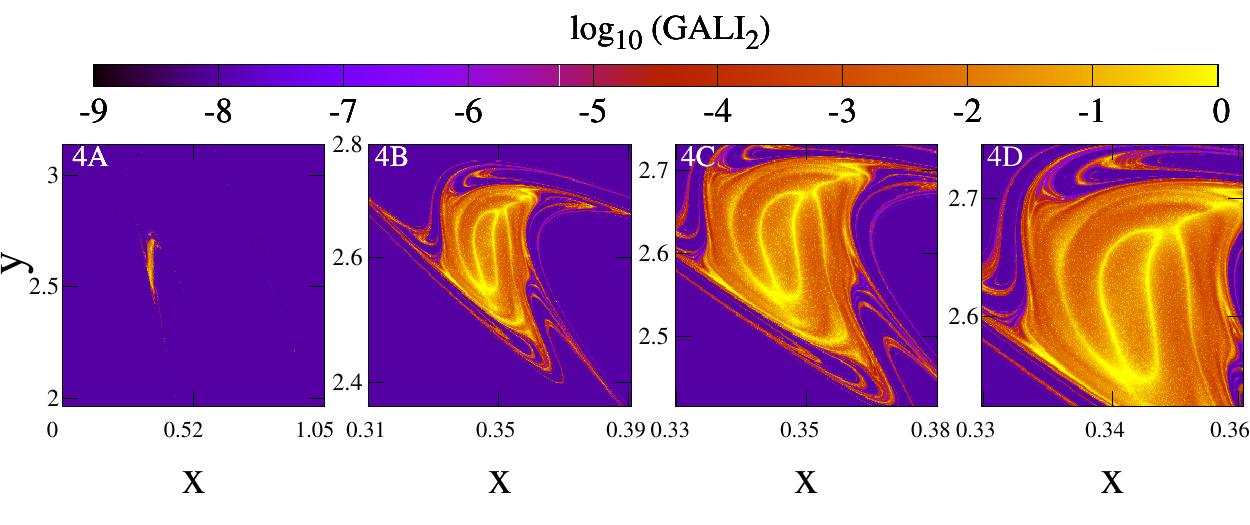}
\vspace{-0.5cm}
\includegraphics[width=\columnwidth]{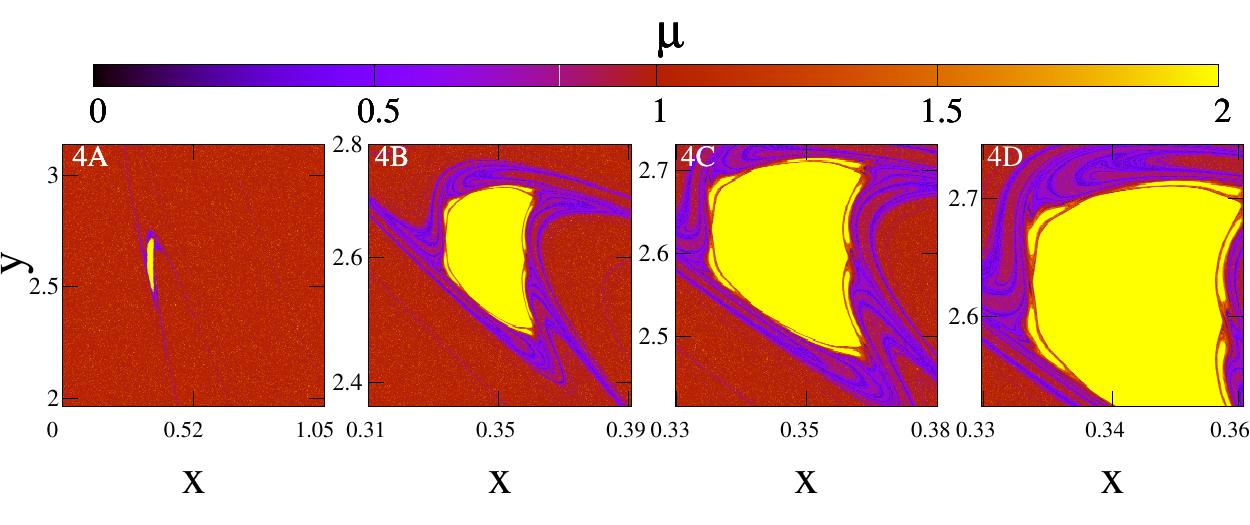}
\caption{Phase space portraits of the 2D SM \eqref{eq:sm} with $K=3.1$ (for which a $p=4$ AM exists; see Table \ref{tb:1}) colored according to the GALI$_2$ \eqref{eq:GALI} values (left four panels), and the diffusion exponent $\mu$ [Eq.~\eqref{eq:yvar}] (right four panels). Panels labelled as 4A, 4B, 4C and 4D (as in Table \ref{tb:1}) refer to ensembles of ICs around the AM with respectively $P_C\approx99\%$, $P_C\approx75\%$, $P_C\approx50\%$ and $P_C\approx25\%$.
}
\label{fig:14}
\end{figure*}

\begin{table*}[h!] \caption{Nomenclature of different coupled SMs arrangements. In Fig.~\ref{fig:15} we present  results for the coupled system \eqref{eq:csm} with $N=5$, having  \textit{mixed} (M) $K$ values: always $K_3=3.1$ (corresponding to the existence of an  AM of period $p=4$) for the  central map ($j=3$), for which ICs are chosen on a grid covering the 4A region of Fig.~\ref{fig:14} (see also Table \ref{tb:1}), while the other four maps have $K_1=K_2=K_4=K_5=6.5$ (corresponding to the appearance of a $p=1$ AM) with ICs lying  on phase space regions 1B, 1C and 1D (see Fig.~\ref{fig:11}, upper row of Fig.~\ref{fig:7} and Table \ref{tb:1}) with $P_C\approx75\%$, $P_C\approx50\%$ and $P_C\approx25\%$ (\textit{Arrangements A}). In Fig.~\ref{fig:16} we show results for another set of arrangements (named \textit{Arrangements B}) for which the $j=3$ map corresponds to a 2D map with $K_3=6.5$ and ICs on its entire phase space (left panels in the two quartets of panels in Fig.~\ref{fig:11}), and the remaining  maps have $K_1=K_2=K_4=K_5=3.1$, belonging to the cases 4B, 4C and 4D (see Fig.~\ref{fig:14}, lower row of Fig.~\ref{fig:7} and Table \ref{tb:1}) with $P_C\approx75\%$, $P_C\approx50\%$ and $P_C\approx25\%$. For all the 6 presented cases the coupling-strength parameter is set to $\beta=0.001$.
} \label{tb:3}
\centering
\begin{tabular}{l|c|c|c|c|c|c}
\hline	& Case name & Map 1 &  Map 2 & Map 3 & Map 4 & Map 5 \\
\hline  & M75A & 1B & 1B & & 1B & 1B \\
Fig.~\ref{fig:15} & M50A & 1C & 1C & 4A &  1C & 1C \\
& M25A & 1D & 1D & ($K=3.1$ - AM period: 4) & 1D & 1D \\
\hline  & M75B & 4B & 4B & & 4B & 4B \\
Fig.~\ref{fig:16} & M50B & 4C & 4C & $[0, 2\pi) \times [0, 2\pi)$ & 4C & 4C \\
& M25B & 4D & 4D & ($K=6.5$ - AM period: 1) & 4D & 4D \\
\hline
\end{tabular}
\end{table*}

Our findings regarding \textit{Arrangement A} are shown in Fig.~\ref{fig:12} where we consider the three cases (referring to different fractions of chaotic motion) described in Table~\ref{tb:2}. For example, the case with name S75A refers to a system of 5 coupled 2D SMs with \textit{single} kick-strength values, namely $K_j=K=6.5$, for which an  AM of period $p=1$ exists. In this case the ICs in the central ($j=3$) map are on a grid covering the entire phase space, while the other four maps are identical having  ICs corresponding to a $P_C \approx75 \%$ level of chaotic orbits. Fig.~\ref{fig:12}(a) shows the evolution of $\sum_{j=1}^{5} \left< (\Delta y^j)^2 \right>$ [Eq.~\eqref{Deff_csm}] for $\approx 100,000$ ICs of the coupled system. We can see that only case S25A  tends to eventually recover a ballistic diffusion rate, while cases S75A and S50A seem to converge to normal diffusion. In the inset we show the evolution of the diffusion exponent $\mu$ [Eq.~\ref{eq:yvarN}] as a function of the system's iterations $n$ for the same 3 cases. The evolution of the average ftMLE $\left< \Lambda \right>$ as a function of the iterations $n$ is given in Fig.~\ref{fig:12}(b). We see that  all $\langle \Lambda \rangle$ values show a tendency to converge towards a similar, positive value indicating chaotic behavior. An important  observation here is that the larger the fraction of chaotic ICs is in the non-central maps the faster the convergence to this asymptotic $\langle \Lambda \rangle$ value takes place. The inset of Fig.~\ref{fig:12}(b) shows the  evolution of the corresponding $\left< \rm{GALI}_{2} \right>$ values which decay exponentially fast to zero as the collective dynamics of this particular ensembles of ICs is chaotic.

The investigation of cases belonging to what we call \textit{Arrangement B} reveals different trends and the corresponding results are presented in Fig.~\ref{fig:13}. We here consider cases S75B, S50B and S25B (see Table \ref{tb:2} for more information). For instance, case S50B refers to a setup for which again  \textit{single} (S) $K$ values ($K=6.5$) are considered with the central ($j=3$) map corresponding to a  $P_C \approx 50 \%$ level of chaos, while the four other maps have ICs on the entire phase space (see also Fig.~\ref{fig:11}). Fig.~\ref{fig:13}(a) depicts results similar to the ones of  Fig.~\ref{fig:12}(a). All cases show a common tendency to converge to normal diffusion rate ($\mu \approx 1$), something  which is also seen in the inset of Fig.~\ref{fig:13}(a), as the diffusion exponent $\mu$ of these particular cases does not show signs of approaching an asymptotic extreme rate (ballistic transport), at least within the considered  time scales. Regarding the average global chaoticity of these ensembles of ICs, we observe a more uniform evolution both for the ftMLEs $\langle \Lambda \rangle$ [Fig.~\ref{fig:13}(b)], which converge to approximately the same non-zero value, and the $\left< \rm{GALI}_{2} \right>$ values [inset of Fig.~\ref{fig:13}(b)], which follow very similar exponential decay rates.

\begin{figure*}[h!]\centering
\includegraphics[width=\columnwidth]{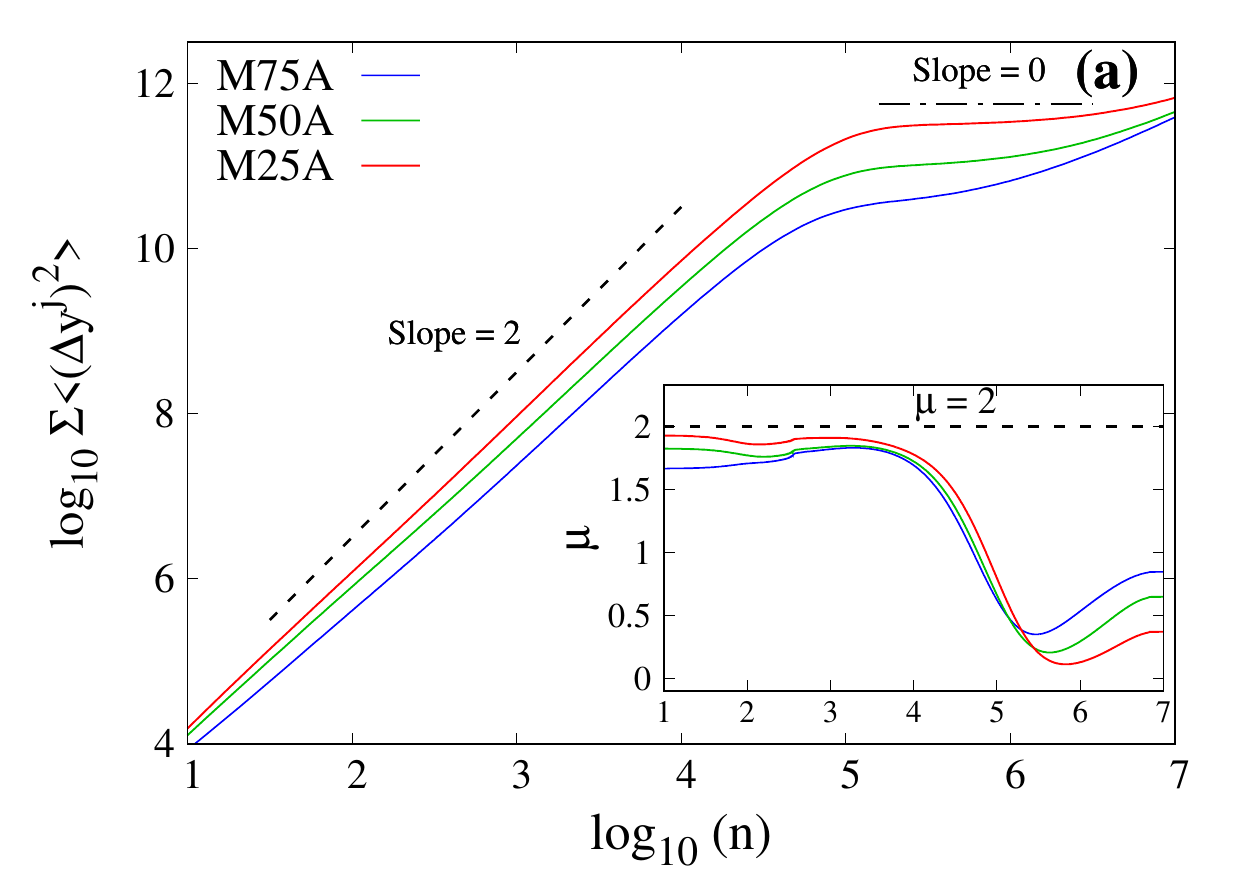}
\includegraphics[width=\columnwidth]{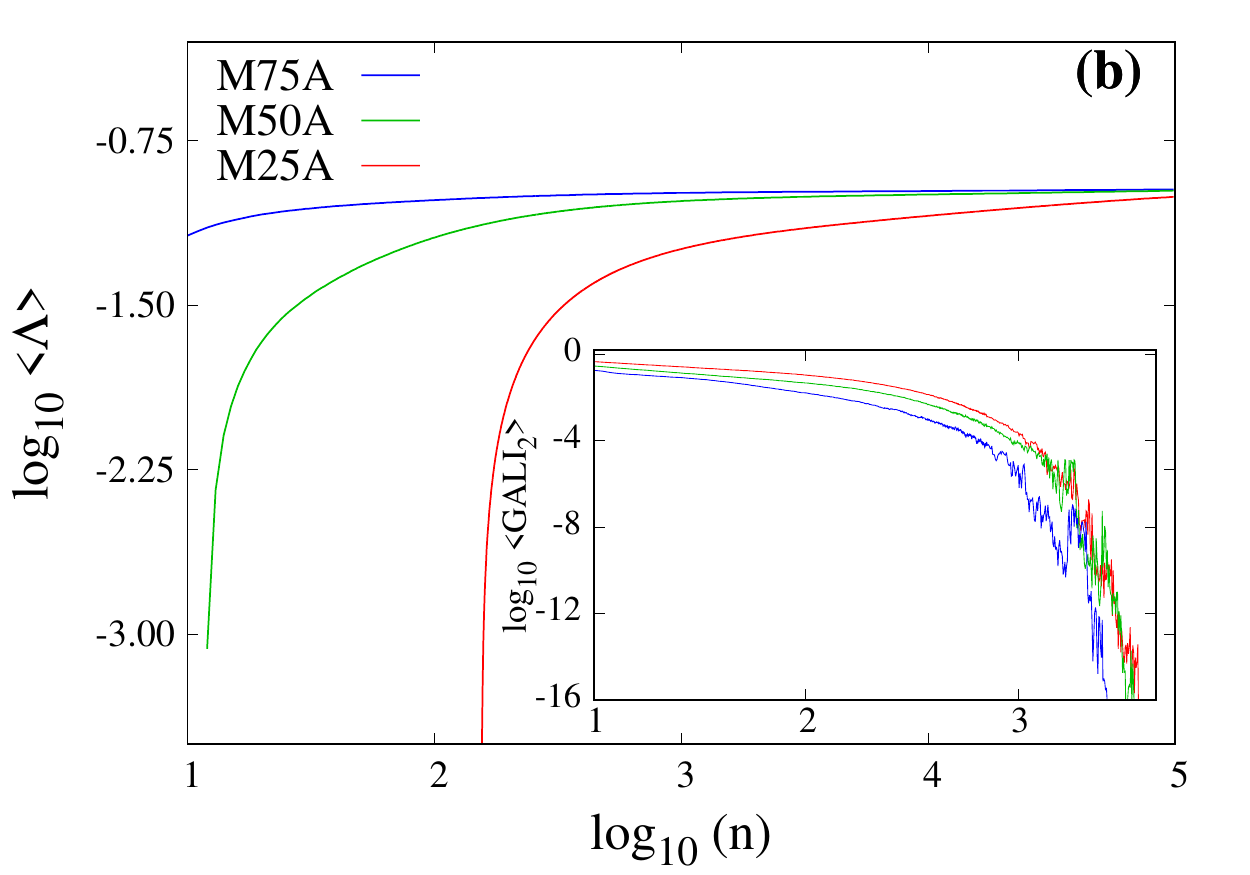}
\caption{Similar to Fig.~\ref{fig:12} but for cases M75A (blue curves), M50A (green curves) and M25A (red curves) of Table \ref{tb:3}.
}
\label{fig:15}
\end{figure*}

\begin{figure*}\centering
\includegraphics[width=\columnwidth]{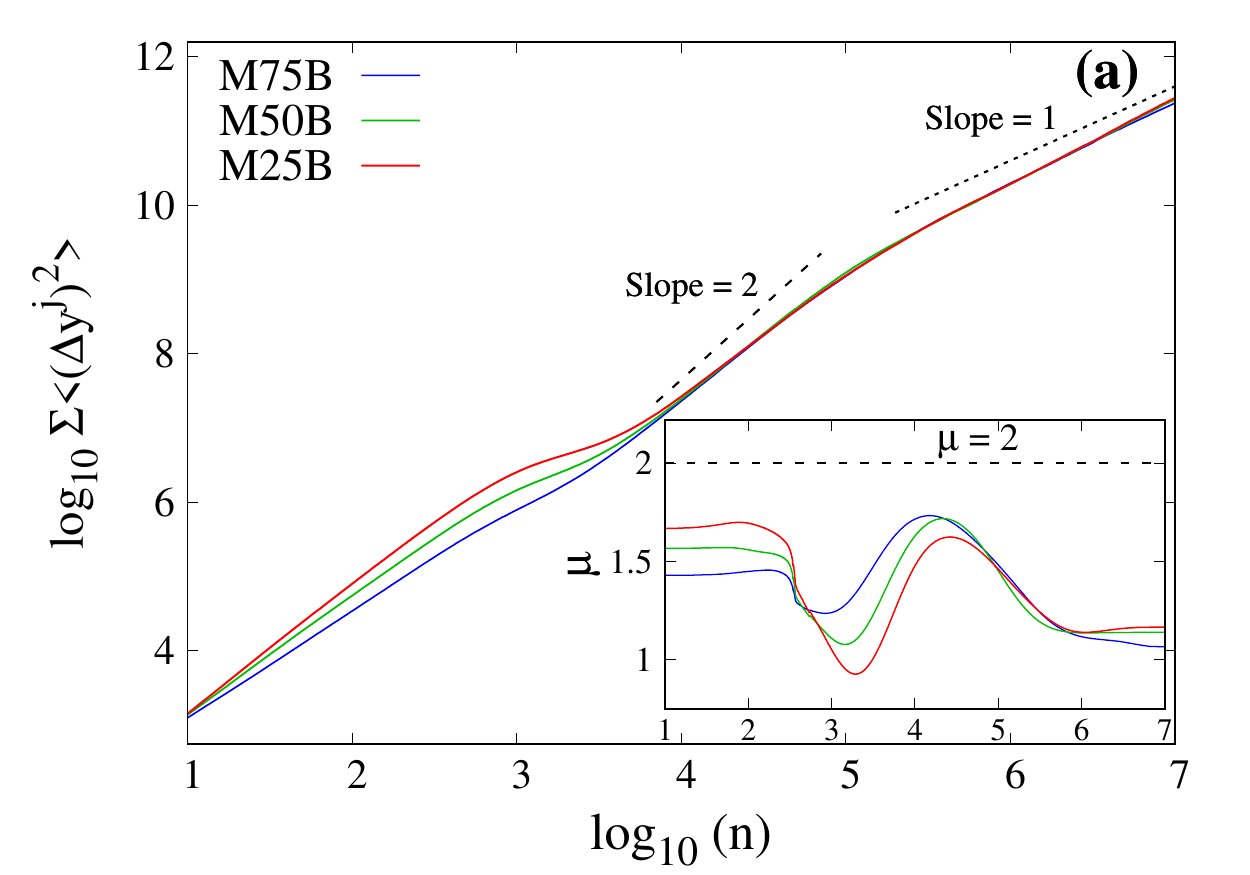}
\includegraphics[width=\columnwidth]{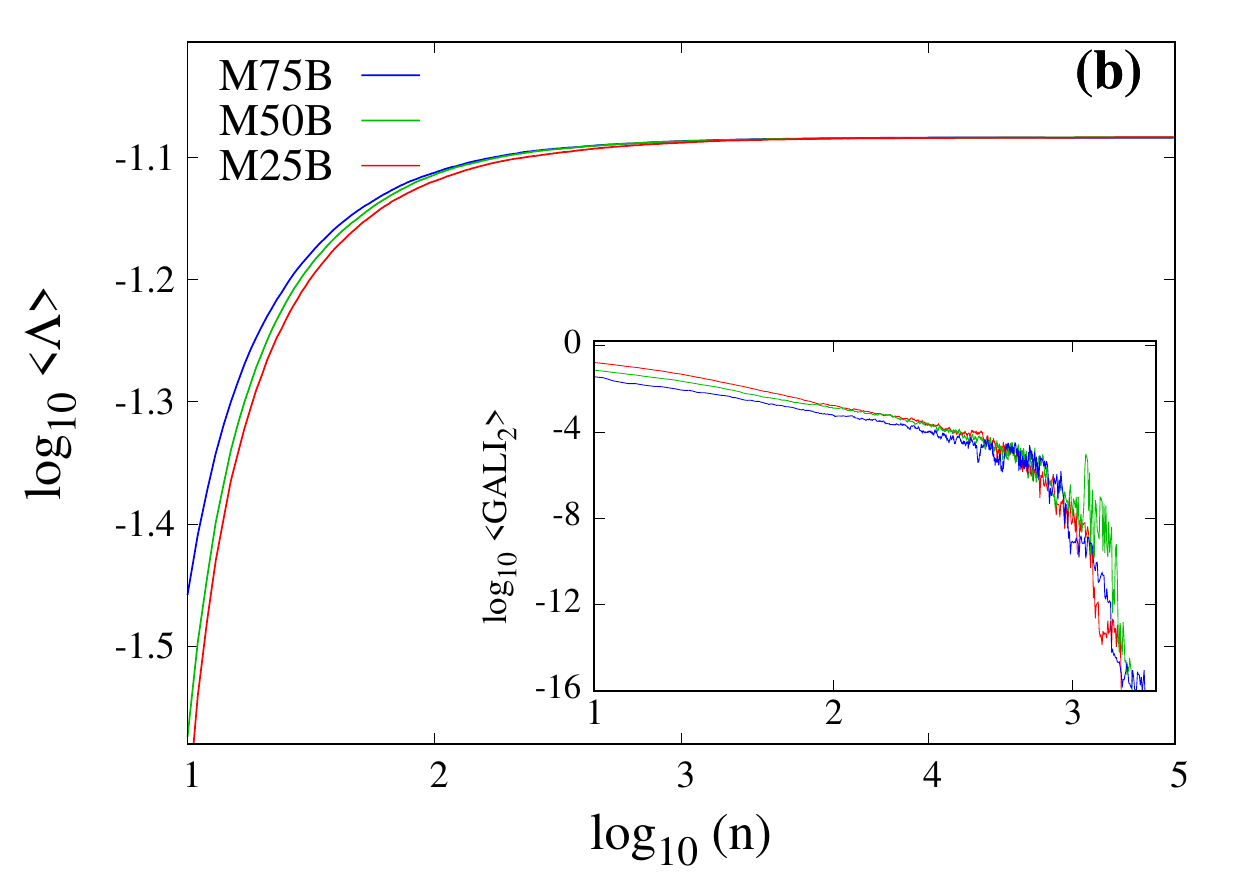}
\caption{Similar to Fig.~\ref{fig:12} but for cases M75B (blue curves), M50B (green curves) and M25B (red curves) of Table \ref{tb:3}
}
\label{fig:16}
\end{figure*}

\subsubsection{Coupled 2D SMs with different kick-strengths and AMs' periods} \label{sec:C_SMs_noneq_K}

Let us now extend our analysis by considering alternative coupled SMs configurations, namely setups with different kick-strength values for individual 2D maps containing AMs of different periods. Similarly to the study  performed in Sect.~\ref{sec:C_SMs_eq_K}, we explore diffusion rate properties and the system's global chaoticity for some cases belonging to what we name  \textit{Arrangements A} and \textit{B}, including SMs with  different fractions of chaotic ICs and various [\textit{mixed} (M)] values of kick-strengths. The details of these configurations are given in Table~\ref{tb:3}. \textit{Arrangements A} contain in the off-center maps ($j=1,2,4,5$) ensembles on regions 1B, 1C and 1D (see Fig.~\ref{fig:11}, upper row of Fig.~\ref{fig:7} and Table \ref{tb:1}), while \textit{arrangements B} include the cases 4B, 4C and 4D presented in the upper row of Fig.~\ref{fig:7} and in Table \ref{tb:1}. In addition, color plots of $\mu$ and GALI$_2$ values for the latter 2D map regions are seen in  Fig.~\ref{fig:14}. More specifically, region 4A corresponds to  an ensemble of ICs within the intervals $[0.0,1.047] \times [1.964,3.142]$, where an AM of period $p=4$ exists, and exhibits extensive chaos ($P_C\approx99\%$). Regions 4B, 4C and 4D refer to ensembles of ICs around the same AM of period $p=4$, respectively containing $P_c\approx75\%$, $P_c\approx50\%$ and $P_c\approx25\%$ of chaotic ICs.

The results for  \textit{Arrangement A} (where the central map  contains an AM of period $p=4$, while the other maps contain AMs of period $p=1$) are presented in Fig.~\ref{fig:15}. From the results of Fig.~\ref{fig:15}(a) we see that all cases initially exhibit superdiffusive spreading ($\mu \approx2$), while after some iterations ($n \approx10^5$) they show a tendency to halt this process, exhibiting very low subdiffusion rates ($\mu \rightarrow 0$). However, towards the end of the simulation the respective rates show a tendency to increase again. These changes are better depicted in the inset of Fig.~\ref{fig:15}(a) where the evolution of the diffusion exponent $\mu$ [Eq.~\eqref{eq:yvarN}] is given as a function of the number $n$ of iterations. The respective ftMLE $\left< \Lambda \right>$  converges to approximately the same positive value for all cases [Fig.~\ref{fig:15}(b)]. Analogous trends for $\langle \Lambda \rangle$ to the ones presented in Fig.~\ref{fig:12}(b) are also observed here: the larger the fraction of chaotic ICs is in the non-central maps, the faster the convergence to the index's asymptotic value happens. In addition, $\left< \rm{GALI}_{2} \right>$  decreases exponentially fast to zero [inset of Fig.~\ref{fig:15}(b)], clearly indicating the chaotic nature of the considered ensembles of ICs.

Cases  M75B, M50B and M25B belong to what we call \textit{Arrangement B} and refer again to  coupled SMs with \textit{mixed} (M) $K$ values:  $K_3=6.5$ and  $K_1=K_2=K_4=K_5=3.1$. The ICs of 2D maps with $K=3.1$ lie on regions 4B, 4C and 4D  (see Fig.~\ref{fig:14}, lower row of Fig.~\ref{fig:7} and Table \ref{tb:1}), while the ICs in the central map with  $K_3=6.5$ cover the whole phase space. For these configurations, we find that all cases initially exhibit superdiffusion rates ($\mu \approx 1.5$) [Fig.~\ref{fig:16}(a) and its inset], which are lower than the ones seen  in \textit{Arrangement A} cases [Fig.~\ref{fig:15}(a)],  while after some iterations ($n\approx 10^5$) they show a tendency to gradually decrease these rates towards  normal diffusion behaviors ($\mu \rightarrow 1$). Fig.~\ref{fig:16}(b) renders the respective time evolution of the average ftMLE $\langle \Lambda \rangle$ for these cases.  Again all $\langle \Lambda \rangle$ values seem to converge to the same asymptotic limit with  the fraction $P_C$ of chaotic ICs of the central map practically not affecting the route to this limiting value. The dynamical proximity of all these cases is also seen in the evolution of $\left< \rm{GALI}_{2} \right>$ in the inset of Fig.~\ref{fig:16}(b).

\section{Summary and discussion}\label{sec:sum}

\noindent In this work, we examined the long-term diffusion  and chaoticity properties of single 2D SMs, as well as systems of 5 coupled 2D maps, focusing our attention on parameter values for which the phase space of the respective systems exhibit anomalous diffusion rates due to the presence of AMs of different periods.

For single 2D SMs we reviewed the most typical diffusion  properties for regular (subdiffusion) and chaotic (normal diffusion) orbits, as well as stable/unstable AMs of different periods (with the former exhibiting ballistic transport). For the considered sets of ICs, we have also quantified  their chaoticity, using the GALI$_2$ \eqref{eq:GALI} and ftMLE \eqref{eq:ftMLE} indices. Furthermore, we investigated the map's diffusion properties, by measuring the diffusion exponent $\mu$ [Eq.~\eqref{eq:yvar}] and the effective diffusion coefficient $D_{\rm{eff}}$ \eqref{eq:Deff_sm},  for a range of kick-strength values $K$ and for different final iteration numbers (Fig.~\ref{fig:3}).  A systematic study of the impact of the number of iterations, in the presence of AMs of period $p=1$,  lead to the conclusion that all AMs of period $p=1$ found in the interval $K\in [0,70]$ asymptotically acquire an extreme diffusion rate (ballistic transport with $\mu \approx 2$), even when for relatively short  times they behave closely to normal diffusion  ($\mu \approx 1$, see Fig.~\ref{fig:5}). We have found that the larger the $K$ value of the SM with a $p=1$ AM  is, the more chaotic is its respective phase space as measured by the average ftMLE $\langle \Lambda \rangle$  for large ensembles of ICs [Fig.~\ref{fig:6}(b)], with $\langle \Lambda \rangle$ being accurately described by the law $\langle \Lambda \rangle = \ln (K/2)$ \cite{S04,HKC2019}. Moreover, we considered SMs with kick-strength values where AM of different periods occur. For these cases, we investigated the impact of the choice of the ensemble of ICs (i.e.~practically changing the  fraction $P_C$ of chaotic ICs around the AM) on the  diffusion exponent $\mu$. We found that the more chaotic the ensemble is, the longer it takes for ballistic transport to occur, i.e.~for the diffusion exponent $\mu$ to converge to $\mu=2$ (Fig.~\ref{fig:7}).

In the second part of our analysis, we extended our investigation by exploring diffusion and chaos properties for coupled SMs containing AMs of different periods. We began our analysis by choosing identical ensembles of ICs on a grid covering the entire phase space of each coupled 2D map, with equal kick-strength values $K$ and moderate couplings  ($\beta=10^{-4}$ and $10^{-3}$) for all maps, and identified the main $K$ intervals  where AMs of different periods appear (Fig.~\ref{fig:8}). We derived scaling laws for the regions where AMs of period $p=1$ occur (pronounced peaks in Fig.~\ref{fig:8}), which describe the dependence of those peaks on $K$ [Figs.~\ref{fig:9}(a)-(c)],  and estimated the respective global diffusion exponent $\mu^{*}$ for a range of coupling parameter values $\beta$ [Fig.~\ref{fig:9}(d)]. For AMs of period $p=1$, the main finding is that, as $\beta$ increases the global diffusion exponent decreases, tending towards $\mu^{*}=1$ with a faster rate as $K$ reaches larger values. By setting the kick-strength values at $K_j=K=6.5$, $j=1,2,\ldots,5$ for all SMs (corresponding to the occurrence of the strongest superdiffusion in each 2D map, with $\mu \approx 2$), we obtained a global asymptotic diffusion for the coupled system characterized by  normal diffusion rates ($\mu \approx 1$) [Fig.~\ref{fig:10}(b)]. Furthermore, this convergence towards normal diffusion rates takes place faster as the coupling parameter $\beta$ increases. The trend of the relatively faster convergence to normal diffusion  is also found to be related to the global chaoticity (as measured by the ftMLE $\langle \Lambda \rangle$ and the $\left< \rm{GALI}_{2} \right>$) of the ensembles of ICs. Namely, the average ftMLE value  increases (indicating stronger chaos) monotonically with the increase of the coupling parameter $\beta$ [Figs.\ref{fig:10}(c) and (d)].

We have also performed a similar analysis for different arrangements of the $5$ coupled SMs system, in order to investigate how different configurations of individual 2D maps affect the respective global diffusion rates (see Tables \ref{tb:2}, and \ref{tb:3}). Namely, we sought to find conditions under which the initially ballistic transport (due to the presence of AM of low periods) can be suppressed by the presence of neighboring maps without AMs and ensembles of chaotic ICs exhibiting normal diffusion rates. In order to achieve this goal we performed extensive numerical simulations for two SMs' arrangements.

In the \textit{Arrangement A} type of coupled maps we considered \textit{single} (presence of AMs of period $p=1$ in all SMs, see Fig.~\ref{fig:12}) and \textit{mixed} (presence of AMs of period $p=4$ for the central, $j=3$, SM and of period $p=1$ for the others, see Fig.~\ref{fig:15}) $K$ values for \textit{fixed} coupling-strength $\beta=0.001$. In both cases the diffusion exponent starts at $\mu \approx 2$ (due to the presence of the AMs), then reaches a plateau characterized by $\mu \approx 0$, and finally recovers the initial rate in the \textit{single} case for ensembles with relatively   low fraction $P_C$ of chaotic ICs (case S25A), while it tends to normal diffusion values for relatively high $P_C$ fractions (cases S75A and S50A). For the \textit{mixed} cases (M75A, M50A, M25A) this trend is not clearly evident  [Fig.~\ref{fig:15}(a)]  for the maximum number of iterations we managed to reach. In terms of global chaoticity (quantified through ftMLE $\langle \Lambda \rangle$ computations), we found that all $\langle \Lambda \rangle$ values show a tendency to converge towards a similar positive value. The larger the fraction $P_C$ of chaotic ICs is in the off-center maps, i.e.~respectively cases S75A, S50A, S25A and M75A, M50A, M25A the faster the convergence to this value takes place.

In the \textit{Arrangement B} type of coupled maps we considered \textit{single} (central SM with an AM of period $p=1$ and varying fractions of chaotic ICs, while the remaining SMs have ICs in the whole phase space, see Fig.~\ref{fig:13}) and \textit{mixed} (central SMs with a $p=1$ AM and ICs in the whole phase space, along with varying fractions $P_C$ for the other SMs, which have AMs of period $p=4$, see Fig.~\ref{fig:16}) $K$ values with $\beta=0.001$. For both cases we found  that the diffusion rate starts again at being strongly anomalous with $\mu \approx 2$ (due to the presence of AMs of period $p=1$), while it gradually converges to a normal diffusion rate with $\mu \approx 1$ (cases S75B, S50B, S25B and M75B, M50B, M25B). Furthermore,  in all cases, the average ftMLE $\langle \Lambda \rangle$  asymptotically tend to similar values  [see Fig.~\ref{fig:13}(b) and Fig.~\ref{fig:16}(b)], but in a more uniform manner than in the cases of \textit{arrangement A}.

In summary, the only configuration we found (up to the maximum number of iterations we managed to reach) in which the system globally recovers extreme diffusion rates is case S25A in  Fig.~\ref{fig:12}(a). In all other explored cases, we observed a suppression of the diffusion rate. Thus, this case deserves further investigation for larger number of iterations, in order to determine possible changes in the diffusion process. This is a task we plan to tackle in the future.

Let us also mention here that in our analysis we did not investigate coupling combinations of SMs that may include weakly chaotic or sticky ICs, which may be confined or bounded by invariant tori, and whose diffusion can be more complex (potentially exhibiting subdiffusion rates). Instead, we focused our attention only on cases where chaotic motion dominates the phase space, i.e.~the chosen 2D SMs have relatively large kick-strength values. Taking into account SMs and ensembles of ICs leading to subdiffusion and coupling them together will most likely result in a much richer global  diffusion behavior, which deserves  separate investigations. Finally, we note that we limited our study to coupled SMs with a relatively small number, $N=5$, of 2D maps. This choice allowed us to analyze rather large  ensembles of ICs, as well as  to perform extensive numerical simulations and calculate diffusion exponents and chaos indicators for large numbers of map iterations.

\section{Acknowledgements}
\noindent H.~T.~M.~acknowledges support by a PhD Fellowship from the Science Faculty of the University of Cape Town and partial funding by the UCT Incoming International Student Award. H.~T.~M.~and Ch.~S.~thank the High Performance Computing facility of the University of Cape Town, as well as the Centre for High Performance Computing (CHPC) of South Africa for providing the computational resources needed for obtaining the numerical results of this work.

\section{Author contributions} \label{sec:auth_contr}
\noindent H.~T.~M.: Software, Formal analysis, Investigation,  Writing - Original draft, Visualization. T.~M.: Conceptualization, Methodology, Software, Validation, Formal analysis, Investigation, Writing - Original draft, Reviewing and Editing. Ch.~S.: Conceptualization, Methodology, Validation, Formal analysis, Investigation, Writing - Reviewing and Editing, Supervision, Project administration, Funding acquisition.

\bibliography{MMS_SMs_2021.bib}

\end{document}